\documentclass[sigconf]{acmart}

\usepackage{booktabs} 
\usepackage[normalem]{ulem}

\usepackage{cleveref}

\crefformat{section}{\S#2#1#3} 
\crefformat{subsection}{\S#2#1#3}
\crefformat{subsubsection}{\S#2#1#3}
\crefformat{appendix}{\S#2#1#3}

\usepackage[export]{adjustbox}

\usepackage{datetime}
\usepackage{url}
\usepackage{hyperref}
\usepackage[ruled, noend]{algorithm2e}
\usepackage{float}
\usepackage{color}
\usepackage{multirow}
\usepackage{soul}
\usepackage{array}
\usepackage{subcaption}
\usepackage{soul}
\usepackage{balance}
\usepackage{enumitem}
\usepackage{graphicx}
\usepackage{amsmath}
\usepackage{kotex}
\usepackage{mathtools}
\usepackage{physics}

\usepackage{stmaryrd}
\usepackage{trimclip}
\usepackage{xcolor}
\usepackage{fontawesome}
\usepackage{makecell}

\makeatletter
\DeclareRobustCommand{\shortto}{%
  \mathrel{\mathpalette\short@to\relax}%
}

\DeclareMathOperator*{\K}{\mathbb{K}}
\DeclareMathOperator*{\I}{\mathbb{I}}
\DeclareMathOperator*{\Z}{\mathbb{Z}}
\DeclareMathOperator*{\E}{\mathbb{E}}
\DeclareMathOperator*{\Var}{\mathbb{V}ar}
\DeclareMathOperator*{\Cov}{\mathbb{C}ov}

\DeclareMathOperator*{\argmax}{arg\,max}
\DeclareMathOperator*{\argmin}{arg\,min}

\newcommand{\short@to}[2]{%
  \mkern2mu
  \clipbox{{.3\width} 0 0 0}{$\m@th#1\vphantom{+}{\shortrightarrow}$}%
  }
\makeatother

\newcolumntype{L}[1]{>{\raggedright\let\newline\\\arraybackslash\hspace{0pt}}m{#1}}

\newcommand{\eat}[1]{}

\usepackage{tikz}
\usetikzlibrary{calc}

\usepackage{algorithmicx}
\algdef{SE}[DOWHILE]{Do}{doWhile}{\algorithmicdo}[1]{\algorithmicwhile\ #1}%

\newtheorem{theorem}{\bf Theorem}

\newtheorem{definition2}{\bf Definition}
\newtheorem{proposition2}{\bf Proposition}
\newtheorem{lemma2}{\bf Lemma}

\newtheorem{p-rule}{\bf Rule}

\newtheorem{example}{\bf Example}

\copyrightyear{2023}
\acmYear{2023}
\setcopyright{acmlicensed}
\acmConference[PODS '23] {Proceedings of the 42nd ACM SIGMOD-SIGACT-SIGAI Symposium on Principles of Database Systems}{June 18--23, 2023}{Seattle, WA, USA.}
\acmBooktitle{Proceedings of the 42nd ACM SIGMOD-SIGACT-SIGAI Symposium on Principles of Database Systems (PODS '23), June 18--23, 2023, Seattle, WA, USA}
\acmPrice{15.00}
\acmISBN{979-8-4007-0127-6/23/06}
\acmDOI{10.1145/3584372.3588676}

\ccsdesc[500]{Information systems~Database query processing}
\ccsdesc[300]{Mathematics of computing~Probability and statistics}

\keywords{Join size estimation, Uniform sampling, Worst-case optimal}

\begin{document}

\let\oldnl\nl
\newcommand{\nonl}{\renewcommand{\nl}{\let\nl\oldnl}}

\DeclarePairedDelimiter{\ceil}{\lceil}{\rceil}

\newcommand{\revtext}[1]{#1}
\newcommand{\todotext}[1]{#1}

\newcommand{\revision}[1]{#1}
\newcommand{\rerevision}[1]{#1}
\newcommand{\coverletter}[1]{#1}
\newcommand{\second}[1]{#1}

\DeclarePairedDelimiter\floor{\lfloor}{\rfloor}

\newcommand{\FuncName}[1]{\textsc{{#1}}}
\newcommand{\QueryNum}{\abs{\mathcal{M}}}
\newcommand{\CondQueryNum}{c(q, g|M)}
\newcommand{\AGM}{AGM(q)}
\newcommand{\SampleSpace}{\revtext{\Omega}}
\newcommand{\sappend}{\revtext{s \uplus s_I}}
\newcommand{\SamplerTree}{\mathcal{T}}
\newcommand{\Queries}{\mathcal{H}}
\newcommand{\Samples}{\mathcal{S}}
\newcommand{\database}{\mathcal{D}}
\newcommand{\TablePair}{\mathcal{T}}
\newcommand{\Matches}{\mathcal{M}}
\newcommand{\Branch}{b \cdot w(i-1)}
\newcommand{\truecard}{card(Q)}
\newcommand{\estcard}{\widehat{card}(Q)}
\newcommand{\truesel}{P(Q)}
\newcommand{\estsel}{\widehat{P}(Q)}
\newcommand{\trueprob}{P(X)}
\newcommand{\estprob}{\widehat{P}(X)}
\newcommand{\hypergraph}{\mathcal{H}}
\newcommand{\hypernodes}{\mathcal{V}}
\newcommand{\samplednodes}{\mathcal{B}}
\newcommand{\outputnodes}{\mathcal{O}}
\newcommand{\hyperedges}{\mathcal{E}}
\newcommand{\reducedgraph}{\mathcal{G}}
\newcommand{\reducedGHD}{\tau}
\newcommand{\leaves}{\mathcal{L}}
\newcommand{\J}{\mathcal{J}}
\newcommand{\Algorithm}{\mathcal{A}}
\newcommand{\GHD}{\mathcal{T}}
\newcommand{\bag}{\chi}
\newcommand{\lambdahat}{\widehat{\lambda}}
\renewcommand{\O}{\tilde{O}}
\newcommand{\OUT}{\text{OUT}}
\newcommand{\IN}{\text{IN}}

\useunder{\uline}{\ul}{}
\newcommand{\veryshortarrow}[1][3pt]{\mathrel{%
   \hbox{\rule[\dimexpr\fontdimen22\textfont2-.2pt\relax]{#1}{.4pt}}%
   \mkern-4mu\hbox{\usefont{U}{lasy}{m}{n}\symbol{41}}}}

\newcommand{\DataGraph}{g}
\newcommand{\DataVertexSet}{V_G}
\newcommand{\DataEdgeSet}{E_G}
\newcommand{\QueryGraph}{q}
\newcommand{\QueryVertex}{u}
\newcommand{\DataVertex}{v}
\newcommand{\Degree}[1]{Deg({#1})}
\newcommand{\Neighbor}[1]{N({#1})}
\newcommand{\Edge}[2]{\left({#1},{#2}\right)}
\newcommand{\VSet}[1]{V({#1})}
\newcommand{\ESet}[1]{E({#1})}
\newcommand{\GESet}[1]{GE({#1})}
\newcommand{\ELabel}[1]{L({#1})}

\newcommand{\QueryO}{O^*_{H}}
\newcommand{\DataO}{\tilde{O}}

\newif\ifFullVersion

\def\FullVersion{\let\ifFullVersion=\iftrue}
\def\ShortVersion{\let\ifFullVersion=\iffalse}
\FullVersion

\newcommand{\OddCycleSampler}{\FuncName{OddCycleSampler}\xspace}
\newcommand{\StarSampler}{\FuncName{StarSampler}\xspace}

\newcommand{\SSTE}{\FuncName{SSTE}\xspace}
\newcommand{\SUST}{\FuncName{SUST}\xspace}
\newcommand{\SSTES}{\FuncName{SSTE*}\xspace}
\newcommand{\DO}{\FuncName{DO}\xspace}
\newcommand{\GenericSampler}{\FuncName{GroupByCardEst}\xspace}
\newcommand{\RecursiveSampler}{\FuncName{RecursiveSampler}\xspace}
\newcommand{\GJSampler}{\FuncName{GJ-Sample}\xspace}
\newcommand{\DRS}{\FuncName{DRS}\xspace}
\newcommand{\GHDSampler}{\FuncName{GHDCardEst}\xspace}
\newcommand{\Enumerator}{\FuncName{Enumerator}\xspace}
\newcommand{\EnumerateTuples}{\FuncName{EnumerateTuples}\xspace}
\newcommand{\SamplingEstimator}{\FuncName{Sampler}\xspace}
\newcommand{\SamplingEstimatorOpt}{\FuncName{SamplingEstimatorOpt}\xspace}
\newcommand{\Yan}{\FuncName{Yannakakis}\xspace}
\newcommand{\ExactWeight}{\FuncName{ExactWeight}\xspace}
\newcommand{\AggYan}{\FuncName{AggroYannakakis}\xspace}
\newcommand{\SimpleAggYan}{\FuncName{SimpleAggroYannakakis}\xspace}
\newcommand{\GenericJoin}{\FuncName{GenericJoin}\xspace}
\newcommand{\GHDJoin}{\FuncName{GHDJoin}\xspace}
\newcommand{\AggGHDJoin}{\FuncName{AggroGHDJoin}\xspace}
\newcommand{\SamplingFramework}{\FuncName{GenericCardEst}\xspace}
\newcommand{\DetermineSampleSpace}{\FuncName{DetermineSampleSpace}\xspace}
\newcommand{\DetermineSampleSize}{\FuncName{DetermineSampleSize}\xspace}
\newcommand{\DetermineSampleSizeIfNone}{\FuncName{DetermineSampleSizeIfNone}\xspace}
\newcommand{\DetermineNextSampleSize}{\FuncName{DetermineNextSampleSize}\xspace}
\newcommand{\DetermineSampleSpaceAndSize}{\FuncName{GetSampleSpaceAndSize}\xspace}
\newcommand{\Distribution}{\FuncName{GetDistribution}\xspace}
\newcommand{\Partition}{\FuncName{PartitionAttributes}\xspace}
\newcommand{\GetSubset}{\FuncName{GetNonEmptySubset}\xspace}
\newcommand{\SampleFrom}{\FuncName{SampleFrom}}
\newcommand{\Reject}{\FuncName{RejectAndAdjust}}

\newcommand{\LW}{\FuncName{LW}\xspace}
\newcommand{\BN}{\FuncName{BN}\xspace}
\newcommand{\NARU}{\FuncName{NARU}\xspace}
\newcommand{\QuickSel}{\FuncName{QuickSel}\xspace}
\newcommand{\DBEstP}{\FuncName{DBEst++}\xspace}
\newcommand{\JCT}{\FuncName{JCT}\xspace}
\newcommand{\PGMJoins}{\FuncName{PGMJoins}\xspace}
\newcommand{\Fauce}{\FuncName{Fauce}\xspace}
\newcommand{\FCN}{\FuncName{FCN}\xspace}
\newcommand{\FCNMSCN}{\FuncName{FCN+MSCN}\xspace}
\newcommand{\FCNPool}{\FuncName{FCN+Pool}\xspace}
\newcommand{\FCNConv}{\FuncName{FCN+Conv}\xspace}
\newcommand{\TreeLSTM}{\FuncName{TreeLSTM}\xspace}
\newcommand{\MSCN}{\FuncName{MSCN}\xspace}
\newcommand{\EtoE}{\FuncName{E2E}\xspace}
\newcommand{\DQM}{\FuncName{DQM}\xspace}
\newcommand{\DQMQ}{\FuncName{DQM-Q}\xspace}
\newcommand{\DQMD}{\FuncName{DQM-D}\xspace}
\newcommand{\SPN}{\FuncName{SPN}\xspace}
\newcommand{\FSPN}{\FuncName{FSPN}\xspace}
\newcommand{\DeepDB}{\FuncName{DeepDB}\xspace}
\newcommand{\DeepDBJCT}{\FuncName{DeepDB+JCT}\xspace}
\newcommand{\DeepDBNARU}{\FuncName{DeepDB+NARU}\xspace}
\newcommand{\DeepDBJCTNARU}{\FuncName{DeepDB+JCT+NARU}\xspace}
\newcommand{\DeepDBBN}{\FuncName{DeepDB+BN}\xspace}
\newcommand{\NeuroCard}{\FuncName{NeuroCard}\xspace}
\newcommand{\FLAT}{\FuncName{FLAT}\xspace}
\newcommand{\BayesNet}{\FuncName{BayesNet}\xspace}
\newcommand{\BayesCard}{\FuncName{BayesCard}\xspace}
\newcommand{\Astrid}{\FuncName{Astrid}\xspace}
\newcommand{\UAE}{\FuncName{UAE}\xspace}
\newcommand{\EMP}{\FuncName{EMP}\xspace}
\newcommand{\AWR}{\FuncName{AWR}\xspace}
\newcommand{\Postgres}{\FuncName{PostgreSQL}\xspace}
\newcommand{\WordVec}{\FuncName{Word2Vec}\xspace}

\newcommand{\Catalog}{\FuncName{Catalog}\xspace}
\newcommand{\CSET}{\FuncName{C-SET}\xspace}
\newcommand{\GCARE}{\FuncName{G-CARE}\xspace}
\newcommand{\Alley}{\FuncName{Alley}\xspace}
\newcommand{\AlleyS}{\FuncName{Alley+}\xspace}
\newcommand{\AlleyPM}{\FuncName{Alley+TPI}\xspace}
\newcommand{\IBJS}{\FuncName{IBJS}\xspace}
\newcommand{\JSUB}{\FuncName{JSUB}\xspace}
\newcommand{\BSK}{\FuncName{BSK}\xspace}
\newcommand{\AlleyD}{\FuncName{AlleyD}\xspace}
\newcommand{\AlleyTPI}{\FuncName{Alley+TPI}\xspace}
\newcommand{\AlleyNaive}{\FuncName{Alley+Naive}\xspace}
\newcommand{\WanderJoin}{\FuncName{WanderJoin}\xspace}
\newcommand{\ST}{\FuncName{SSTE}\xspace}
\newcommand{\Estimator}{\FuncName{OurMethod}\xspace}
\newcommand{\LogicBlox}{\FuncName{LogicBlox}\xspace}
\newcommand{\EmptyHeaded}{\FuncName{EmptyHeaded}\xspace}
\newcommand{\Graphflow}{\FuncName{Graphflow}\xspace}
\newcommand{\Recursive}{\FuncName{Recur}}
\newcommand{\ChooseRandomVertices}{\FuncName{ChooseRandomVerticesWOR}}
\newcommand{\ChooseRandomVertex}{\FuncName{ChooseRandomVertex}}
\newcommand{\TurboHOM}{\FuncName{TurboHom++}\xspace}
\newcommand{\TurboH}{\FuncName{TH}\xspace}
\newcommand{\ChooseSamplingOrder}{\FuncName{ChooseSamplingOrder}}
\newcommand{\SamplePotentialEmbedding}{\FuncName{SamplePotentialEmbedding}}
\newcommand{\SamplePotentialEmbeddings}{\FuncName{RandomWalkWithIntersect}}
\newcommand{\SamplePotentialEmbeddingsAndEstimate}{\FuncName{SampleAndEstimate}}
\newcommand{\SampleOnePotentialEmbedding}{\FuncName{SampleOnePotentialEmbedding}}
\newcommand{\SampleOneEdge}{\FuncName{SampleOneEdge}}
\newcommand{\SampleOneVertex}{\FuncName{SampleOneVertex}}
\newcommand{\TangledPatternMining}{\FuncName{TangledPatternMining}}
\newcommand{\ExtendPattern}{\FuncName{ExtendPattern}}
\newcommand{\CalculateDomains}{\FuncName{CalculateDomains}}
\newcommand{\HasUnindexedSubgraph}{\FuncName{HasSubgraphWithEmptyDomain}}
\newcommand{\GetMininumDomainsFromSubgraphs}{\FuncName{GetMininumDomainsFromSubgraphs}}
\newcommand{\GetFailureRate}{\FuncName{GetFailureRate}}
\newcommand{\SearchDomainsRecursive}{\FuncName{SearchDomainsRecursive}}
\newcommand{\SearchDomains}{\FuncName{SearchDomains}}
\newcommand{\PriorityFirstSearchFrom}{\FuncName{PriorityFirstSearchFrom}}
\newcommand{\Intersect}{\FuncName{Intersect}}
\newcommand{\RandomWalkWithDO}{\FuncName{RandomWalkWithDO}}
\newcommand{\HalfIntegralCover}{\FuncName{HalfIntegralOptimalCover}}

\newcommand{\GraphGrep}{\FuncName{GraphGrep}\xspace}
\newcommand{\GRAMI}{\FuncName{GraMi}\xspace}
\newcommand{\VF}{\FuncName{VF2}\xspace}
\newcommand{\QuickSI}{\FuncName{QuickSI}\xspace}
\newcommand{\GraphQL}{\FuncName{GraphQL}\xspace}
\newcommand{\gIndex}{\FuncName{gIndex}\xspace}
\newcommand{\SPath}{\FuncName{SPath}\xspace}
\newcommand{\TurboISO}{\FuncName{TurboIso}\xspace}
\newcommand{\CFL}{\FuncName{CFL-Match}\xspace}
\newcommand{\DAF}{\FuncName{DAF}\xspace}
\newcommand{\RDFX}{\FuncName{RDF-3X}\xspace}
\newcommand{\Trinity}{\FuncName{Trinity.RDF}\xspace}
\newcommand{\gStore}{\textsf{gStore}\xspace}
\newcommand{\gMark}{\textsf{gMark}\xspace}
\newcommand{\WJ}{\textsf{WJ}\xspace}
\newcommand{\CS}{\textsf{CS}\xspace}

\newcommand{\Chau}{Chaudhuri's Method}
\newcommand{\SumRDF}{\FuncName{SumRDF}\xspace}
\newcommand{\Bernoulli}{\FuncName{BernoulliSampling}\xspace}
\newcommand{\EndBiased}{\FuncName{End-biasedSampling}\xspace}
\newcommand{\Correlated}{\FuncName{CorrelatedSampling}\xspace}
\newcommand{\IMPR}{\FuncName{IMPR}\xspace}

\newcommand{\AIDS}{\textsf{AIDS}\xspace}
\newcommand{\HPRD}{\textsf{HPRD}\xspace}
\newcommand{\Youtube}{\textsf{Youtube}\xspace}
\newcommand{\LUBM}{\textsf{LUBM}\xspace}
\newcommand{\DBLP}{\textsf{DBLP}\xspace}
\newcommand{\Human}{\textsf{Human}\xspace}
\newcommand{\YAGO}{\textsf{YAGO}\xspace}
\newcommand{\Epinions}{\textsf{Epinions}\xspace}
\newcommand{\Amazon}{\textsf{Amazon}\xspace}
\newcommand{\Google}{\textsf{Google}\xspace}
\newcommand{\WatDiv}{\textsf{WatDiv}\xspace}
\newcommand{\UniProt}{\textsf{UniProt}\xspace}
\newcommand{\TPCDS}{\textsf{TPC-DS}\xspace}
\newcommand{\TPCH}{\textsf{TPC-H}\xspace}
\newcommand{\IMDB}{\textsf{IMDB}\xspace}
\newcommand{\DMV}{\textsf{DMV}\xspace}
\newcommand{\Census}{\textsf{Census}\xspace}
\newcommand{\Synthetic}{\textsf{Synthetic}\xspace}

\newcommand{\JOBsmall}{\textsf{IMDB-small}\xspace}
\newcommand{\JOBmedium}{\textsf{IMDB-medium}\xspace}
\newcommand{\JOBlarge}{\textsf{IMDB-large}\xspace}
\newcommand{\JOBbench}{\textsf{JOB-bench}\xspace}
\newcommand{\TPCDSbench}{\textsf{TPC-DS-bench}\xspace}
\newcommand{\TPCDSlarge}{\textsf{TPC-DS-large}\xspace}

\newcommand{\JOBlight}{\textsf{JOB-light}\xspace}
\newcommand{\JOBsmalls}{\textsf{IMDB-{small}-syn}\xspace}
\newcommand{\JOBM}{\textsf{JOB-M}\xspace}
\newcommand{\JOBmediums}{\textsf{IMDB-{medium}-syn}\xspace}
\newcommand{\JOBlarges}{\textsf{IMDB-{large}-syn}\xspace}
\newcommand{\TPCDSsub}{\textsf{TPC-DS-sub}\xspace}
\newcommand{\TPCDSlarges}{\textsf{TPC-DS-{large}-syn}\xspace}

\newcommand{\SynSingle}{\textsf{Syn-Single}\xspace}
\newcommand{\SynMulti}{\textsf{Syn-Multi}\xspace}

\newcommand{\Lfun}{\mathrm{L}}
\newcommand{\Vset}{\mathrm{V}}

\newcommand{\vertex}{v}
\newcommand{\none}{-}
\newcommand{\QueryTree}{q'}
\newcommand{\convert}{convert}
\newcommand{\precondition}{precondition}
\SetKw{Continue}{continue}

\newcommand{\GMO}{mo_{g}}
\newcommand{\PartialOrders}{PO}

\newcommand{\SWITCH}[1]{\STATE \textbf{switch} (#1)}
\newcommand{\ENDSWITCH}{\STATE \textbf{end switch}}
\newcommand{\CASE}[1]{\STATE \textbf{case} #1\textbf{:} \begin{ALC@g}}
\newcommand{\ENDCASE}{\end{ALC@g}}
\newcommand{\CASELINE}[1]{\STATE \textbf{case} #1\textbf{:} }
\newcommand{\DEFAULT}{\STATE \textbf{default:} \begin{ALC@g}}
\newcommand{\ENDDEFAULT}{\end{ALC@g}}
\newcommand{\DEFAULTLINE}[1]{\STATE \textbf{default:} }

\def\QEDmark{\ensuremath{\square}}
\def\endproof{\hfill\QEDmark}

\newcommand{\bluecomment}[1]{}
\newcommand{\redcomment}[1]{\tcc{#1}}

\newcommand{\spellcheck}[1]{#1}
\newcommand{\redtext}[1]{#1}
\newcommand{\bluetext}[1]{#1}
\newcommand{\newredtext}[1]{#1}
\newcommand{\newbluetext}[1]{#1}

\newcommand{\profwshan}[1]{{#1}}
\newcommand{\profwshangreen}[1]{{#1}}
\newcommand{\profwshangr}[1]{#1} 

\newcommand{\full}[1]{#1}

\ifFullVersion
\newcommand{\todelete}[1]{#1}
\newcommand{\release}[1]{#1}
\else
\newcommand{\todelete}[1]{\textcolor{red}{#1}}
\newcommand{\release}[1]{#1}
\fi

\newcommand{\iseo}[1]{#1} 

\LinesNumbered
\SetAlgoCaptionSeparator{.}
\SetKwProg{Fn}{Function}{}{end}
\SetKwFor{uForEach}{foreach}{do}{}
\SetStartEndCondition{ (}{) }{)}
\SetNlSty{texttt}{}{:}
\SetArgSty{}

\def\checkmark{\tikz\fill[scale=0.4](0,.35) -- (.25,0) -- (1,.7) -- (.25,.15) -- cycle;}

\def\ojoin{\setbox0=\hbox{$\bowtie$}%
  \rule[-.0ex]{.25em}{.4pt}\llap{\rule[\ht0]{.25em}{.4pt}}}
\def\leftouterjoin{\mathbin{\ojoin\mkern-5.8mu\bowtie}}
\def\rightouterjoin{\mathbin{\bowtie\mkern-5.8mu\ojoin}}
\def\fullouterjoin{\mathbin{\ojoin\mkern-5.8mu\bowtie\mkern-5.8mu\ojoin}}

\title{Guaranteeing the $\O\big(\frac{AGM}{\OUT}\big)$ Runtime for Uniform Sampling and Size Estimation over Joins}

\author{Kyoungmin Kim}
\email{kmkim@dblab.postech.ac.kr}
\affiliation{%
  \institution{Pohang University of Science and Technology (POSTECH)}
  \city{Pohang}
  \country{Republic of Korea}
}

\author{Jaehyun Ha}
\email{jhha@dblab.postech.ac.kr}
\affiliation{%
  \institution{Pohang University of Science and Technology (POSTECH)}
  \city{Pohang}
  \country{Republic of Korea}
}

\author{George Fletcher}
\email{g.h.l.fletcher@tue.nl}
\affiliation{%
  \institution{Eindhoven University of Technology (TU/e)}
  \city{Eindhoven}
  \country{Netherlands}
}

\author{Wook-Shin Han}
\authornote{corresponding author}
\authornotemark[1]
\email{wshan@dblab.postech.ac.kr}
\affiliation{%
  \institution{Pohang University of Science and Technology (POSTECH)}
  \city{Pohang}
  \country{Republic of Korea}
}

\begin{abstract}
\second{We propose a new method for estimating the number of answers $\OUT$ of a small join query $Q$ in a large database $D$, and for uniform sampling over joins. Our method is the first to satisfy all the following statements.}
\vspace*{-0.1cm}
\second{
\begin{itemize}
  \item Support arbitrary $Q$, which can be either acyclic or cyclic, and contain binary and non-binary relations.
  \item Guarantee an arbitrary small error with a high probability always in $\O\big(\frac{AGM}{\OUT}\big)$ time, where $AGM$ is the AGM bound (an upper bound of $\OUT$), and $\O$ hides the polylogarithmic factor of input size.
\end{itemize}
}

\second{We also explain previous join size estimators in a unified framework. All methods including ours rely on certain indexes on relations in $D$, which take linear time to build offline. Additionally,} we extend our method using generalized hypertree decompositions \second{(GHDs)} to achieve \second{a lower complexity than $\O\big(\frac{AGM}{\OUT}\big)$ when $\OUT$ is small}, and present optimization techniques for improving estimation efficiency and accuracy.
\end{abstract}

\maketitle



\section{Introduction}\label{sec:introduction}




\second{The evaluation of join queries is one of the most fundamental problems in databases \cite{LeapFrog, NPRR, GenericJoin, WanderJoin, ExactWeight, EmptyHeaded}.
In theory, reducing the time complexity of evaluation algorithms has been the main goal \cite{Yannakakis, GenericJoin, gottlob2016hypertree, PANDA}. This also applies to the counting problem \cite{AJAR}, which we study in this paper. Given a join query $Q$ and a database $D$, an \emph{answer} of $Q$ in $D$ is a mapping from free variables of $Q$ to attribute values in $D$ (or inversely as stated in \cite{conjunctive2}) constrained by the relations in $Q$. For example, if a database $D$ of a social network has a binary relation $R(A,B)$ of friendship and a tuple $(a,b) \in R(A,B)$ indicates that two people $a, b$ are friends, finding all triples $(a, b, c)$ of people who are all friends, is equivalent to finding the answers of the triangle join query ($R(B,C)$ and $R(A,C)$ are renamed relations of $R(A,B)$):}

\vspace*{-0.5cm}
\begin{equation}\label{eq:example:query}
\begin{split}
\second{Q(A,B,C) = R(A,B) \wedge R(B,C) \wedge R(A,C)}
\end{split}
\end{equation}
\vspace*{-0.5cm}

The complexity of evaluating join queries is commonly expressed as $\O(\IN^w + \OUT)$ where $\IN$ is the \emph{input size} and $w (\geq 1)$ is a \textit{width} of $Q$ \second{(see \cref{sec:background} for definition of $\IN$ and example widths). While acyclic queries can be evaluated with $w = 1$ \cite{Yannakakis}, cyclic queries have higher $w$ values, i.e., more difficult to solve. If $w$ is \emph{fractional edge cover number} $\rho$ (see Definition \ref{def:agm}), $\IN^w$ becomes $AGM$, the AGM bound \cite{AGM} which is an upper bound of $\OUT$. Instead of evaluating join queries, the counting problem is to compute $|Q|$, the join size or the number of answers (e.g., triples above). It has its own application for answering COUNT queries in DBMS. Then, the complexity can be reduced from $\O(\IN^w + \OUT)$ to $\O(\IN^w)$ \cite{AJAR}.}

\second{Approximate counting problem is to approximate $|Q|$. Approximate algorithms are alternatively used whenever the exact counting is expensive and approximation is enough. In practice, a famous application is cost-based query optimization which requires efficient approximation of thousands of sub-queries \cite{JOB}. In theory, going beyond the polynomial time algorithms, approximate algorithms established the complexity inversely proportional to $\OUT$, under the requirement that guarantees an arbitrary small error with a high probability (see \cref{sec:background} for formal definitions). In a limited setting where every relation in $Q$ is binary and identical (e.g., as in (\ref{eq:example:query})), the complexity of $\O(\IN^w)$ has reduced to $\O\big(\frac{\IN^w}{\OUT}\big)$ \cite{SSTE, SUST}. However, in a general setting where $Q$ contains higher-arity relations, the complexity has reduced to $\O\big(\frac{\IN^{w+1}}{\OUT}\big)$ or $\O\big(\frac{\IN^{w}}{\OUT} + \IN\big)$ only \cite{GJSampler}, with an additional multiplicative or additive factor of $\IN$. Furthermore, all these methods have $w = \rho$ only (i.e., $\O\big(\frac{\IN^w}{\OUT}\big)$ = $\O\big(\frac{AGM}{\OUT}\big)$), and are not extended to other widths such as \emph{fractional hypertree width} (see Definition \ref{def:fhtw}) smaller than $\rho$. In this paper, we propose a new method that achieves $\O\big(\frac{AGM}{\OUT}\big)$ complexity for $Q$ containing any-arity relations, under the same $w = \rho$. As a corollary, ours achieve a linear or even constant complexity for a sufficiently large $\OUT$. We also extend ours to incorporate fractional hypertree width by using generalized hypertree decompositions (GHDs), achieving a lower complexity than $\O\big(\frac{AGM}{\OUT}\big)$ when $\OUT$ is small.} 

\second{A closely related problem to approximate counting is uniform sampling, which has its own important applications such as training set generation for machine learning \cite{ExactWeight}. Our method also supports uniform sampling over joins. We define the problems and technical concepts in \cref{sec:background}, explain related work in \cref{sec:related} and overview of our results in \cref{sec:ours}. We present our method of size estimation and uniform sampling over joins in \cref{sec:sampling} including a unified framework of previous work, extend our method using GHDs in \cref{sec:generalized}, propose optimization techniques in \cref{sec:optimization} and future work in \cref{sec:future}. More details of existing algorithms and extension of our algorithms to join-project (a.k.a. conjunctive) queries are presented in Appendix.}

\section{\second{Technical Background}} \label{sec:background}

\noindent\underline{\second{Hypergraph.}} \second{A join query $Q$ is commonly represented as a \emph{hypergraph} $\hypergraph(\hypernodes, \hyperedges)$}, where \second{\emph{hypernodes}} $\hypernodes$ is the set of \second{variables in $Q$} and \second{\emph{hyperedges}} $\hyperedges \subseteq 2^{\hypernodes}$ \cite{GenericJoin}.
\second{Each hyperedge $F \in \hyperedges$ is a subset of $\hypernodes$ and specifies the relation $R(F)$ (or $R_F$ for brevity) in $Q$.
For example, the query in (\ref{eq:example:query}) has $\hypernodes = \{A, B, C\}$ and $\hyperedges = \{\{A,B\}, \{B,C\}, \{A,C\}\}$.}
\second{For any $I \subset \hypernodes$, we define two notations used throughout the paper: $\hyperedges_{I} \coloneqq \{F \in \hyperedges | F \cap I \neq \emptyset\}$ and $\hyperedges_{I, \pi} \coloneqq \{ \pi_I{F} | F \in \hyperedges_I \}$, i.e., the set of hyperedges in $\hyperedges$ that intersect with $I$ and the projection of each edge in $\hyperedges_I$ onto $I$.}

\second{In relational algebra, $Q$ is equivalent to} $\Join_{F \in \hyperedges} R_F = {R_{F_1} \Join ... \Join R_{F_{\abs{\hyperedges}}}}$. {Here, join ($\Join$) is a \textit{commutative} binary operator between relations, allowing us to use $\Join_{F \in \hyperedges}$.}
Table \ref{table:operators} summarizes the operators used in this paper.

\vspace*{-0.2cm}
\begin{table}[htp]
\scriptsize
\vspace*{-0.2cm}
\caption{\revision{Operators used in the paper.}}
\vspace{-0.3cm}
\small
\begin{tabular}{ |c|l| }
 \hline
 $\Join$ & natural join operator \\ \hline
 $\ltimes$ & semi-join operator \\ \hline
 $\pi$ & projection operator \\ \hline
 $\sigma$ & selection operator \\ \hline
 $\uplus$ & list concatenation operator \\ \hline
 
\end{tabular}\label{table:operators}
\end{table}

\vspace*{-0.2cm}
\second{Since all variables in $\hypernodes$ are free (output) variables in join queries, join queries are subsumed by \emph{full} conjunctive queries. For a more general class of join-project (a.k.a. conjunctive) queries, the set of output variables is a subset of $\hypernodes$. We assume $Q$ is a join query and extend to join-project queries in \cref{appendix:conjunctive}.}






\noindent\underline{\second{Input and output size.}} We denote $\revision{\IN(\hypergraph, D)}$ = {$\max_{F \in \hyperedges} \abs{R_F}$} as the input size \second{($|R_F|$ is the number of tuples in $R_F$)} and \revision{$\OUT(\hypergraph, D) = |Q|$}, the number of query \second{answers}, as the output size of \second{$Q$ (or $\hypergraph$) in $D$}. We drop $\hypergraph$ \revision{and $D$} if the context is clear, \second{as well as in upcoming notations.}


\noindent\underline{\second{Complexity of join.}} The \second{time complexity} of join \second{evaluation} algorithms is commonly expressed as $\O(\IN^w + \OUT)$ where $w$ is a \emph{width} of $Q$, and $\O$ hides a polylogarithmic factor of IN \cite{GenericJoin}. \second{We assume that $Q$ is extremely small compared to $D$} and regard $|\hypernodes|$ and $|\hyperedges|$ as constants as in \cite{GenericJoin, SSTE, GJSampler}.
\emph{Worst-case optimal} \revision{(WCO)} algorithms \bluetext{such as \GenericJoin \cite{GenericJoin} \revision{(Algorithm \ref{alg:genericjoin})}} achieve $w = \rho(\hypergraph, \second{D})$, \second{called \emph{fractional edge cover number}:}

\begin{definition2}
\label{def:agm}
 \emph{\cite{GenericJoin}:} The fractional edge cover number ${\rho(\hypergraph, \second{D})}$ is the optimal solution of the linear program (LP): $\min_{{\{x_F | F \in \hyperedges\}}}$ $\sum_{F \in \hyperedges} x_F $ $\cdot \log_{\IN} \abs{R_F}$ s.t. $\sum_{F \ni v} x_F \geq 1 : \forall v \in \hypernodes$ and $0 \leq x_F \leq 1 : \forall F \in \hyperedges$.
\end{definition2}


It is well known that $\prod_{F \in \hyperedges} \abs{R_F}^{x_F}$ is an upper bound of $\OUT$ for any fractional edge cover $x$ satisfying the constraints of the LP {\cite{AGM, AGMOLD}}.
The optimal value, $\IN^{\rho} = \min_{x} \prod_{F \in \hyperedges} \abs{R_F}^{x_F}$, is called the AGM bound of $\hypergraph$ \second{on $D$} denoted as \second{$AGM(\hypergraph, D)$} {\cite{AGM}}. 
\second{For example, the query in (\ref{eq:example:query}) has $\IN = |R_{\{A,B\}}| = |R_{\{B,C\}}| = |R_{\{A,C\}}|$ and $\rho = 3/2$ from the optimal solution $x_{\{A,B\}} = x_{\{B,C\}} = x_{\{A,C\}} = 1/2$ for the LP in Definition \ref{def:agm}. Therefore, its AGM bound is $\IN^{\rho} = \IN^{3/2}$.}



\noindent\underline{\second{Generalized hypertree decomposition.}} As a special case, the \Yan algorithm \cite{Yannakakis} \second{achieves $w = 1$ if $Q$ is acyclic.}
It builds a \emph{join tree} by assigning each hyperedge $F \in \hyperedges$ to a tree node and performs dynamic programming (DP)-like join over the nodes' outputs (i.e., $R_F$'s) in a bottom-up manner.

To bridge the gap between $w = {\rho}$ and $w = 1$ \second{when $Q$ is cyclic, generalized hypertree decompositions (GHDs, see Definition \ref{def:ghd}) \cite{GHD} extend} the tree for arbitrary joins by assigning multiple hyperedges to a tree node, where the DP-like join is performed over these nodes' outputs (i.e., \second{answers of the corresponding sub-query}) computed by \revision{WCO} algorithms. This way, $w$ decreases from ${\rho}$ to \second{\emph{fractional hypertree width}} ${fhtw}$ (see Definition \ref{def:fhtw}). We refer to a comprehensive survey \cite{GenericJoin} for more details about these concepts and \cref{appendix:algorithm} for our explanations of algorithms.


\begin{definition2}
\label{def:ghd}
\emph{\bluetext{\cite{GHD}}:} A GHD of $\hypergraph = (\hypernodes, \hyperedges)$ is a pair ($\GHD, \bag$) where
1) $\GHD$ is a tree of $(\hypernodes_\GHD, \hyperedges_\GHD)$,
2) $\bag : \hypernodes_\GHD \rightarrow 2^\hypernodes$, 
3) for each $F \in \hyperedges$, there exists $t \in \hypernodes_\GHD$ s.t. $F \subseteq \bag(t)$, 
and 4) for each $v \in \hypernodes$, $\{t \in \hypernodes_\GHD | v \in \bag(t)\}$ forms a non-empty connected subtree of $\GHD$.
\end{definition2}

Here, $\bag(t)$ is called the \emph{bag} of $t$, and 4) is called the \emph{running intersection property}.
Each node $t$ corresponds to a sub-hypergraph \revision{$\hypergraph_{\bag(t)}$} of $\hypergraph$ where $\revision{\hypergraph_{\bag(t)}} \coloneqq (\bag(t), \hyperedges_{\bag(t), \pi})$.
We then define the fractional hypertree width \bluetext{\cite{Fhtw}}.

\begin{definition2}
\label{def:fhtw}
\emph{\bluetext{\cite{Fhtw}}:} The fractional hypertree width 1) of a GHD ($\GHD, \bag$) of $\hypergraph$, $fhtw(\GHD, \hypergraph)$, is defined as $\max_{t \in \hypernodes_\GHD} \rho(\revision{\hypergraph_{\bag(t)}})$, and 2) of $\hypergraph$, $fhtw(\hypergraph)$, is defined as $\min_{(\GHD, \bag)} fhtw(\GHD, \hypergraph)$.
\end{definition2}

Given a GHD ($\GHD, \bag$), \GHDJoin \bluetext{\cite{AJAR}} performs \GenericJoin for each $\revision{\hypergraph_{\bag(t)}} (t \in \hypernodes_\GHD)$ and \Yan \bluetext{\cite{Yannakakis}} on these results along the join tree \bluetext{(see Algorithms \ref{alg:simpleaggyan}-\ref{alg:ghdjoin} in \cref{appendix:algorithm:yan}-\ref{appendix:algorithm:ghdjoin})}. Since \Yan runs in linear - $\O(\IN + \OUT)$ - time for $\alpha$-acyclic join queries \bluetext{\cite{AlphaAcyclic}}, the total runtime of \GHDJoin is $\O(\IN^{fhtw(\GHD, \hypergraph)} + \OUT)$, since {the input size for \Yan} is $\O\big(\IN^{\max_{t \in \hypernodes_\GHD} \rho(\revision{\hypergraph_{\bag(t)}})}\big)$ after \GenericJoin.
It is sufficient to execute a brute-force algorithm to find $fhtw(\hypergraph)$ since the number of possible GHDs is bounded by the query size. 

\noindent\underline{\second{Counting.}} \second{For the counting problem}, the \second{complexity} is reduced to $\O(\IN^{\second{w}})$ without the +OUT term. This is a direct result of \revtext{Joglekar et al.} \cite{AJAR}, solving a more general problem of \second{evaluating aggregation queries} (e.g., count, sum, max). \second{We explain in detail in \cref{subsec:generalized:aggregations}.}

\noindent\underline{\second{Approximate counting.}} \second{Approximate counting achieves even smaller complexities.} 
\second{The primary goal of approximate counting, especially using sampling, is to obtain} an arbitrarily small error with a high probability, formally defined as \cite{SSTE, GJSampler}:


\vspace*{-0.2cm}
\begin{definition2}
\label{def:approximation}
\emph{\cite{SSTE}:} 
For a given $\hypergraph$, error bound $\epsilon \in (0,1)$, and probability threshold $\delta \in (0,1)$, if $P(\abs{Z-\OUT} \geq \epsilon \OUT) \leq \delta$ for a random variable $Z$ approximating $\OUT$, then $Z$ is a $(1\pm\epsilon)$-approximation of $\OUT$.
\end{definition2}
\vspace*{-0.2cm}


\second{The challenge is, how far can we reduce the complexity while achieving the above goal. Assume that we approximate $\OUT$ using an \emph{unbiased} estimator with a random variable $Y$, i.e., $\E[Y] = \OUT$, where $\E[Y]$ is the \emph{expectation} of $Y$. Assume that the \emph{variance} $\Var[Y]$ of $Y$ is bounded and an upper bound $U_{var}$ is known, and the time complexity of \emph{instantiating} (or computing) $Y$ is upper-bounded by $U_{time}$. Then, we can trade off the approximation accuracy and efficiency by controlling the number of runs $N$ for sampling. Formally, let $Z = (Y_1 + Y_2 + ... + Y_N)/N$ where every $Y_i$ ($1 \leq i \leq N$) is an equivalent random variable to $Y$. Then, $\E[Z] = \E[Y] = \OUT$ and $\Var[Z] = \Var[Y]/N$. Hence, by setting $N = \frac{U_{var}}{\epsilon^2 \delta \OUT^2}$, we can achieve the following from the Chebyshev's inequality:}

\vspace*{-0.2cm}
\begin{equation}\label{eq:chebyshev}
\begin{split}
\second{P(|Z - \OUT| \geq \epsilon \OUT) \leq \frac{\Var[Z]}{\epsilon^2 \E[Z]^2} = \frac{\Var[Y]}{N \epsilon^2 \OUT^2} \leq \frac{U_{var}}{N \epsilon^2 \OUT^2} = \delta}
\end{split}
\end{equation}
\vspace*{-0.2cm}

\second{Since the complexity of computing each $Y_i$ is bounded by $U_{time}$, the complexity of computing $Z$ is bounded by $N \cdot U_{time} = \frac{U_{var} U_{time}}{\epsilon^2 \delta \OUT^2}$. Therefore, the key challenge is to implement an unbiased estimator with $U_{var} U_{time}$ as smallest as possible. 
$\epsilon$ and $\delta$ are regarded as constants as in $\O$ \cite{SSTE, GJSampler}.
One can easily replace $\frac{1}{\delta}$ in $N$ with $\log \frac{1}{\delta}$ using the median trick \cite{MedianTrick}.}

\noindent\underline{\second{Uniform sampling.}} \second{If every query answer has the same probability $p$ to be sampled, then the sampler is a uniform sampler \cite{GJSampler, SUST}. Since there are $|Q| = \OUT$ answers, $p$ can be at most $1/\OUT$. If $p < 1/\OUT$, the sampling has the probability to \emph{fail} (to sample any of the query answers) which is $1 - p \cdot \OUT$.} 

\second{For acyclic queries, uniform sampling can be done in $\O(1)$ time after $\O(\IN)$-time preprocessing \cite{ExactWeight}, by additionally storing the intermediate join sizes of tuples during the bottom-up DP in \Yan (see \cref{appendix:algorithm:yan} for more details).}

\vspace*{-0.2cm}
\section{\second{Related Work}} \label{sec:related}

\second{In the context of approximate counting (and uniform sampling, which is closely related), there have been two lines of work: 1) determining the classes of queries of tractable and intractable cases and 2) reducing the degree of polynomial-time complexity for tractable cases, especially for join or join-project queries.}

\second{The first class of work widens the class of queries that admit \emph{fully polynomial-time randomized approximation scheme} (FPRAS) and \emph{fully polynomial-time almost uniform sampler} (FPAUS) \cite{conjunctive1}. Arenas et al. \cite{conjunctive1}  showed that every class of conjunctive queries with bounded hypertree width admits FPRAS and FPAUS. Focke et al. \cite{conjunctive2} extended this to conjunctive queries with disequalities and negations, under some relaxations from FPRAS. These works focus on showing the existence of a polynomial-time algorithm over a wider class of queries, but not on reducing the exact degree of the polynomial complexity.}

\second{The second class of work \cite{SSTE, SUST, GJSampler, CycleCounting, StarCounting, CliqueCounting} focuses on reducing the degree for a specific range of queries, and even further to making the complexity inversely proportional to $\OUT$, e.g., $\O\big(\frac{\IN^w}{\OUT}\big)$.
However, they all consider $w = \rho$ only (i.e., $\O\big(\frac{\IN^w}{\OUT}\big)$ is $\O\big(\frac{AGM}{\OUT}\big)$), and the complexity $\O\big(\frac{AGM}{\OUT}\big)$ has been achieved for limited classes of queries.}
Assadi et al. \cite{SSTE} proposed a \underline{S}imple \underline{S}ublinear-\underline{T}ime \underline{E}stimator (\SSTE for short) \second{with the complexity of $\O\big(\frac{AGM}{\OUT}\big)$ for \emph{unlabeled graph queries}, i.e., where $Q$ contains \emph{identical binary} relations only as in the query in (\ref{eq:example:query}).}
However, we found that a missing factor in the proof by Assadi et al. \cite{SSTE} prevented \SSTE from satisfying the $\O\big(\frac{AGM}{\OUT}\big)$ bound and thus notified the author to modify the proof 
\second{
\ifFullVersion
(see \cref{appendix:proof}).
\else
(see \S{E} in our full paper \cite{AAA}).
\fi
}
They also proposed some conjectures on labeled graphs {(i.e., removing the identity assumption)}, but 1) no concrete algorithms were given and 2) {the proposed bound for labeled graphs is larger than $\O\big(\frac{AGM}{\OUT}\big)$ (see \cref{appendix:decomposition:labeled})}. Therefore, it is not known whether these methods can be extended to labeled graphs {with $\O\big(\frac{AGM}{\OUT}\big)$ time.}
Fichtenberger et al. \cite{SUST} extended \SSTE to sampling \underline{S}ubgraphs \underline{U}niformly in \underline{S}ublinear \underline{T}ime (\SUST for short) for uniform sampling while achieving the same complexity. Here, they said sublinear from the assumption that $\OUT$ is large as $\Omega(\IN^{\rho-1})$.
Unlike join processing, it is unnecessary to scan the inputs for each query that requires $\O(\IN)$ time. 

Kim et al. \cite{Alley} proposed \Alley, a hybrid method that combines sampling and synopsis. We \revision{analyze} the sampling part of \Alley since analyzing the synopsis is out of scope. \Alley solves the approximate subgraph counting problem (but not uniform sampling) for \emph{labeled} graphs, \second{where the relations in $Q$ are binary but not necessarily identical.} \second{The complexity, however, is $\O(AGM)$ instead of $\O\big(\frac{AGM}{\OUT}\big)$, which we explain why in \cref{sec:sampling}.}


\second{For a more general setting where relations in $Q$ can have higher arities, Chen \& Yi \cite{GJSampler} proposed \GJSampler that also performs uniform sampling, and achieved $\O\big(\frac{\IN \, AGM}{\OUT}\big)$ complexity.}
The additional factor of $\IN$ \second{compared to \SSTE and \SUST} results from \second{computing sample probabilities (or weights) prior to sampling, where the number of candidates is $\O(\IN)$. We explain in detail in \cref{sec:sampling}}.
Therefore, they do not achieve $\O\big(\frac{AGM}{\OUT}\big)$ complexity for arbitrary join queries and leave it as an open problem \cite{GJSampler}.
Note that it is important to remove the $\IN$ factor, for $\IN$ can be higher than $\frac{AGM}{\OUT}$ when relation sizes are highly skewed. For example, for a triangle query $\hypergraph$ with $\hyperedges = \{F_1, F_2, F_3\}$, $|R_{F_1}| = N^{10}, |R_{F_2}| = |R_{F_3}| = N$, and $\OUT = N$, then $\IN$ is $N^{10}$ and $\frac{AGM}{\OUT}$ is $\frac{N^2}{N} = N$ for $\rho = \frac{1}{5}$.

\second{We notice that Deng et al. \cite{concurrent} have independently pursued the same primary objective as ours, which is to attain $\O\big(\frac{AGM}{\OUT}\big)$ for uniform sampling and size estimation over joins. Both papers have been accepted in this PODS. The authors recursively 1) partition the input space into a constant number of sub-spaces and 2) sample a sub-space, to reduce the number of sample candidates from $\O(\IN)$ to $\O(1)$. Our approach differs from theirs since we do not compute probabilities for candidates prior to sampling nor modify the set of candidates. Instead, we exploit known probability distributions to sample a candidate.}

\vspace*{-0.2cm}
\section{Our Results} \label{sec:ours}



We first analyze {existing} sampling-based \second{approximation algorithms} in a new unified framework in \cref{sec:sampling}.
We present a new algorithm \second{that bounds $U_{var} U_{time}$ by $\O(AGM \cdot \OUT)$ for arbitrary join queries for the first time, achieving the complexity of $\O\big(\frac{AGM}{\OUT}\big)$ (see \cref{sec:background}).}
To \second{avoid pre-sampling overheads} in \GJSampler, we propose \emph{degree-based rejection sampling} (\DRS) \second{that first samples a sample space from a \emph{meta} sample space and then samples a value from the sampled space, where the value can be \emph{rejected} based on its \emph{degree}. We explain the details in \cref{sec:sampling}.}
The results of \second{Chen \& Yi} then directly hold for \second{DRS}. The removed overhead reduces the complexity down to $\O\big(\frac{AGM}{\OUT}\big)$ for arbitrary joins. We also extend \DRS using GHDs in \cref{sec:generalized}. \second{The following theorems state our results which we prove in the subsequent sections.}

\begin{theorem}
\label{th:drs}
\second{\DRS performs $(1\pm\epsilon)$-approximation of $\OUT$ in $\O\big(\frac{AGM}{\OUT}\big)$ time.}
\end{theorem}



\begin{theorem}
\label{th:drs_ghd}
\second{If $\OUT$ is small enough, \DRS with GHDs can perform $(1\pm\epsilon)$-approximation of $\OUT$ with a lower complexity than $\O\big(\frac{AGM}{\OUT}\big)$. A sufficient condition is when $\OUT$ is $\O(\IN^{\rho - fhtw})$.}
\end{theorem}



Additionally, we extend \Alley to support arbitrary join queries in \cref{sec:sampling} and \revision{a core lemma} of \SSTE/\SUST to hold for labeled graphs in \cref{appendix:decomposition}. \second{We discuss our extension to join-project (a.k.a. conjunctive queries) in \cref{appendix:conjunctive} and extension of the $\O(1)$-time uniform sampler \cite{ExactWeight} in \cref{sec:background} to support cyclic queries using our framework in \cref{appendix:algorithm:yan}.}

\section{\revtext{Achieving $\O\big(\frac{AGM}{\OUT}\big)$ Bound}} \label{sec:sampling}

\subsection{Generic Sampling-based Framework} \label{subsec:sampling:framework}


{We can classify} existing sampling-based methods into three groups: \second{variable}-at-a-time (a.k.a., vertex-at-a-time) \cite{Alley, GJSampler}, edge-at-a-time \cite{WanderJoin}, and {component-at-a-time} \cite{SSTE, SUST}.
The first/second group samples a \second{variable}/hyperedge at a time until every \second{variable}/hyperedge in $\hypergraph$ is sampled, while the third group samples a component, either a cycle with an odd number of edges or a star, at a time, having a larger sample granularity than the other two groups.


To analyze these groups in a unified manner, we propose a powerful sampling-based estimation framework \SamplingFramework \revtext{(Algorithm \ref{alg:sampler})}. 
It returns an unbiased estimate of \second{$\OUT(\hypergraph_s)$}, where $\hypergraph_s$ is the \emph{residual hypergraph} given {a current sample} $s$ which is initially empty. \second{Formally, $\hypergraph_s \coloneqq \Join_{F \in \hyperedges_{\rerevision{\outputnodes}}} \pi_{\rerevision{\outputnodes}} (R_F \ltimes s)$ and $\hypergraph_\emptyset = \hypergraph$ for $s = \emptyset$.}

\begin{algorithm} [htb]
\caption{{\SamplingFramework}(\revision{$\hypergraph(\hypernodes, \hyperedges), \rerevision{\outputnodes}, s$})} \label{alg:sampler}
\small{
    \KwIn{\revision{Query hypergraph $\hypergraph$, \rerevision{\second{variables} $\outputnodes$ to sample}, and current sample $s$}}

    \If{\rerevision{$\outputnodes = \emptyset$}}
    { \label{alg:sampler:if}
        \Return 1 \label{alg:sampler:return1} \\
    }
    $I \leftarrow \GetSubset(\rerevision{\outputnodes})$ \redcomment{\mbox{$I \neq \emptyset$}} \label{alg:sampler:I}
    
    $\SampleSpace_I, k \leftarrow \DetermineSampleSpaceAndSize(I, \hypergraph, s)$ \label{alg:sampler:samplespace}
    
    $P \leftarrow {\Distribution}(\SampleSpace_I)$ \label{alg:sampler:P} \redcomment{$P(\perp) = 1 - \sum_{s_I \in \SampleSpace_I} P(s_I)$}
    
    $S_I \leftarrow {\SampleFrom}(\SampleSpace_I, k, P)$ \redcomment{$|S_I| = k$} \label{alg:sampler:sample}
    
    ${S_I, k (= |S_I|), \revision{\{P(s_I) | s_I \in S_I\}} \leftarrow \Reject(S_I, P, \hypergraph, s)}$ \label{alg:sampler:reject}
    
    \Return $\frac{1}{k} \sum_{s_I \in S_I} \frac{1}{P(s_I)} \cdot {\I[s_I \in \, \Join_{F \in \hyperedges_I} \pi_{I} \revision{(R_F \ltimes s)}]} \cdot {\SamplingFramework(\rerevision{\hypergraph, \outputnodes \setminus I, \sappend)}}$ \label{alg:sampler:return2}
}
\end{algorithm}

\second{An invariance is that} $s \in \, \Join_{F \in \hyperedges_{\rerevision{\hypernodes \setminus \outputnodes}}} \pi_{\rerevision{\hypernodes \setminus \outputnodes}} R_F$, i.e., $s$ is an answer to the sub-query of $Q$ on the \emph{bound} variables $\hypernodes \setminus \outputnodes$. This is a key to prove Lemma \ref{lemma:unbiased}. 
\ifFullVersion
\second{Due to space limitations, we prove all lemmas and propositions in \cref{appendix:proof}.}
\else
\second{Due to space limitations, we prove all lemmas and propositions in \S{E} of our full paper \cite{AAA}.}
\fi


\begin{lemma2}\label{lemma:unbiased}
\SamplingFramework returns an unbiased estimate of $\abs{\Join_{F \in \hyperedges} R_F}$ \second{for $\outputnodes = \hypernodes$ and $s = \emptyset$}. 
\end{lemma2}

\second{We now explain each line of Algorithm \ref{alg:sampler}.}
\rerevision{If there is no \second{variable} to sample for, just return one in Lines \ref{alg:sampler:if}-\ref{alg:sampler:return1}.}
{Otherwise}, Line \ref{alg:sampler:I} \rerevision{takes a subset $I$ of $\outputnodes$ to sample at this step}. $I$ is 1) a singleton set for \second{variable}-at-a-time methods, 2) an edge or its projection for edge-at-a-time methods, or 3) a part of a component for component-at-a-time methods.
Line \ref{alg:sampler:samplespace} defines the \emph{sample space} $\SampleSpace_I$ and \emph{sample size} $k$.
Line \ref{alg:sampler:P} computes $P$, a probability distribution over \revision{$\SampleSpace_I \cup \{\perp\}$; $\perp$ is a null tuple that does not belong to any join \second{answer}.}
{Line \ref{alg:sampler:sample} samples $S_I$ from \revision{$\SampleSpace_I$} where $|S_I| = k$.}
{Line \ref{alg:sampler:reject} optionally rejects samples in $S_I$. Here, $S_I$, $k$, and \revtext{$\{P(s_I) | s_I \in S_I\}$} are adjusted according to the accepted samples; \revtext{$P(s_I)$} is multiplied by the probability of accepting $s_I$.}
{\SamplingFramework in Line \ref{alg:sampler:return2} returns an unbiased estimate of the residual query $\hypergraph_{\sappend}$. The unbiasedness of the final return value at Line \ref{alg:sampler:return2} is guaranteed by the inverse probability factor, $\frac{1}{P(s_I)} \I[\cdot]$, which is the key idea of the Horvitz-Thompson estimator \cite{Horvitz}.} $\I[c]$ denotes the indicator which is 1 if $c$ is true and 0 otherwise.
\subsection{\second{Query Model}} \label{subsec:ours:query_model}



\second{\SamplingFramework relies on a certain query model used in previous WCO join and sampling algorithms \cite{GenericJoin, LeapFrog, NPRR, SSTE, GJSampler}, following the property testing model \cite{PropertyTesting}. To avoid any confusion with the join query $Q$, we refer to the queries here as operations. For a relation $R_F$, the query model allows that $\pi_{I}(R_F \ltimes s)$ \second{in \SamplingFramework} can be readily obtained in $\O(1)$ time. We explain the underlying index structures that enable this later in \cref{subsec:ours:data_model} for brevity.}

\second{The same indexes provide the following $\O(1)$-time operations as well.} Here, $T$ can be a projection/selection of another relation, which is also a relation.
\begin{itemize}
\setlength{\itemindent}{-.3in}
  \item Degree($T, s$): $|T \ltimes s|$ for a relation $T$ and a tuple $s$
  \item \mbox{Access($T, I, i$): $i$-th element of $\pi_I(T)$ for a relation $T$ and attributes $I$}
  \item Exist($T, r$): test whether or not a row $r$ exists in $T$
  \item Sample($T$): uniformly sample a tuple $r$ from $T$
\end{itemize}
Eden et al. \cite{TriCounting} have developed their algorithms using operations on binary tables. 
The above four operations are generalizations of theirs to $n$-ary tables.  

\vspace*{-0.2cm}
\subsection{\revtext{State-of-the-art Sampling-based Estimators}} \label{subsec:sampling:instance}
 
\revtext{We explain state-of-the-art sampling-based estimators as instances of \SamplingFramework.}

\noindent {\textbf{\WanderJoin}} \cite{WanderJoin} is a practical edge-at-a-time method. 
The edges in $\hyperedges$ are first ordered by an index $i \in [1,|\hyperedges|]$. At the $i$-th step, $I = \rerevision{F_i \cap \outputnodes}$, 
$\SampleSpace_I = \revision{\pi_I (R_{F_i} \ltimes s)}$, $k = 1$, and $P(s_I) = \frac{1}{|\SampleSpace_I|}$. That is, one tuple $s_I$ is uniformly sampled from $\SampleSpace_I$. 
\revision{Note that \rerevision{$\outputnodes$ can reduce to $\emptyset$} before proceeding to the last edge, e.g., the triangle query. However, \cite{WanderJoin} proceeds first with sampling for every edge and then checks the join predicates, resulting in unnecessary computations.}



\noindent {\textbf{\AlleyS.}} \Alley \cite{Alley} is a \second{variable}-at-a-time method \revision{(i.e., $|I| = 1$)} and assumes that every $R_F$ is binary. {We here explain \AlleyS, our extension of \Alley to $n$-ary relations:} $\SampleSpace_I = \cap_{F \in \hyperedges_I} \pi_I \revision{(R_F \ltimes s)}$, {$k = \ceil{b \cdot |\SampleSpace_I|}$ for some fixed $b \in (0,1]$}, and $P(s_I) = \frac{1}{|\SampleSpace_I|}$. Here, the sampling is done \emph{without} replacement. \second{Due to the heavy intersection ($\cap$) operation, the runtime of \AlleyS is $\O(AGM(\hypergraph))$ (Proposition \ref{prop:time:alley}).}
\second{At an extreme case of} $b = 1$, \AlleyS becomes an instance of \GenericJoin.


\noindent \textbf{\SSTE} \cite{SSTE} and \textbf{\SUST} \cite{SUST} {are {component-at-a-time} methods.} Assuming that every $R_F$ is binary and identical, 
{Lemma \ref{lemma:decomposition} in \cref{appendix:decomposition}} decomposes $\hypergraph$ into a set of {components, where each component is either an odd cycle or a star.}
\SSTE and \SUST slightly differ in sampling odd cycles and stars, \revision{and \SUST performs uniform sampling over joins. We explain the details in \cref{appendix:decomposition} due to space limitations.}

\noindent {\textbf{\GJSampler}} \cite{GJSampler} \revision{is a \second{variable}-at-a-time method (i.e., $|I| = 1$).} Here, $\SampleSpace_I = \pi_I \revision{(R_{\second{F^{GJ}}} \ltimes s)}$ where $\second{F^{GJ}} = \argmin_{F \in \hyperedges_I} \revision{|\pi_I (R_F \ltimes s)|}$. This \second{avoids} the intersection operation in \AlleyS, allowing \GJSampler to achieve $\O\big(\frac{AGM(\hypergraph)}{\OUT}\big)$ runtime for sampling instead of $\O(AGM(\hypergraph))$ \cite{GJSampler}.
$k = 1$ and $P(s_I) = {\frac{AGM(\hypergraph_{\sappend})}{AGM(\hypergraph_s)}}$ where $AGM(\hypergraph_s) = \prod_{F \in \hyperedges_{\rerevision{\outputnodes}}} |R_F \ltimes s|^{x_F}$ and $AGM(\hypergraph_{\sappend}) = \prod_{F \in \hyperedges_{\rerevision{\outputnodes} \setminus I}} |R_F \ltimes (\sappend)|^{x_F}$, where $x$ is a fractional edge cover of the original query $\hypergraph$ \cite{GJSampler}.
Then, due to {Lemma \ref{lemma:decomposition:gjsampler} in \cref{appendix:lemmas}}, $\sum_{s_I \in \SampleSpace_I} P(s_I) \leq 1$. 


More importantly, setting $P(s_I) = {\frac{AGM(\hypergraph_{\sappend})}{AGM(\hypergraph_s)}}$ enables \GJSampler to perform uniform sampling \rerevision{over $Q$}, with $\frac{1}{AGM(\hypergraph)}$ probability for each \second{answer of} $Q$.
Let $s$ be a final sample that reaches Line \ref{alg:sampler:return1}, and {$\hypergraph_{s_1}, \hypergraph_{s_2}, ..., \hypergraph_{s_n}$} be the series of hypergraphs {$\hypergraph_{\sappend}$} at Line \ref{alg:sampler:return2}. Then, \revision{$AGM(\hypergraph_{s_n}) = 1$ from $\second{\hyperedges_{\outputnodes}} = \hyperedges_{\rerevision{\emptyset}} = \emptyset$ \second{for $\hypergraph_{s_n}$}, and}

\vspace*{-0.4cm}
\begin{equation} \label{eq:gjsample}
\begin{split}
P(s) = \frac{AGM(\hypergraph_{s_1})}{AGM(\hypergraph)} ... \frac{AGM(\hypergraph_{s_n})}{AGM(\hypergraph_{s_{n-1}})} = \frac{AGM(\hypergraph_{s_n})}{AGM(\hypergraph)} = \frac{1}{AGM(\hypergraph)}.
\end{split}
\end{equation}
\vspace*{-0.4cm}

Since $|Q| = \OUT$, a call to \SamplingFramework succeeds to sample \second{any answer of} $Q$ with $\frac{\OUT}{AGM(\hypergraph)}$ probability and fails with $1 - \frac{\OUT}{AGM(\hypergraph)}$ probability. 
Note that if $P(s_I)$ is set to $\OUT(\hypergraph_{\sappend})/\OUT(\hypergraph_s)$, then $P(s) = \frac{1}{\OUT}$, {so every call to \SamplingFramework will succeed.} However, computing $\OUT(\hypergraph_{\sappend})$ for every $s_I \in \SampleSpace_I$ is intractable, which is the reason for \GJSampler to use ${\frac{AGM(\hypergraph_{\sappend})}{AGM(\hypergraph_s)}}$ that can be computed in $\O(1)$ time.


However, computing {${\frac{AGM(\hypergraph_{\sappend})}{AGM(\hypergraph_s)}}$} values at Line \ref{alg:sampler:P} takes $\O(|\SampleSpace_I|) = \O(|\pi_I \revision{(R_{\second{F^{GJ}}} \ltimes s)}|) = \O(\IN)$ time. This results in an $\IN$ factor in \GJSampler's runtime, $\O\big(\frac{\IN \cdot AGM(\hypergraph)}{\OUT}\big)$ \cite{GJSampler}.
In order to remove this $\IN$ factor, \revtext{Chen \& Yi} \cite{GJSampler} separated out \emph{sequenceable} queries, where all $P(s_I)$ values required in the sampling procedure can be precomputed in $\O(\IN)$ time.
This results in $\O\big(\IN + \frac{AGM(\hypergraph)}{\OUT}\big)$ runtime for sequenceable queries and still $\O\big(\frac{\IN \cdot AGM(\hypergraph)}{\OUT}\big)$ for non-sequenceable queries.
Determining whether a query is sequenceable or not requires a brute-force algorithm \revtext{of $\O(1)$ time,} since the search space is bounded by the query size \cite{GJSampler}.


\vspace*{-0.2cm}
\subsection{\revtext{Beyond the State of the Art with Degree-based Rejection Sampling}} \label{subsec:sampling:ours}

We propose \emph{degree-based rejection sampling} (DRS) to \second{avoid} computing ${\frac{AGM(\hypergraph_{\sappend})}{AGM(\hypergraph_s)}}$ values in \GJSampler while achieving \second{a similar $P(s_I)$, only} a constant factor smaller. While we set $|I| = 1$ and $k = 1$ as in \GJSampler, we sample $F^*$ \second{(a counterpart of $F^{GJ}$ in \GJSampler) uniformly from $\hyperedges_I$ and let $\SampleSpace_I = \pi_I(R_{F^*} \ltimes s)$. Therefore, $\{\pi_I(R_F \ltimes s) \,|\, F \in \hyperedges_I\}$ is our \emph{meta} sample space.}
For each $s_I \in \SampleSpace_I$ and $F \in \hyperedges_I$, we define the \emph{relative degree} of $s_I$ in \revision{$R_F \ltimes s$} as \revision{$rdeg_{F,s}(s_I) \coloneqq \frac{|R_F \ltimes (\sappend)|}{|R_F \ltimes s|}$}.

In order to use $P(s_I) = rdeg_{F^*,s}(s_I)$ in Line \ref{alg:sampler:P}, we 1) uniformly sample a row $t$ from $R_{F^*} \ltimes s$ and 2) let $s_I = \pi_I t$ in Line \ref{alg:sampler:sample}, without computing $P(s_I)$ for every $s_I \in \SampleSpace_I$.
Then, $rdeg_{F,s}(s_I)$ for every $F \in \hyperedges_I$ is computed, and $s_I$ is \emph{rejected} if $F^* \neq \argmax_{F \in \hyperedges_I} \revision{rdeg_{F,s}(s_I)}$ (Line \ref{alg:sampler:reject}).
We first assume that one $F$ has a higher \revision{$rdeg_{F,s}(s_I)$} than all other edges in $\hyperedges_I$.
Then, \revision{every $P(s_I)$} is multiplied by $\frac{1}{|\hyperedges_I|}$ at Line \ref{alg:sampler:reject}, resulting in $P(s_I) = \frac{1}{|\hyperedges_I|} \revision{rdeg_{F^*,s}(s_I)}$.
Line \ref{alg:sampler:reject} further uses $p = \frac{AGM(\hypergraph_{\sappend})}{\revision{rdeg_{F^*,s}(s_I)} \cdot AGM(\hypergraph_s)}$ ($\leq 1$ from Lemma \ref{lemma:psmall}) as the keeping probability of $s_I$ to make the final $P(s_I) = \frac{1}{|\hyperedges_I|} \frac{AGM(\hypergraph_{\sappend})}{AGM(\hypergraph_s)}$.


\vspace*{-0.1cm}
\begin{lemma2}\label{lemma:psmall}
$\frac{AGM(\hypergraph_{\sappend})}{AGM(\hypergraph_s)} \leq \max_{F \in \hyperedges_I} rdeg_{F,s}(s_I) = rdeg_{F^*,s}(s_I)$ \second{if $s_I \in \Join_{F \in \hyperedges_I} \pi_{I} (R_F \ltimes s)$}.
\end{lemma2}
\vspace*{-0.1cm}

\revtext{From (\ref{eq:gjsample})}, any final sample $s$ that reaches Line \ref{alg:sampler:return1} has $P(s) = \frac{1}{AGM(\hypergraph)} \prod_{I \subset \hypernodes, |I| = 1} \frac{1}{|\hyperedges_I|}$ probability to be sampled, indicating a uniform sampling \rerevision{over $Q$}. Note that the added term $\prod_{I} \frac{1}{|\hyperedges_I|}$ is regarded as a constant since it depends on the query size only. If we break our initial assumption so that $m > 1$ edges in $\hyperedges_I$ can have the same maximum relative degree, the $P(s_I)$ above will increase by $m$ times. Therefore, we simply decrease the keeping probability $p$ by $m$ times to preserve $P(s_I) = \frac{1}{|\hyperedges_I|} \frac{AGM(\hypergraph_{\sappend})}{AGM(\hypergraph_s)}$.

\vspace*{-0.1cm}
\begin{example}
{Let $N$ and $M$ be two arbitrary numbers. Let $R(A,B) = \{(i, 1) \,|\, i \in [1,M]\} \cup \{(1, j) \,|\, j \in [2,N]\}$ be a binary relation and $T(A,C)$ is obtained by renaming $B$ to $C$ in $R$. Suppose we have a join query $Q(A,B,C) = R(A,B) \Join T(A,C)$. For ease of explanation, we use a hyperedge $F$ and its corresponding relation $R_F$ interchangeably. Then, we have an optimal fractional edge cover of $Q$ as $x_R = x_T = 1$.
We choose $I = \{A\}$, $I = \{B\}$, and $I = \{C\}$ in turn for \second{DRS}.
For $I = \{A\}$ and $s = \emptyset$, $\hyperedges_I = \{R, T\}$, so we sample a relation from $\hyperedges_I$. Assume that $R$ is sampled, and \revtext{then} a row $t = (1,1)$ is sampled from $R$. Then, $s_I = 1$ and $\revision{rdeg_{R,s}(s_I)} = \frac{N}{N+M-1} = \revision{rdeg_{T,s}(s_I)}$, so $R$ is not rejected. 
Since $R$ and $T$ tie w.r.t. $s_I$, we keep $s_I$ with probability $p = \frac{1}{2} \frac{1}{\revision{rdeg_{R,s}(s_I)}} \frac{AGM(\hypergraph_{\sappend})}{AGM(\hypergraph_s)} = \frac{1}{2} \revtext{\frac{1}{\revision{rdeg_{R,s}(s_I)}} \frac{N^2}{(N+M-1)^2}}$. Then, $P(s_I) = P(s_{\{A\}}) = \revision{rdeg_{R,s}(s_I)} \frac{1}{2} \frac{1}{\revision{rdeg_{R,s}(s_I)}} \frac{N^2}{(N+M-1)^2}$ $= \frac{1}{2} \frac{N^2}{(N+M-1)^2}$. 
Next, for $I = \{B\}$ and $s = \{A = 1\}$, $\hyperedges_I = \{R\}$, and $\revision{rdeg_{R,s}(s_I)} = \frac{1}{N}$ for any $s_I \in \pi_I \revision{(R \ltimes s)}$. Then, $P(s_{\{B\}}) = \frac{1}{|\hyperedges_I|} \frac{AGM(\hypergraph_{\sappend})}{AGM(\hypergraph_s)} = \frac{N}{N^2} = \frac{1}{N}$. Similarly, $P(s_{\{C\}}) = \frac{1}{N}$ for $I = \{C\}$. In total, a final sample $s$ that reaches Line \ref{alg:sampler:return1} of \SamplingFramework has $P(s) = P(s_{\{A\}}) P(s_{\{B\}}) P(s_{\{C\}}) = \frac{1}{2} \frac{1}{(N+M-1)^2} = \prod_{I} \frac{1}{|\hyperedges_I|} \frac{1}{AGM(\hypergraph)}$.
}
\end{example}

\vspace*{-0.3cm}
\subsection{Unified Analysis} \label{subsec:sampling:analysis}

This section analyzes the bounds of variance and runtime \second{($U_{var}$ and $U_{time}$ in \cref{sec:background})} of sampling-based estimators in a unified manner.
Note that Lemma \ref{lemma:unbiased} already states the unbiasedness of estimators. 
\second{Theorem \ref{th:drs} can be proved from Propositions \ref{prop:variance:ours} and \ref{prop:time:ours}; $U_{var} U_{time}$ is $\O(AGM \cdot \OUT)$.}



We first define two random variables, 1) \revtext{$Z_{\hypergraph_s}$} for the output of $\SamplingFramework$ given $s$ and 2) $Z = \sum_{1 \leq i \leq N} \frac{Z^i_\hypergraph}{N}$ for our final estimate. Here, $Z^i_\hypergraph$'s are independent and identical to $Z_\hypergraph$ \second{$= Z_{\hypergraph_\emptyset}$}, and $N$ is the number of initial calls to $\SamplingFramework$ with $s = \emptyset$.
Let \revtext{$T_{\hypergraph_s}$} be the random variable for the number of core operations \revision{(including the four operations in \cref{subsec:ours:query_model})} in \SamplingFramework and $T$ for the total runtime of obtaining $Z$.
Then, $\Var[Z] = \frac{\Var[Z_\hypergraph]}{N}$ and $\E[T] = N \cdot \E[T_\hypergraph]$.




\begin{proposition2}\label{prop:variance:sste}
\todotext{$\Var[Z_\hypergraph] \leq 2^{d} \cdot AGM(\hypergraph) \cdot \OUT$ for \SSTE \revtext{and \SUST} where $d$ is the number of odd cycles and stars in $\hypergraph$.}
\end{proposition2}

\begin{proposition2}\label{prop:variance:alley}
$\Var[Z_\hypergraph] \leq \frac{t^{|\hypernodes|} - t}{t - 1} \cdot \OUT^2$ for \AlleyS where $t = \frac{2(1-b)}{b}$.
\end{proposition2}


The upper bound $U_{var}$ of \AlleyS explains the unique property of \AlleyS, that $\Var[Z_\hypergraph]$ approaches to 0 as $b$ approaches to 1.
That is, $\lim_{b \to 1} U_{var} = 0$ since $\lim_{b \to 1} t = 0$. In fact, this $U_{var}$ is \revision{tighter} than the \revtext{original} bound $U_{org} = \frac{b}{1-b}(\frac{1}{b^{|\hypernodes|}} - 1) \cdot \OUT^2$ proved in \cite{Alley}, where $\lim_{b \to 1} U_{org} = \lim_{b \to 1} \frac{1+b+b^2+...+b^{|\hypernodes|-1}}{b^{|\hypernodes|-1}}$ $\OUT^2 = |\hypernodes| \cdot \OUT^2$.




\begin{proposition2}\label{prop:variance:gjsampler}
$\Var[Z_\hypergraph] \leq |\hypernodes| \cdot AGM(\hypergraph) \cdot \OUT$ for \GJSampler.
\end{proposition2}


\begin{proposition2}\label{prop:variance:ours}
$\Var[Z_\hypergraph] \leq |\hypernodes| \cdot \revtext{\prod_{I} |\hyperedges_I|} \cdot AGM(\hypergraph) \cdot \OUT$ for \second{DRS}. \revtext{Note that $\prod_{I} |\hyperedges_I|$ is $\O(1)$.}
\end{proposition2}









\begin{proposition2}\label{prop:time:sste}
\todotext{$\E[T_\hypergraph]$ is $\O(1)$ for \SSTE.}
\end{proposition2}

\begin{proposition2}\label{prop:time:alley}
$\E[T_\hypergraph]$ is $\O(b^{|\hypernodes|} AGM(\hypergraph))$ for \AlleyS.
\end{proposition2}


\begin{proposition2}\label{prop:time:gjsampler}
For \GJSampler, $T_\hypergraph$ is $\O(1)$ for sequenceable queries and $\O(\IN)$ for non-sequenceable queries.
\end{proposition2}

\begin{proposition2}\label{prop:time:ours}
{$T_\hypergraph$} is $\O(1)$ for {\SUST and} \second{DRS}.
\end{proposition2}



\vspace*{-0.1cm}
Until now, we have assumed that $\OUT$ is given in computing $N = \frac{U_{var}}{\epsilon^2 \delta \OUT^2}$ before \revtext{invoking} \SamplingFramework, which is unrealistic since our goal is to estimate $\OUT$ itself. To tackle this, \revtext{we use the geometric search by Assadi et al. \cite{SSTE}.} They first assume that $\OUT$ is large as $AGM(\hypergraph)$ and run estimation with a small $N$ \cite{SSTE}. Then, they perform a geometric search on $\OUT$; assume a smaller $\OUT$ (decreasing {$\OUT$} by 2), increase $N$, and run the estimation again. They repeat this until the assumed $\OUT$ becomes consistent with the algorithm output. \second{While we implicitly assume that $\OUT > 0$, we can detect when $\OUT = 0$ in real applications whenever the assumed $\OUT$ falls below 1, in total $\O(AGM)$ time.}

This geometric search adds a constant factor (= 4) in the $\O$ bound of $T_\hypergraph$, not a $\log\IN$ factor explained \revtext{by Chen \& Yi} \cite{GJSampler}.
{Starting with an assumption $1 \leq \frac{AGM(\hypergraph)}{\OUT} < 2$, assume that the geometric search stops at $2^k \leq \frac{AGM(\hypergraph)}{\OUT} < 2^{k+1}$.} Then, \revision{the total sample size used so far is} $\sum_{1 \leq i \leq k+1} \frac{2^{i}}{\epsilon^2 \delta} \leq \frac{2^{k+2}}{\epsilon^2 \delta} \leq \frac{4 AGM(\hypergraph)}{\epsilon^2 \delta \OUT}$, i.e., only a constant factor (= 4) is introduced.



{Since \GJSampler can perform uniform sampling, it instead uses a simpler approach \cite{GJSampler}, by repeating the sampling until a constant number $c$ of trials succeed.} Then, the number of total trials becomes a random variable, having $\frac{AGM(\hypergraph) \cdot c}{\OUT}$ as its expectation \cite{GJSampler}. Therefore, $\E[T] = \O\big(\IN + \frac{AGM(\hypergraph) \cdot c}{\OUT}\big)$ for sequenceable queries and $\O\big(\frac{\IN \cdot AGM(\hypergraph) \cdot c}{\OUT}\big)$ for non-sequenceable queries. {Using the same approach, \second{DRS} requires $\prod_{I} |\hyperedges_I| \frac{AGM(\hypergraph) \cdot c}{\OUT}$ trials in expectation and \revtext{thus} $\E[T] = \O\big(\frac{AGM(\hypergraph)}{\OUT}\big)$, asymptotically faster than \GJSampler.}



Finally, \revtext{we give a simple explanation of why $T$ (or $\E[T]$)} of \SSTE, \revtext{\SUST}, \GJSampler, and \second{DRS}, is \emph{inversely} \bluetext{proportional} to $\OUT$. 
Intuitively, if $\OUT$ is as large as $AGM(\hypergraph)$, a set of samples $S_I$ are highly likely to be the actual join \second{answers of} $Q$. In other words, the join \second{answers} are spread over a \bluetext{dense} database. A small number of samples would give us enough information about the distribution of the join \second{answers} and estimating $\OUT$ accurately. In contrast, if $\OUT$ is small, the join \second{answers} would be skewed and sparsely distributed in certain regions of a database, and most samples would not be the join \second{answers}. Therefore, \bluetext{a small number of} samples cannot effectively decrease the uncertainty of estimation, which may require a lot more sampling for an accurate estimation.

\vspace*{-0.2cm}
\subsection{\second{Underlying Index Structure}} \label{subsec:ours:data_model}

\second{We explain the underlying index $Index_F$ for each relation $R_F$ to evaluate the operation $\pi_I(R_F \ltimes s)$ in $\O(1)$ time, as mentioned in \cref{subsec:ours:query_model}. We define $F_I = F \cap I$ and $F_s = F \cap Attr(s)$, where $Attr(s)$ is the set of attributes of $s$. Note that $F_I$ and $F_s$ are disjoint in \SamplingFramework. 
If $F_s = \{A_1, A_2, ..., A_m\}$, then $R_F \ltimes s = R_F \ltimes \pi_{F_s}(s) = \sigma_{A_1 = \pi_{A_1} s} \sigma_{A_2 = \pi_{A_2} s} ... \sigma_{A_m = \pi_{A_m} s} (R_F)$.}

\second{A mild assumption in \cite{GenericJoin} is to build a B-tree-like index (e.g., a trie index in \cite{EmptyHeaded}) under a global attribute order over the whole database.}
\second{Then, if all attributes in $F_s$ precede all attributes in $F_I$, $R_F \ltimes s$ is readily available as an index lookup of $Index_F(s) = Index_F(\pi_{F_s}(s))$. Furthermore, if no attribute in $F$ lies between $F_s$ and $F_I$, $\pi_I(R_F \ltimes s)$ is an index lookup $Index_F(s)$ up to depth $|F_I|$.
To exploit these, the selection of $I$ in each step of \SamplingFramework should be consistent with the global order. Instead, we can build $|F|!$ indexes for $R_F$, one for each possible attribute order as in \cite{EmptyHeaded}, to enable arbitrary selection of $I$.}


\noindent\underline{\second{Remark.}} \second{The complexity of building an index for $R_F$ is linear, i.e., $\O(|R_F|)$ \cite{EmptyHeaded}. However, in contrast to the pre-computing overheads in \GJSampler, the indexing occurs once per database instead of per query. Therefore, its complexity is ignored from the analysis of $T_\hypergraph$ in \cref{subsec:sampling:analysis}. Due to their simple structures, indexes can be easily extended to dynamic setting where tuples can be inserted and deleted, with a constant or logarithmic update time.}


\vspace*{-0.2cm}
\section{Achieving a Generalized Bound}\label{sec:generalized}




In this section, we generalize our $\O\big(\frac{AGM(\hypergraph)}{\OUT}\big)$ bound \revision{for join size estimation} using generalized hypertree decompositions (GHDs, \second{see \cref{sec:background}}) and achieve a better bound than $\O(\IN^{fhtw(\hypergraph)})$, which is the bound of an exact aggregation algorithm \AggGHDJoin using GHDs (see \cref{appendix:algorithm:ghdjoin}).
\rerevision{We leave uniform sampling using GHDs as a future work in \cref{sec:future}.}
Our new bound may be better than our previous bound $\O\big(\frac{AGM(\hypergraph)}{\OUT}\big)$, especially when $\OUT$ is small (see Example \ref{example:better}).
\revtext{We first provide background about GHDs and related algorithms.}

\vspace*{-0.2cm}
\subsection{Aggregations over GHDs} \label{subsec:generalized:aggregations}

Instead of performing joins, $\OUT$ can be more efficiently computed by solving an AJAR (Aggregations and Joins over Annotated Relations) query \cite{AJAR}. AJAR queries assume that each relation $R_F$ is annotated; $R_F = \{(r, \lambda(r))\}$ where each tuple $r$ has an \emph{annotation} $\lambda(r)$ (from some domain $\K$). {If two tuples $r_1$ and $r_2$ are joined, the joined tuple $r_1 \Join r_2$ has $\lambda(r_1) \otimes \lambda(r_2)$ as its annotation.} Hence, joining annotated relations results in another annotated relation: 

\begin{equation}\label{eq:ajarjoin}
\Join_{F \in \hyperedges} R_F = \{(r, \lambda) \,|\, \second{r \text{ is a join answer}}, \lambda = \otimes_{F \in \hyperedges} \lambda(\pi_{F} r)\}.
\end{equation}

Furthermore, we can define an aggregation over an annotated relation $R$ \revision{with attributes $\mathcal{A}$} by a pair of attribute $A$ and sum operator $\oplus$ \cite{AJAR}. Let $\revision{G} = \revision{\mathcal{A}} \setminus \{A\}$. Then, the ($A,\oplus$)-aggregation of $R$ generates an annotated relation $R_\revision{G}$ of attributes \revision{$G$}, {where \revision{$G$} is the \emph{grouping/output attributes}, and $\lambda_\revision{G}$ is an aggregated result for $r_\revision{G}$:}

\begin{equation}\label{eq:ajaragg}
\sum_{A, \oplus} R = \{(r_{\revision{G}}, \lambda_{\revision{G}}) : r_{\revision{G}} \in \pi_{\revision{G}} R \text{ and } \lambda_{\revision{G}} = \bigoplus_{(r, \lambda) \in R : \pi_{\revision{G}} r = r_{\revision{G}}} \lambda \}.
\end{equation}

\rerevision{By definition, no tuple $r_G$ is duplicated in the aggregation result.}
If multiple aggregation attributes $M = \{A_1, A_2, ..., A_n\}$ share the same $\oplus$ operator, we simply write $\sum_{A_1,\oplus} \sum_{A_2,\oplus} ... \sum_{A_n,\oplus}$ as $\sum_{M, \oplus}$. Here, attributes in $M$ are marginalized out and removed from the resulting relation. 
\revision{The output attributes $G$ becomes $\mathcal{A} \setminus M$.}
Altogether, if ($\K, \oplus, \otimes$) forms a commutative semiring \cite{SemiRing}, we can aggregate over joins (of annotated relations) \cite{AJAR} as $\sum_{M,\oplus} \Join_{F \in \hyperedges} R_F$. If $\K = \Z$, $\second{\lambda(r) = 1 \, \forall r \in R_F}$, $\oplus = +$, $\otimes = \times$, and $M = \hypernodes$, \revision{the query $\sum_{\hypernodes, +} \Join_{F \in \hyperedges} R_F$} generates an annotated relation with a single tuple $r$ without any attribute. Here, $\lambda(r) = \OUT = \abs{\Join_{F \in \hyperedges} R_F}$. {We hereafter omit $\oplus$ whenever possible without ambiguity.}
{In addition, AJAR queries are known to be more general than FAQ queries \cite{FAQ}, for a query can have multiple aggregation operators \cite{AJAR}.}

\revtext{Joglekar et al.} \cite{AJAR} present \AggGHDJoin and \AggYan to solve the AJAR queries in $\O(\IN^{w^*} + \OUT_{agg})$ time, where $w^*$ is their width of algorithms, and $\OUT_{agg}$ is the output size of the aggregation results which is typically smaller than $\OUT$, e.g., 1 in our case.
We prove in {\cref{appendix:algorithm:ghdjoin}} that $w^* \geq fhtw$ for general AJAR queries and $w^* = fhtw$ for computing $\OUT$. {Since we show in \cref{subsec:generalized:analysis} that our new bound using GHDs is smaller than $\O(\IN^{fhtw})$, it is also smaller than $\O(\IN^{w^*})$.}

\vspace*{-0.2cm}
\subsection{\revtext{Sampling over GHDs}} \label{subsec:generalized:sampling}

From now, \revision{we use the same setting from \cref{subsec:generalized:aggregations} to compute the AJAR query $\sum_{\hypernodes} \Join_{F \in \hyperedges} R_F$.}
For a set of grouping attributes $G$, \GenericSampler (Algorithm \ref{alg:genericsampler}) returns an approximate answer to the AJAR query $\second{\hypergraph^G \coloneqq} \revision{\sum_{\hypernodes \setminus G} \hypergraph} = \sum_{\hypernodes \setminus G} \Join_{F \in \hyperedges} R_F$,
i.e., an annotated relation $\{(g, Z[g]) | (g, \lambda(g)) \in \second{\hypergraph^G}\}$ (Lines \ref{alg:genericsampler:foreach}-\ref{alg:genericsampler:return});
$\lambda(g)$ and $Z[g]$ represent the exact and approximate value of $\OUT(\second{\hypergraph_g})$, where $\second{\hypergraph_g}$ is the residual hypergraph of $\hypergraph$ given $g$ (see \cref{subsec:sampling:framework}).
\second{Therefore, $\sum_{(g, \lambda(g)) \in \hypergraph^G} \lambda(g) = \OUT(\hypergraph)$.}


\vspace*{-0.2cm}
\begin{algorithm} [htb]
\caption{\mbox{\GenericSampler({$\hypergraph(\hypernodes, \hyperedges), G$})}} \label{alg:genericsampler}
\small{
    \KwIn{Hypergraph $\hypergraph$ and a set of attributes $G \subseteq \hypernodes$}
    
    $R \leftarrow \emptyset$ \label{alg:genericsampler:R}
    
    \redcomment{\revision{compute $\second{\hypergraph^G} \coloneqq \sum_{\hypernodes \setminus G} \hypergraph$}, \rerevision{skip any duplicated $g$}} 
    \ForEach{$\revision{g \in \GenericJoin(\hypergraph, G, \emptyset)}$}  
    {\label{alg:genericsampler:foreach}
        $Z[g] \leftarrow \revision{\SamplingFramework(\hypergraph, \rerevision{\hypernodes \setminus G}, g)}$ \redcomment{\revision{approximation of the output size of the residual query given $g$}} \label{alg:genericsampler:sampler}

        $R \leftarrow R \cup \{(g, Z[g])\}$
    }    
    \Return $R$ \label{alg:genericsampler:return}
}
\end{algorithm}
\vspace*{-0.2cm}

\GHDSampler (Algorithm \ref{alg:ghdsampler}) is our entry function \revision{given a GHD ($\GHD, \bag$)}. For each node $t \in \hypernodes_\GHD$, \second{we define $\hypergraph_t \coloneqq (\bag(t), \hyperedges_{\bag(t), \pi})$, i.e., sub-query of $\hypergraph$ on $t$ (Line \ref{alg:ghdsampler:H}), and} $G(t)$ as the set of grouping attributes of $t$, \revision{determined as the shared attributes across the nodes (Line \ref{alg:ghdsampler:g}).}
\GenericSampler at Line \ref{alg:ghdsampler:generic} takes this $G(t)$ as the grouping attributes $G$ and returns an annotated relation $R_t$ \second{$\coloneqq \{ (g_t, Z[g_t]) | (g_t, \lambda(g_t)) \in \hypergraph^{G(t)}_{t} \}$},
\second{where $\hypergraph^{G(t)}_{t} = \sum_{\bag(t) \setminus G(t)} \hypergraph_t$. Therefore, $R_t$ is an approximation of the aggregation of $\hypergraph_t$ with output attributes $G(t)$.}
Finally, \SimpleAggYan at Line \ref{alg:ghdsampler:yan} takes these $R_t$'s and runs a simple version of \AggYan (see Algorithm \ref{alg:simpleaggyan} in \cref{appendix:algorithm:yan}) to compute $\sum_{G(\GHD)} \Join_{t \in \hypernodes_\GHD} R_t$ for $G(\GHD) \coloneqq \cup_{t \in \hypernodes_\GHD} G(t)$, \second{the union of grouping attributes of all GHD nodes. $G(\GHD)$ is used as join attributes between $R_t$'s in Line \ref{alg:ghdsampler:yan}. Finally, \second{Lemma \ref{lemma:ghdsampler:unbiased} stats the unbiasedness of \GHDSampler.}}


\vspace*{-0.2cm}
\begin{algorithm} [htb]
\caption{{\GHDSampler}($\hypergraph(\hypernodes, \hyperedges)$, ($\GHD(\hypernodes_\GHD, \hyperedges_\GHD)$, $\bag$))} \label{alg:ghdsampler}
\small{
    \KwIn{Query hypergraph $\hypergraph$ and a GHD ($\GHD$, $\bag$) of $\hypergraph$}
    
    $S_\GHD \leftarrow \emptyset$
    
    \ForEach{$t \in \hypernodes_\GHD$}
    {\label{alg:ghdsampler:iterate}
        $\second{\hypergraph_{t}} \leftarrow (\bag(t), \bluetext{\hyperedges_{\bag(t), \pi}})$ \label{alg:ghdsampler:H}

        $G(t) \leftarrow \{v \in \bag(t) | \exists t' \in \hypernodes_\GHD \text{ s.t., } t' \neq t, v \in \bag(t') \}$ \label{alg:ghdsampler:g}
        
        $R_t \leftarrow \GenericSampler(\second{\hypergraph_{t}}, G(t))$ \redcomment{\revision{approximate annotated relation of $\second{\hypergraph^{G(t)}_{t}} \coloneqq \sum_{\bag(t) \setminus G(t)} \second{\hypergraph_{t}}$}}\label{alg:ghdsampler:generic}

        $S_\GHD \leftarrow S_\GHD \cup R_t$
    }    
    \Return $\SimpleAggYan((\GHD, \bag), S_\GHD)$ \label{alg:ghdsampler:yan}
}
\end{algorithm}


\vspace*{-0.2cm}
\begin{lemma2}\label{lemma:ghdsampler:unbiased}
\revtext{\GHDSampler is an unbiased estimator of $\abs{\Join_{F \in \hyperedges} R_F}$.}
\end{lemma2}

\begin{figure}[h!]
\centering
\vspace*{-0.2cm}
\hspace*{-0.28cm}
\includegraphics[width=1.0\columnwidth]{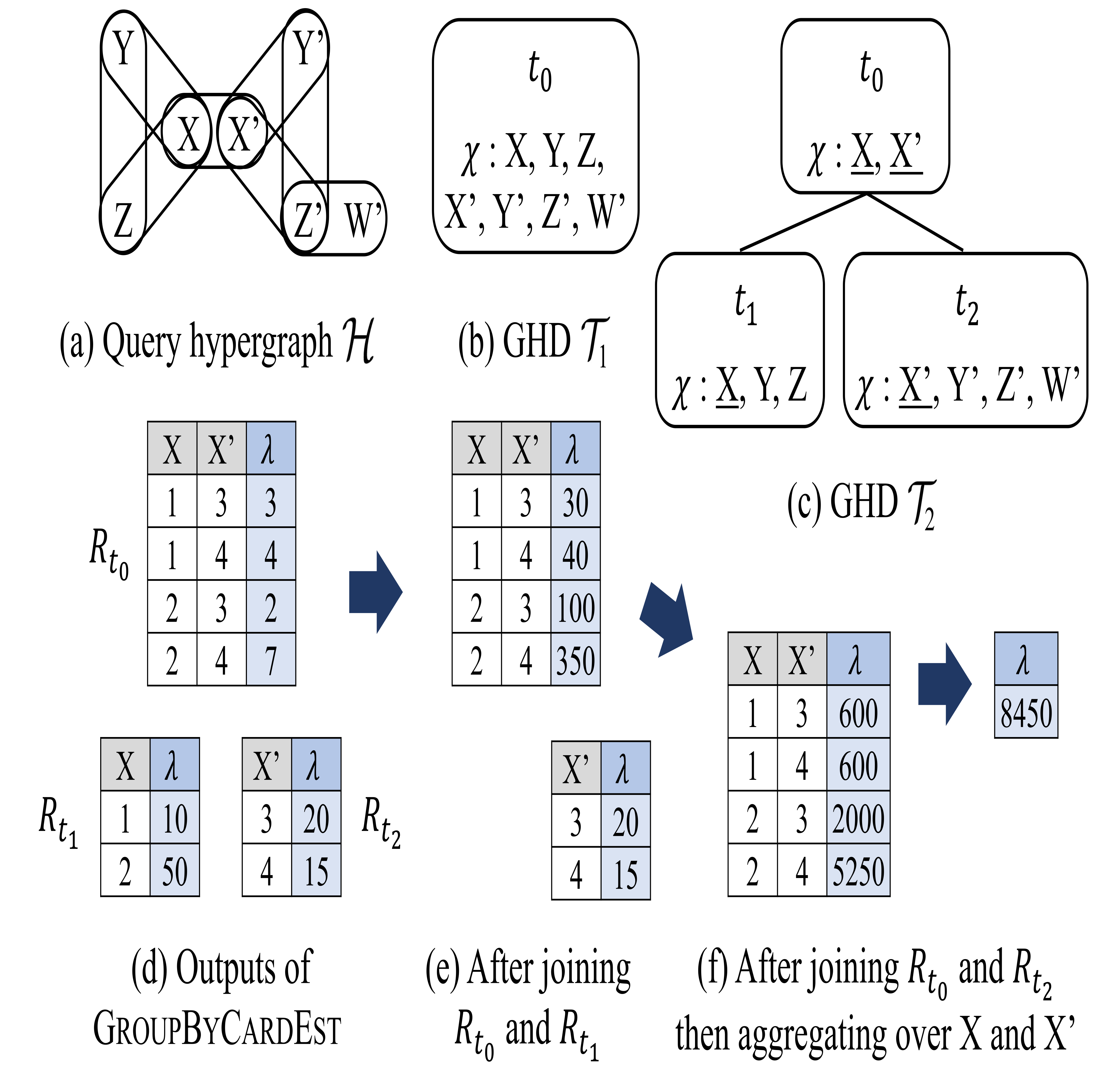}
\vspace*{-0.3cm}
\caption{Example of \GHDSampler using GHD $\GHD_2$ in (c). Attributes in $G(t)$ are underlined in (c).} 
\label{fig:example}
\vspace*{-0.2cm}
\end{figure}

\vspace*{-0.2cm}
\begin{example}
We use $\hypergraph$ and $\GHD_2$ in Figure \ref{fig:example} as an example. Here, $G(t_0) = \{X, X'\}, G(t_1) = \{X\}$, and $G(t_2) = \{X'\}$ are grouping attributes. (d) shows examples of annotated relations $R_t$'s, which are the outputs of \GenericSampler on the three nodes. \second{Since we allow duplicate tuples in any base relation $R_F : F \in \hyperedges$, annotations for $R_{t_0}$ can be larger than one, even if $\bag(t_0)$ is covered by a single hyperedge and $\bag(t_0) = G(t_0)$.} (e) and (f) show the join and aggregation on $R_t$'s performed in \SimpleAggYan. Note that the annotations are multiplied in joins, and added in aggregations. The final result, 8450, is an approximation of $\OUT$.
\end{example}

\second{The key idea of \GHDSampler is to pushdown the partial sum} $\sum_{\bag(t) \setminus G(t)}$ \revision{from $\sum_{\hypernodes}$} to each node $t$ \revision{in computing $R_t$}, which is similar to the key idea of \AggGHDJoin that pushes down the aggregation attributes in GHD as deeply as possible. However, it is slightly different since we shrink each relation obtained from a node \emph{before} the join phase in \AggYan; \AggYan aggregates \emph{after} joining each pair of parent-child relations in GHD.

We also argue that our definition of $G(t)$ is minimal; if an attribute in $G(t)$ is excluded from $R_t$, the node $t$ loses a join relationship with another node. \revision{Our $G(t)$ also minimizes the runtime of a non-sampling procedure, \GenericJoin at Line \ref{alg:genericsampler:foreach} of Algorithm \ref{alg:genericsampler}. Its runtime $\O(AGM(\second{\hypergraph^{G(t)}_{t}}))$ increases with the output attributes $G(t)$ due to the \emph{node-monotone} property of the AGM bound \cite{AJAR}.}

\vspace*{-0.2cm}
\subsection{Analysis of \GHDSampler} \label{subsec:generalized:analysis}

Using the analyses in \cref{subsec:sampling:analysis} as building blocks, we analyze the variance and runtime of \GHDSampler. \revision{For each $(g_t, Z[g_t]) \in R_t$ for a GHD node $t$, we attach two additional annotations} $\Var[Z[g_t]]$ and $T[g_t]$ (the runtime of obtaining $Z[g_t]$), and let $\second{\hypergraph_{t, g_t}}$ \second{(or simply $\hypergraph_{g_t}$)} denote the residual hypergraph of $\second{\hypergraph_t}$ given $g_t$ in Line \ref{alg:genericsampler:sampler} of Algorithm \ref{alg:genericsampler}. 
\revision{Then, $Z[g_t]$ is an unbiased estimate of $\lambda(g_t) = \OUT(\second{\hypergraph_{g_t}})$ from Lemma \ref{lemma:unbiased}.}
We also define \revision{$(f, Z[f])$ as an annotated tuple in} $\Join_{t \in \hypernodes_\GHD} R_t$ so $Z[f] = \prod_{t \in \hypernodes_\GHD} Z[\pi_{G(t)} f]$. 


\revision{Recall that} our primary goal \revision{for join size estimation} is to perform ($1\pm\epsilon$)-approximation of $\OUT$. To use Chebyshev's inequality, we should \second{bound $\Var[Z]$ by} $\epsilon^2\delta\OUT^2$ \revision{as in \cref{subsec:sampling:analysis}.} Then the following questions arise: 1) Can we make this inequality hold for any $\epsilon$ and $\delta$? 2) How would the runtime $T$ be expressed?
\revision{To answer these questions,}
\revtext{we first express $Z$ using $Z[f]$ \second{values}.}

\vspace*{-0.2cm}
\begin{equation}\label{eq:ghdsampler:z}
Z = \sum_{(f, Z[f]) \in \Join_{t \in \hypernodes_\GHD} R_t} Z[f].
\end{equation}


\vspace*{0.1cm}
We arbitrarily order $\hypernodes_\GHD$ and regard each $f$ as an ordered set $\{\pi_{G(t)} f | t \in \hypernodes_\GHD\}$.
Since $Z[g_t]$ \second{values} for $g_t \in f$ are mutually independent, we have


\vspace*{-0.2cm}
\begin{equation}\label{eq:multiexp:prod}
\begin{split}
\E[Z[f]] &= \E\Big[\prod_{g_t \in f} Z[g_t]\Big] = \prod_{g_t \in f} \E[Z[g_t]],
\end{split}
\end{equation}

\vspace*{-0.2cm}
\begin{equation}\label{eq:multivar:prod}
\begin{split}
\Var&[Z[f]] = \Var\Big[\prod_{g_t \in f} Z[g_t]\Big] \\
            &= \prod_{g_t \in f} (\Var[Z[g_t]] + \E[Z[g_t]]^2)  -  \prod_{g_t \in f} \E[Z[g_t]]^2.
\end{split}
\end{equation}
\vspace*{-0.2cm}

While $\E[Z[f]]$ can be simply decomposed into $\prod_{g_t \in f} \E[Z[g_t]]$, 
$\Var[Z[f]]$ has the internal term $\E[Z[g_t]]^2$ which prevents a simple decomposition.
For any two different tuples $f_1, f_2 \in  \, \Join_{t} R_t$, $Z[f_1]$ and $Z[f_2]$ are not independent, since they can have the same sub-tuple, i.e., $g_t \in f_1$ and $g_t \in f_2$ for some $t \in \hypernodes_\GHD$. Therefore, we have to consider the covariance $\Cov(Z[f_1], Z[f_2])$ when expanding $\Var[Z]$ below.


\vspace*{-0.2cm}
\begin{equation}\label{eq:ghdsampler:var}
\begin{split}
\Var[Z] =& \sum_{f} \Var[Z[f]]
        + \sum_{f_1 \neq f_2} \Cov(Z[f_1], Z[f_2]) 
\end{split}
\end{equation}

From the analysis in \cref{subsec:sampling:analysis}, we can arbitrarily control $\Var[Z[g_t]]$ and \revtext{$T[g_t]$} with the sample size, under a condition that $\Var[Z[g_t]] \cdot \revtext{T[g_t]} = \O(AGM(\second{\hypergraph_{g_t}}) \cdot \OUT(\second{\hypergraph_{g_t}}))$. 
In particular, we set $\Var[Z[g_t]]$ = $\O(\OUT(\second{\hypergraph_{g_t}})^2)$ and $\revtext{T[g_t]}$ = $\O\big(\frac{AGM(\second{\hypergraph_{g_t}})}{\OUT(\second{\hypergraph_{g_t}})}\big)$.

\vspace*{-0.1cm}
\begin{lemma2}\label{lemma:varz}
$\Var[Z[f]]$ = $\O(\E[Z[f]]^2)$ if $\Var[Z[g_t]]$ = $\O(\E[Z[g_t]]^2)$ = $\O(\OUT(\second{\hypergraph_{g_t}})^2)$ for every $g_t \in f$.
\end{lemma2}
\vspace*{-0.1cm}

\begin{lemma2}\label{lemma:covz}
$\Cov(Z[f_1], Z[f_2]) =  \O\big(\prod_{g_t \in f_1} \E[Z[g_t]] \cdot \prod_{g_t \in f_2} \\ \E[Z[g_t]]\big)$ if the same condition of Lemma \ref{lemma:varz} holds for $f_1$ and $f_2$.
\end{lemma2}
\vspace*{-0.1cm}

Now we are ready to answer our first question. By applying Lemmas \ref{lemma:varz} and \ref{lemma:covz} to (\ref{eq:ghdsampler:var}), we have

\vspace*{-0.3cm}
\begin{equation}\label{eq:varosquare}
\begin{split}
&\Var[Z] = \sum_{f} \O\Big(\prod_{g_t \in f} \E[Z[g_t]]^2\Big) \\ 
 & \;\;\;\;\;\;\;\;\;\;\;\;\;\;\; + \sum_{f_1 \neq f_2} \O\Big(\prod_{g_t \in f_1} \E[Z[g_t]] \prod_{g_t \in f_2} \E[Z[g_t]]\Big) \\
&= \O\Big(\sum_{f} \prod_{g_t \in f} \E[Z[g_t]]^2 + \sum_{f_1 \neq f_2} \prod_{g_t \in f_1} \E[Z[g_t]] \prod_{g_t \in f_2} \E[Z[g_t]]\Big) \\
&= \O\Big(\Big(\sum_{f} \prod_{g_t \in f} \E[Z[g_t]]\Big)^2\Big) = \O(\OUT^2).
\end{split}
\end{equation}
\vspace*{-0.2cm}


The last equality holds from $\sum_{f} \prod_{g_t \in f} \E[Z[g_t]] = \sum_{f} \E[Z[f]] = \E[Z] = \OUT$.
As a result, $\Var[Z] = \O(\OUT^2)$.

We can make $\Var[Z]$ arbitrarily small, even less than the $\epsilon^2\delta \OUT^2$ we desire, by setting the constant factor of $\Var[Z[g_t]] = \O(\E[Z[g_t]]^2)$ arbitrary small for every $g_t$. We omit setting the constants which is \revtext{trivial}.


\begin{proposition2}\label{prop:converge}
If \, $\Var[Z[g_t]]$ approaches to 0 for every $g_t$, $\Var[Z]$ approaches to 0.
\end{proposition2}


We next answer our second question. 
\GenericSampler for each node $t$ takes \revision{$\O(AGM(\second{\hypergraph^{G(t)}_{t}}))$} time at Line \ref{alg:genericsampler:foreach} and $\sum_{g_t \in R_t} T[g_t]$ at Line \ref{alg:genericsampler:sampler} of Algorithm \ref{alg:genericsampler}.
\SimpleAggYan at Line \ref{alg:ghdsampler:yan} of Algorithm \ref{alg:ghdsampler} takes \revision{$\O(\max_{t \in \hypernodes_\GHD} AGM(\second{\hypergraph^{G(t)}_{t}}))$} time. Therefore, $T$ is $\O\Big(\max_{t \in \hypernodes_\GHD} AGM(\second{\hypergraph^{G(t)}_{t}})$ + $\sum_{t \in \hypernodes_\GHD} \sum_{(g_t, Z[g_t]) \in R_t} T[g_t]\Big)$. By setting $\revtext{T[g_t]} = \O\big(\frac{AGM(\second{\hypergraph_{g_t}})}{\OUT(\second{\hypergraph_{g_t}})}\big)$, $T$ becomes $\O\Big(\max_{t} AGM(\second{\hypergraph^{G(t)}_{t}})$ + $\sum_t \sum_{(g_t, Z[g_t]) \in R_t} \frac{AGM(\second{\hypergraph_{g_t}})}{\OUT(\second{\hypergraph_{g_t}})}\Big)$.



We now prove that our new bound is smaller than $\O(\IN^{fhtw(\hypergraph)})$ = $\O(\max_t \revision{AGM(\second{\hypergraph_t})})$; $AGM(\second{\hypergraph^{G(t)}_{t}}) \leq \revision{AGM(\second{\hypergraph_t})}$ since $AGM$ is node-monotone, and
$\sum_t \sum_{g_t} \frac{AGM(\second{\hypergraph_{g_t}})}{\OUT(\second{\hypergraph_{g_t}})}$ 
$\leq$
$\sum_t \sum_{g_t} AGM(\second{\hypergraph_{g_t}})$ 
$\leq$ 
$\revision{AGM(\second{\hypergraph_t})}$ from Lemma \ref{lemma:decomposition:genericjoin} in \cref{appendix:lemmas}.
\second{Therefore, if $\OUT$ is $\O(\IN^{\rho - fhtw})$, $\IN^{fhtw}$ is asymptotically smaller than $\frac{AGM}{\OUT}$ ($AGM = \IN^{\rho}$), proving Theorem \ref{th:drs_ghd}.} We recommend using \SamplingFramework if $\OUT$ is expected to be large, and using \GHDSampler if $\OUT$ is expected to be small.

\vspace*{-0.2cm}
\begin{example}\label{example:better}
We use the $\hypergraph$ in Figure \ref{fig:example} \revtext{without an edge $\{Z', W'\}$} and assume that every base relation has size $N$. Then, our previous bound $\O\big(\frac{AGM(\hypergraph)}{\OUT}\big)$ becomes $\O\big(\frac{N^3}{\OUT}\big)$, from an optimal fractional edge cover $x$: $x_F = 0$ for $F = \{X, X'\}$ and $0.5$ otherwise. Since our new bound is smaller than $\O(\IN^{fhtw(\hypergraph)}) = N^{1.5}$, it is also smaller than $\O\big(\frac{N^3}{\OUT}\big)$ if $\OUT$ is asymptotically smaller than $N^{1.5}$.
\end{example}

\section{Optimizations} \label{sec:optimization}


\revtext{This section explains two optimization techniques to enhance estimation efficiency or accuracy.}

\vspace*{-0.2cm}
\subsection{Increasing Sampling Probabilities} \label{subsec:optimization:increasing}

If we focus on the join size estimation without persisting uniform sampling, we can further reduce the variance. Since $\Var[Z_\hypergraph] = \sum_{t \in Q} \frac{1}{P(t)} - \OUT^2$ \cite{Alley, GJSampler} for our method \second{in \cref{subsec:sampling:ours}}, increasing $P(t) : t \in Q$ reduces the variance.
First, when $m$ edges tie and have the same maximum $rdeg$ in \cref{subsec:sampling:ours}, we do not decrease the keeping probability $p$ by $m$ times. This increases the final $P(t)$ by $m$ times.
\revision{In fact, we can use any sampled $F^*$ from $\hyperedges_I$; $P(s_I)$ increases from $\frac{1}{|\hyperedges_I|} \cdot \max_{F' \in \hyperedges_I} rdeg_{F',s}(s_I) \cdot \frac{AGM(\hypergraph_{\sappend})}{\max_{F' \in \hyperedges_I} rdeg_{F',s}(s_I) AGM(\hypergraph_s)}$ to $\sum_{F \in \hyperedges_I} \frac{1}{|\hyperedges_I|} \cdot rdeg_{F,s}(s_I) \cdot \frac{AGM(\hypergraph_{\sappend})}{\max_{F' \in \hyperedges_I} rdeg_{F',s}(s_I) AGM(\hypergraph_s)}$.}
\revision{Second}, we can interleave \GJSampler to remove the $\frac{1}{|\hyperedges_I|}$ factor in $P(s_I)$. If the $|\SampleSpace_I|$ of \GJSampler is small enough as a constant, e.g., $\prod_{I} |\hyperedges|$, we can compute $P(s_I) = \frac{AGM(\hypergraph_{\sappend})}{AGM(\hypergraph_s)}$ for every $s_I \in \SampleSpace_I$ as \GJSampler. 


\vspace*{-0.2cm}
\subsection{Skipping Attributes and GHD Nodes} \label{subsec:optimization:skipping}



In \SamplingFramework, sampling $s_I$ for any non-join attribute is unnecessary and thus, can be safely skipped. For our example query in Figure \ref{fig:example}, $W'$ is the only non-join attribute. Let the current sample $s = \{x', y', z'\}$ (attributes are $X', Y', Z'$) and $I = \{W'\}$. Then, sampling any $s_I \in \SampleSpace_I = \pi_{W'}(R_{\{Z',W'\}} \ltimes z')$ does not affect the sample space and sample size for the remaining vertices, i.e., $X$, $Y$, and $Z$. Hence, we skip $W'$ and just return $\frac{1}{P(s_{\{W'\}})} = |\SampleSpace_{\{W'\}}| = |\pi_{W'}(R_{\{Z',W'\}} \ltimes z')|$ at Line \ref{alg:sampler:return1} instead of 1. If there are multiple non-join attributes $X_1, X_2, ..., X_n$, we skip all of these, sample for all join attributes only, then return $\prod_{1 \leq i \leq n} |\SampleSpace_{\{X_i\}}|$ at Line \ref{alg:sampler:return1}.

\revtext{In \GHDSampler, calling \GenericSampler for single-edge GHD nodes can be safely skipped.} If a GHD node $t$ is covered by a single edge, i.e., $\exists F \in \hyperedges : \bag(t) \subseteq F$, we can directly obtain $\E[Z[g_t]] = |R_F \ltimes g_t|$ for every $g_t$ in $R_t$ without sampling in \GenericSampler. \rerevision{To be consistent, we regard $R_t$ as $\pi_{\bag(t)} R_F$ (with annotations) in order to remove any duplicates in $R_t$ as mentioned in \cref{subsec:generalized:aggregations}.}
\revision{
$t_0$ of $\GHD_2$ in Figure \ref{fig:example} is an example of a single-edge node, where $Z[g_{t_0}] = |R_{\{X, X'\}} \ltimes g_{t_0}|$ for $g_{t_0} \in R_{t_0}$. Therefore, $\Var[Z[g_{t_0}]] = 0$ and $T[g_{t_0}] = \O(1)$, reducing both $\Var[Z]$ and $T$.} 

\vspace*{-0.2cm}
\section{Research Opportunities} \label{sec:future}




We present \rerevision{six} research opportunities based on our study. 
First, we could express our bound $\O\Big(\max_{t} AGM(\second{\hypergraph^{G(t)}_{t}}) + \sum_t \sum_{(g_t, Z[g_t]) \in R_t} \\ \frac{AGM(\second{\hypergraph_{g_t}})}{\OUT(\second{\hypergraph_{g_t}})}\Big)$ in \cref{sec:generalized} in a more succinct form by removing internal $g_t$ terms, \revtext{e.g., to $\O\big(\frac{\IN^{fhtw(\hypergraph)}}{\OUT}\big)$.} Then, we could more clearly characterize the gap between this bound and $\O\big(\frac{AGM(\hypergraph)}{\OUT}\big)$ or $\O(\IN^{fhtw(\hypergraph)})$. 

Second, we could dynamically change the sampling approach based on the guess of $\OUT$. Our bound $\O\big(\frac{AGM(\hypergraph)}{\OUT}\big)$ in \cref{sec:sampling} can be small as a constant if $\OUT$ is large as $AGM(\hypergraph)$, while the bound in \cref{sec:generalized} has advantages over small $\OUT$ (e.g., Example \ref{example:better}). As stated in \cref{subsec:sampling:analysis}, we could start with \SamplingFramework assuming a large $\OUT$, increase the sample size, adjust the assumption, and change to \GHDSampler if the estimated $\OUT$ becomes small.









Third, we could even change between sampling and join, then apply the same analysis.
While the join algorithms have zero error but +OUT term in their runtime, sampling algorithms have non-zero errors but $\frac{1}{\OUT}$ factor in their runtime. Therefore, it might be possible to balance between sampling and join to reduce the runtime and error for both large- and small-$\OUT$ cases, i.e., select a sampling-like algorithm for large $\OUT$ and a join-like algorithm \mbox{for small $\OUT$.}

\revtext{Fourth, we could develop join size estimators with lower complexities using more information about relations, e.g., degree constraints or functional dependencies \bluetext{\cite{PANDA}} other than the cardinality constrains in Definition \ref{def:agm}.}

\rerevision{Fifth, we could extend \GHDSampler to perform uniform sampling over arbitrary GHDs from the observations in \cref{appendix:algorithm:yan} that 1) \ExactWeight \cite{ExactWeight} performs uniform sampling over certain types of GHDs for acyclic joins, and 2) \GHDSampler can be easily modified for uniform sampling over the same GHDs. In Algorithm \ref{alg:exactweight}, the initial value of $W(g_t)$ (sampling weight for a tuple $g_t$ in $R_t$) is set to $\E[Z[g_t]] = |R_{F_t} \ltimes g_t|$. Therefore, we could extend \ExactWeight and \GHDSampler for arbitrary GHDs for even cyclic joins, starting from initializing $W(g_t)$ with an estimate $Z[g_t]$ instead of $\E[Z[g_t]]$.}


\rerevision{Sixth, we could extend our algorithms to more general problems, approximate counting and uniform sampling for conjunctive queries \cite{conjunctive1, conjunctive2} or join-project queries $\pi_O(Q)$, where $O$ is the set of output attributes \cite{GJSampler}. We explain in \cref{appendix:conjunctive} that our degree-based rejection sampling can be easily extended for join-project queries, following the same analysis in Section 3 of \cite{GJSampler}. We achieve $\O\big(\frac{AGM(Q)}{|\pi_O(Q)|}\big)$ runtime which is again smaller than that of \GJSampler.}







\section{Conclusion} \label{sec:conclusion}
\vspace*{-0.1cm}

In this paper, we have presented a new sampling-based method for join size estimation \revtext{and uniform sampling}. Our method solves an open problem in the literature \cite{GJSampler}, achieving (1$\pm\epsilon$)-approximation of $\OUT$ in $\O\big(\frac{AGM}{\OUT}\big)$ time for {arbitrary} join queries. We presented a unified approach that explains and analyzes \revtext{four} known \second{methods: \SSTE}, \revtext{\SUST}, \Alley, and \GJSampler. 
We then extended our method using GHDs to achieve a better bound than $\O\big(\frac{AGM}{\OUT}\big)$ for small $\OUT$ \revtext{and optimized the efficiency and accuracy using \todotext{two approaches.}}
Finally, we have highlighted several interesting research opportunities building on our results.
We believe that our work may facilitate studies on achieving lower bounds on the \revtext{uniform sampling and size estimation over joins.}






\section{Acknowledgement}\label{sec:acknowledgement}
\second{We thank Yufei Tao for pointing out the condition of Lemma \ref{lemma:psmall}.}

This work was supported by Institute of Information \& communications Technology Planning \& Evaluation (IITP) grant funded by the Korea government (MSIT) (No.2021-0-00859, Development of a distributed graph DBMS for intelligent processing of big graphs, 34\%), Institute of Information \& communications Technology Planning \& Evaluation (IITP) grant funded by the Korea government (MSIT) (No.2018-0-01398, Development of a Conversational, Self-tuning DBMS, 33\%), and the National Research Foundation of Korea (NRF) grant funded by the Korea government (MSIT) (No.NRF-2021R1A2B5B03001551, 33\%).

\bibliographystyle{ACM-Reference-Format}
\balance
\bibliography{main}

\ifFullVersion
\else
\vspace*{-0.2cm}
\fi
\section*{Appendix}
\appendix

\newtheorem{definition3}{\bf Definition}
\newtheorem{proposition3}{\bf Proposition}
\newtheorem{lemma3}{\bf Lemma}
\newtheorem{corollary3}{\bf Corollary}

\section{Algorithms} \label{appendix:algorithm}

\subsection{\GenericJoin} \label{appendix:algorithm:genericjoin}

\GenericJoin (Algorithm \ref{alg:genericjoin}) \cite{GenericJoin} is a representative worst-case optimal join algorithm. From the set of \revision{output attributes $\outputnodes$ (initially $\hypernodes$)}, it selects a subset $I$ where $1 \leq |I| < \revision{|\outputnodes|}$ (Line \ref{alg:genericjoin:I}). 
For each join \second{answer} $s_I$ of \revision{an induced hypergraph of $\hypergraph$ projected onto $I$}
(Line \ref{alg:genericjoin:foreach}), it bounds attributes $I$ to $s_I$ and proceeds to match the residual hypergraph 
(Line \ref{alg:genericjoin:recursive}). If $\revision{|\outputnodes|} = 1$, the results are obtained from the intersection ($\cap$) operation (Lines \ref{alg:genericjoin:if}-\ref{alg:genericjoin:return}).
The runtime of \GenericJoin is $\O(AGM(\hypergraph))$, which can be proven from Lemma \ref{lemma:decomposition:genericjoin} \cite{GenericJoin}.

\ifFullVersion
\else
\vspace*{-0.3cm}
\fi
\begin{algorithm} [htb]
\caption{{\GenericJoin}($\hypergraph(\hypernodes, \hyperedges)$, \revision{$\outputnodes$, $s$})} \label{alg:genericjoin}
\small{
    \KwIn{{Query hypergraph $\hypergraph$, \revision{output attributes $\outputnodes \subseteq \hypernodes$, and current tuple $s$}}} 
    
    \If{\revision{$|\outputnodes| = 1$}}
    {\label{alg:genericjoin:if}
        \Return \revision{$\cap_{F \in \hyperedges_{\outputnodes}} \pi_{\outputnodes} (R_F \ltimes s)$} \label{alg:genericjoin:return} \\
    }
    
    
    
    $R \leftarrow \emptyset$ \\
    $I \leftarrow $ a non-empty proper subset of \revision{$\outputnodes$} \label{alg:genericjoin:I} \\
    
    
    

    \ForEach{$\revision{s_I \in \GenericJoin(\hypergraph, I, s)}$}
    {\label{alg:genericjoin:foreach}
        $R[s_I] \leftarrow \revision{\GenericJoin(\hypergraph, \outputnodes \setminus I, s \uplus s_I)}$ \label{alg:genericjoin:recursive} \\

        $R \leftarrow R \cup \{s_I\} \times R[s_I]$ \label{alg:genericjoin:R} \\
    }
    
    \Return $R$ \\
}
\end{algorithm}

\ifFullVersion
\else
\vspace*{-0.5cm}
\fi
\subsection{\revision{\Yan, \SimpleAggYan, and \ExactWeight}} \label{appendix:algorithm:yan}

\rerevision{In this section and in Algorithms \ref{alg:simpleaggyan}-\ref{alg:exactweight}, we assume that the root node $r$ of any GHD $\GHD$ has a virtual parent node $pr$ where 1) $pr \not\in \hypernodes_\GHD$, 2) $\bag(pr) = \emptyset$, and 3) $R_{pr}$ (input relation for $pr$) contains a single tuple $g_{pr}$ that joins with all tuples in $R_r$, for ease of explanation. As mentioned in \cref{subsec:generalized:aggregations}, we assume no duplicate tuples in each $R_t$.}

Given a join tree $\GHD$ of a $\alpha$-acyclic query $Q$ \bluetext{\cite{AlphaAcyclic}}, \Yan \cite{Yannakakis} performs the join in $\O(\IN + \OUT)$ time using dynamic programming.
Algorithm \ref{alg:simpleaggyan} without Lines \ref{alg:simpleaggyan:beta}-\ref{alg:simpleaggyan:rprime} \revtext{and setting $R' = R_t$} is \Yan, where the bottom-up and top-down semi-join reductions (Lines \ref{alg:simpleaggyan:bottom}-\ref{alg:simpleaggyan:top}) are followed by the bottom-up join (Lines \ref{alg:simpleaggyan:joinstart}-\ref{alg:simpleaggyan:joinend}).
Since the semi-join reductions remove dangling tuples (that do not participate in the final join \second{answers}), \second{the intermediate join size} monotonically increases up to $\OUT$ in the bottom-up join. Therefore, the total runtime is $\O(\IN + \OUT)$.

\begin{algorithm} [htb]
\caption{{\SimpleAggYan}($(\GHD(\hypernodes_\GHD, \hyperedges_\GHD), \bag)$, $\{R_t|t \in \hypernodes_\GHD\}$)} \label{alg:simpleaggyan}
\small{
    \KwIn{GHD ($\GHD$, $\bag$) and relations $R_t$ for each $t \in \hypernodes_\GHD$}
    
    \ForEach{$t \in \hypernodes_\GHD$ in some bottom-up order}
    { \label{alg:simpleaggyan:bottom}
        $p \leftarrow $ {parent of} $t$ \\
        $R_p \leftarrow R_p \ltimes R_t$ \label{alg:simpleaggyan:bottomend} \\
    }
    
    \ForEach{$t \in \hypernodes_\GHD$ in some top-down order}
    {
        $p \leftarrow $ {parent of} $t$ \\
        $R_t \leftarrow R_t \ltimes R_p$ \label{alg:simpleaggyan:top} \\
    }
    
    \ForEach{$t \in \hypernodes_\GHD$ in some bottom-up order}
    { \label{alg:simpleaggyan:joinstart}
        $\beta \leftarrow \{A \in G(t) | TOP_\GHD(A) = t\}$ \label{alg:simpleaggyan:beta} \\
        $R' \leftarrow \sum_\beta {R_t}$ \label{alg:simpleaggyan:rprime} \\
        
        {
            $p \leftarrow $ {parent of} $t$ \\
            $R_p \leftarrow R_p \Join R'$ \label{alg:simpleaggyan:joinend} \\
        }
    }
    
    \Return $R_r$ for the root $r$ \\
}
\end{algorithm}




\SimpleAggYan (Algorithm \ref{alg:simpleaggyan}) is a simplified version of \AggYan \cite{AJAR}, \bluetext{where 1) all attributes are used for aggregation and 2) sum is the only aggregation operator.} \AggYan can handle a more general class of aggregation queries, e.g., sum and max operators are used for different attributes. 
In Line \ref{alg:simpleaggyan:beta}, $TOP_\GHD(A)$ denotes the \emph{top} node of an attribute $A$, which is the closest node to the root among $\{t \in \hypernodes_\GHD | \bag(t) \ni A\}$. 
\bluetext{$G(t)$ is the set of output attributes of $R_t$.}
Line \ref{alg:simpleaggyan:rprime} aggregates on the attributes \bluetext{in $G(t)$} having $t$ as their top node, since these attributes do not appear in ancestors and are marginalized out. This early marginalization is the key idea of maintaining the runtime of \SimpleAggYan to be $\O(\IN + \OUT_{agg})$ \cite{AJAR}, where $\OUT_{agg}$ is the output size of aggregation. For example, $\OUT_{agg} = 1$ if all attributes are \bluetext{marginalized out} in computing the join size.

\ifFullVersion
\else
\vspace*{-0.2cm}
\fi
\begin{algorithm} [htb]
\caption{\mbox{\revision{{\ExactWeight}($(\GHD(\hypernodes_\GHD, \hyperedges_\GHD), \bag)$, $\{R_t|t \in \hypernodes_\GHD\}$)}}} \label{alg:exactweight}
\small{
    \KwIn{GHD ($\GHD$, $\bag$) where $\exists F_t \in \hyperedges : \bag(t) \subseteq F_t$ for every $t \in \hypernodes_\GHD$ and relations $R_t$ for each $t \in \hypernodes_\GHD$}
    
    $W(g_t) \leftarrow |R_{F_t} \ltimes g_t|$ for every $t \in \hypernodes_\GHD$ and $g_t \in R_t$ \label{alg:exactweight:initialize} \\

    \ForEach{$t \in \hypernodes_\GHD$ in some bottom-up order}
    { \label{alg:exactweight:bottom}
    
        
        $p \leftarrow $ {parent of} $t$ \\
        
        \ForEach{$g_p \in R_p$}
        {
            $W(g_p, R_t) \leftarrow \sum_{g_t \in R_t \ltimes g_p} W(g_t)$ \\
            $W(g_p) \leftarrow W(g_p) \cdot W(g_p, R_t)$ \\
        }
        
        $R_p \leftarrow R_p \ltimes R_t$ \label{alg:exactweight:bottomend} \\
    }
    
    $s \leftarrow g_{pr}$ \label{alg:exactweight:top}
    
    \ForEach{$t \in \hypernodes_\GHD$ in some top-down order}
    {
        $p \leftarrow $ {parent of} $t$
        
        $s_p \leftarrow \pi_{\bag(p)} s$
        
        $\{s_t\} \leftarrow {\SampleFrom}(R_t \ltimes s_p, 1, \{\frac{W(g_t)}{W(s_p, R_t)} \,|\, g_t \in R_t \ltimes s_p\})$ \redcomment{sample a tuple $s_t$ from $R_t \ltimes s_p$ with probability $\frac{W(s_t)}{W(s_p, R_t)}$} \label{alg:exactweight:sample}
        
        $s \leftarrow s \uplus s_t$ \label{alg:exactweight:topend}
    }
    \ForEach{$t \in \hypernodes_\GHD$ in some order}
    { \label{alg:exactweight:some}
        Replace $s_t \in s$ in-place with a uniform sample from $R_{F_t} \ltimes s_t$ \label{alg:exactweight:replace}
    }
    \Return $s$ \label{alg:exactweight:return}
}
\end{algorithm}
\ifFullVersion
\else
\vspace*{-0.2cm}
\fi

\rerevision{We explain \ExactWeight \cite{ExactWeight} in our context of using GHDs (Algorithm \ref{alg:exactweight}), \second{but assume that $R_t$'s are not annotated and compute each weight $W$ explicitly.} 
\ExactWeight performs uniform sampling over acyclic joins, where the GHDs of single-edge nodes are given.
\ExactWeight computes the sampling weights $W(g_p)$ and $W(g_p, R_t)$ for each tuple $g_p$ and a child relation $R_t$ bottom-up (Lines \ref{alg:exactweight:bottom}-\ref{alg:exactweight:bottomend}) and samples tuples proportional to their weights top-down (Lines \ref{alg:exactweight:top}-\ref{alg:exactweight:topend}). The sample is then replaced with the tuples sampled from the base relations (Lines \ref{alg:exactweight:some}-\ref{alg:exactweight:replace}).
The bottom-up part is the preprocessing step that takes $\O(|\hypernodes_\GHD| \sum_{t \in \hypernodes_\GHD} |R_t|) = \O(\max_{t} |R_{F_t}|) = \O(\IN)$ time, and the top-down sampling and replacement takes $\O(1)$ for each sample and can be performed multiple times after the preprocessing.}

\rerevision{We now briefly explain why \ExactWeight returns a uniform sample.
$W(g_p, R_t)$ corresponds to the join size between $g_p$ and the child subtree of $p$ rooted at $t$, and $W(g_p)$ corresponds to the join size between $g_p$ and all child subtrees of $p$.
Hence, $W(g_{pr}) = \OUT$ which is the join size of the whole tree.
For a node $p$, its children $\{t_1, t_2, ..., t_m\}$ and its parent $pp$, $P(s_p) = \frac{W(s_p)}{W(s_{pp}, R_p)}$ and $P(s_{t_i}) = \frac{W(s_{t_i})}{W(s_p, R_{t_i})}$ for $i \in [1,m]$. Therefore, $P(s_p, s_{t_1}, ..., s_{t_m}) = \frac{W(s_p)}{W(s_{pp}, R_p)} \prod_{1 \leq i \leq m} \frac{W(s_{t_i})}{W(s_p, R_{t_i})} = \frac{|R_{F_p} \ltimes s_p| \prod_i W(s_p, R_{t_i})}{W(s_{pp}, R_p)} \frac{\prod_{i} W(s_{t_i})}{\prod_{i} W(s_p, R_{t_i})} = \frac{|R_{F_p} \ltimes s_p| \prod_{i} W(s_{t_i})}{W(s_{pp}, R_p)}$.
We have seen a similar elimination of probability terms in \cref{eq:gjsample}, which leads to $P(s) = \frac{\prod_{t} |R_{F_t} \ltimes s_t|}{W(g_{pr}, R_r)} = \frac{\prod_{t} |R_{F_t} \ltimes s_t|}{\OUT}$ after Line \ref{alg:exactweight:topend}. 
By uniformly sampling a tuple from each $R_{F_t} \ltimes s_t$ at Line \ref{alg:exactweight:replace}, $P(s)$ becomes $\frac{1}{\OUT}$ at Line \ref{alg:exactweight:return}.
}

\rerevision{We can also make \GHDSampler in Algorithm \ref{alg:ghdsampler} perform uniform sampling for the same GHD by 1) iterating $t$ bottom-up at Line \ref{alg:ghdsampler:iterate}, 2) adding the same initialization and updates of sampling weights, and 3) adding the same sampling part. Then, the preprocessing step of \GHDSampler also takes $\O(\IN)$ as \ExactWeight. From this perspective, \ExactWeight is similar to a specific instance of \GHDSampler for GHDs of single-edge nodes and acyclic queries. Extending the above modifications of \GHDSampler to work for arbitrary GHDs and cyclic queries, would be an interesting future work.}
\rerevision{We could also extend our work to unions of acyclic conjunctive queries and sampling \emph{without} replacement, as what \cite{ExactWeightExtend} is to \cite{ExactWeight}.
}


\subsection{\GHDJoin and \AggGHDJoin} \label{appendix:algorithm:ghdjoin}

Given a generalized hypertree decomposition (GHD) of a query, \GHDJoin (Algorithm \ref{alg:ghdjoin}) \bluetext{\cite{AJAR}} uses $\GHD$ as a join tree and performs \GenericJoin to obtain the join \second{answers} of each node (Lines \ref{alg:ghdjoin:foreach}-\ref{alg:ghdjoin:generic}). \second{$\hypergraph_t$} is the subquery defined on node $t$ (Line \ref{alg:ghdjoin:h}).
Then, it performs \Yan over these join \second{answers} to evaluate the original query (Line \ref{alg:ghdjoin:yan}). Therefore, the runtime of \GHDJoin is $\O(\max_{t \in \hypernodes_\GHD} \IN^{\rho(\second{\hypergraph_t})} + \OUT) = \O(\IN^{fhtw(\GHD, \hypergraph)} + \OUT)$ from Definition \ref{def:fhtw}.

\begin{algorithm} [htb]
\caption{\mbox{\GHDJoin($\hypergraph(\hypernodes, \hyperedges)$, $(\GHD(\hypernodes_\GHD, \hyperedges_\GHD), \bag)$)}} \label{alg:ghdjoin}
\small{
    \KwIn{Query hypergraph $\hypergraph$ and a GHD ($\GHD$, $\bag$) of $\hypergraph$}
    
    $S_\GHD \leftarrow \emptyset$ \\
    \ForEach{$t \in \hypernodes_\GHD$}
    { \label{alg:ghdjoin:foreach}
        $\second{\hypergraph_t} \leftarrow (\bag(t), \hyperedges_{\bag(t), \pi})$ \label{alg:ghdjoin:h} \\


        \redcomment{\revision{compute $\second{\hypergraph_t}$}}
        $S_\GHD \leftarrow S_\GHD \cup \revision{\GenericJoin(\second{\hypergraph_t}, \bag(t), \emptyset)}$ \label{alg:ghdjoin:generic} \\
    }
    \Return $\Yan((\GHD, \bag), S_\GHD)$ \label{alg:ghdjoin:yan} \\
}
\end{algorithm}

\AggGHDJoin \cite{AJAR} is an aggregation version of \GHDJoin. It is different from \GHDJoin in two aspects: 1) calling \AggYan instead of \Yan and 2) including some extra work to ensure that each annotation of a tuple is passed to exactly one call of \GenericJoin \cite{AJAR}. Note that a hyperedge $F \in \hyperedges$ can be visited in multiple nodes, thus, its annotations may be aggregated multiple times, leading to an incorrect output.

From the runtime of \GenericJoin and \AggYan, the runtime of \AggGHDJoin is $\O(\IN^{fhtw(\GHD, \hypergraph)} + \OUT_{agg})$. In \cite{AJAR}, the runtime of \AggGHDJoin is expressed as $\O(\IN^{w^*} + \OUT_{agg})$, where $w^* \coloneqq \min_{(\GHD,\bag) \in \revtext{Val}} fhtw(\GHD, \hypergraph)$. \revtext{Here, $Val$ denotes the set of valid GHDs of $\hypergraph$, that preserve the aggregation ordering of the specified aggregation $\sum_{A_1,\oplus_1} \sum_{A_2,\oplus_2} ... \sum_{A_n,\oplus_n}$; changing the order of $\sum_{A_i, \oplus_i}$ and $\sum_{A_j, \oplus_j}$ might vary the aggregation result \cite{AJAR}.}
Since $fhtw(\hypergraph) = \min_{(\GHD, \bag)} fhtw(\GHD, \hypergraph)$ from Definition \ref{def:fhtw} and \revtext{$Val$ is a subset of the set of all GHDs}, $w^* \geq fhtw(\hypergraph)$.
However, as in our case where \revtext{all aggregation operators are the same ($\oplus_i = +$), thus commutative, any aggregation order gives the same aggregation result. Therefore, $Val$ becomes the set of all GHDs}. This leads to $w^* = fhtw(\hypergraph)$.

\section{Query Decomposition Lemmas} \label{appendix:lemmas}

This section explains two similar query decomposition lemmas proved by \revtext{Ngo et al.} \cite{GenericJoin} and \revtext{Chen \& Yi} \cite{GJSampler}.

\begin{lemma2} \label{lemma:decomposition:genericjoin}
\emph{\cite{GenericJoin}:} Given a query hypergraph $\hypergraph(\hypernodes, \hyperedges)$, let $x$ be any fractional edge cover of $\hypergraph$. Let $I$ be an arbitrary non-empty proper subset of $\hypernodes$, $J = \hypernodes \setminus I$, and $L = \, \Join_{F \in \hyperedges_I} \pi_I R_F$. Then, 

\begin{equation}
    \vspace{-2pt}
    \sum_{s_I \in L} \prod_{F \in \hyperedges_J} |R_F \ltimes s_I|^{x_F} \leq \prod_{F \in \hyperedges} |R_F|^{x_F}. 
    \vspace{-1pt}
\end{equation}

\end{lemma2}


\revision{Using this lemma, Ngo et al. \cite{GenericJoin} prove that \GenericJoin has $\O(\prod_{F \in \hyperedges} |R_F|^{x_F})$ runtime (i.e., the RHS above) using an induction hypothesis on $|\hypernodes|$. If $x$ is an optimal fractional edge cover, the RHS becomes $AGM(\hypergraph)$.}


\revision{We briefly explain this with Algorithm \ref{alg:genericjoin}. If we replace 1) $\hypernodes$ in the lemma with $\outputnodes$ in Algorithm \ref{alg:genericjoin} and 2) $R_F$ with $R_F \ltimes s$, we obtain $L = \, \Join_{F \in \hyperedges_I} \pi_{I} (R_F \ltimes s)$ and}

\begin{equation}
    \vspace{-2pt}
    \revision{\sum_{s_I \in L} \prod_{F \in \hyperedges_{\outputnodes \setminus I}} |R_F \ltimes s \uplus s_I|^{x_F} \leq \prod_{F \in \hyperedges_\outputnodes} |R_F \ltimes s|^{x_F}.}
    \vspace{-1pt}
\end{equation}

\revision{If $|\outputnodes| = 1$, Line \ref{alg:genericjoin:return} takes $\O(\min_{F \in \hyperedges_{\outputnodes}} |\pi_{\outputnodes} (R_F \ltimes s)|)$ time where $\min_{F \in \hyperedges_{\outputnodes}} |\pi_{\outputnodes} (R_F \ltimes s)| \leq \prod_{F \in \hyperedges_{\outputnodes}} |\pi_{\outputnodes} (R_F \ltimes s)|^{x_F} \leq \prod_{F \in \hyperedges_{\outputnodes}} |R_F \ltimes s|^{x_F}$ for $\sum_{F \in \hyperedges_{\outputnodes}} x_F \geq 1$. The last term is equivalent to the RHS above.}

If $|\outputnodes| > 1$, Lines \ref{alg:genericjoin:foreach}-\ref{alg:genericjoin:recursive} take $\O(|L| + \sum_{s_I \in L} \prod_{F \in \hyperedges_{\outputnodes \setminus I}} |\pi_{\outputnodes \setminus I} (R_F \ltimes s \uplus s_I)|^{x_F})$ time where $|L| \leq \prod_{F \in \hyperedges_I} |R_F \ltimes s|^{x_F}$ from the induction hypothesis ($\because |I| < |\outputnodes|$, and $\prod_{F \in \hyperedges_I} |R_F \ltimes s|^{x_F} \leq \prod_{F \in \hyperedges_\outputnodes} |R_F \ltimes s|^{x_F}$) and 
$\sum_{s_I \in L} \prod_{F \in \hyperedges_{\outputnodes \setminus I}} |\pi_{\outputnodes \setminus I} (R_F \ltimes s \uplus s_I)|^{x_F} \leq \sum_{s_I \in L} \prod_{F \in \hyperedges_{\outputnodes \setminus I}} |R_F \ltimes s \uplus s_I|^{x_F} \leq \prod_{F \in \hyperedges_\outputnodes} |R_F \ltimes s|^{x_F}$ from the lemma, which is again the RHS above.

\revision{Therefore, in both cases, the runtime of Algorithm \ref{alg:genericjoin} is the big-Oh of the RHS above.}


\begin{lemma2} \label{lemma:decomposition:gjsampler}
\emph{\cite{GJSampler}:} Given a query hypergraph $\hypergraph(\hypernodes, \hyperedges)$, let $x$ be any fractional edge cover of $\hypergraph$. Let $I = \{A\}$ for an arbitrary attribute $A \in \hypernodes$, $J = \hypernodes \setminus I$, and $L' = \pi_{I} R_{\second{F^{GJ}}}$ for $\second{F^{GJ}} = \argmin_{F \in \hyperedges_{I}} \revision{|\pi_I R_F|}$. Then, 

\begin{equation}
    \sum_{s_I \in L'} \prod_{F \in \hyperedges_J} |R_F \ltimes s_I|^{x_F} \leq \prod_{F \in \hyperedges} |R_F|^{x_F}. 
\end{equation}

\end{lemma2}

Since \revtext{Chen \& Yi} \cite{GJSampler} explain Lemma \ref{lemma:decomposition:gjsampler} without a name, we also call it the query decomposition lemma since it is similar to Lemma \ref{lemma:decomposition:genericjoin}. However, it is different from Lemma \ref{lemma:decomposition:genericjoin} in that 1) $|I|$ must be 1 and 2) $L'$ is obtained without an actual join or intersection in \GenericJoin. If $|I| = 1$ in Lemma \ref{lemma:decomposition:genericjoin}, then $L = \, \Join_{F \in \hyperedges_I} \pi_I R_F = \cap_{F \in \hyperedges_I} \pi_I R_F$. Since $\second{F^{GJ}} \in \hyperedges_I$, $L' \supseteq L$. Therefore, Lemma \ref{lemma:decomposition:gjsampler} is stronger than Lemma \ref{lemma:decomposition:genericjoin} if $|I| = 1$.

\revtext{Lemma \ref{lemma:decomposition:gjsampler} is the key to show that $\sum_{s_I \in \SampleSpace_I} P(s_I) \leq 1$ for \GJSampler.}
\revision{By replacing 1) $\hypernodes$ in the above lemma with \rerevision{$\outputnodes$} in \GJSampler in \cref{subsec:sampling:instance} and 2) $R_F$ with $R_F \ltimes s$, we obtain $L' = \pi_I (R_{\second{F^{GJ}}} \ltimes s) = \SampleSpace_I$ for $\second{F^{GJ}} = \argmin_{F \in \hyperedges_I} |\pi_I (R_F \ltimes s)|$ and}

\begin{equation}
    \revision{\sum_{s_I \in L'} \prod_{F \in \hyperedges_{\rerevision{\outputnodes} \setminus I}} |R_F \ltimes s \uplus s_I|^{x_F} \leq \prod_{F \in \hyperedges_{\rerevision{\outputnodes}}} |R_F \ltimes s|^{x_F},}
\end{equation}

\noindent\revision{resulting in}

\ifFullVersion
\else
\vspace*{-0.3cm}
\fi
\begin{equation}
\begin{split}
    \revision{\sum_{s_I \in \SampleSpace_I} P(s_I)} & \revision{= \sum_{s_I \in L'} \frac{AGM(\hypergraph_{\sappend})}{AGM(\hypergraph_s)} \\
    &= \sum_{s_I \in L'} \frac{\prod_{F \in \hyperedges_{\rerevision{\outputnodes} \setminus I}} |R_F \ltimes s \uplus s_I|^{x_F}}{\prod_{F \in \hyperedges_{\rerevision{\outputnodes}}} |R_F \ltimes s|^{x_F}}} \revision{\leq 1.}
\end{split}
\end{equation}

\ifFullVersion
\vspace*{-0.2cm}
\else
\fi
\section{\rerevision{Extension to Conjunctive Queries}} \label{appendix:conjunctive}

\rerevision{From the introduction, the join query we have considered so far is a full conjunctive query where all attributes are output attributes. In this section, we easily extend our degree-based rejection sampling to approximate counting and uniform sampling for join-project queries (a.k.a. conjunctive queries) \cite{conjunctive1, conjunctive2, GJSampler}, following the same analysis of Section 3 in \cite{GJSampler}.}

\rerevision{Let $\pi_O(Q)$ be a join-project query, where $O$ is the set of projection/output attributes and $Q$ is a join query. We can use the two-step approach in \cite{GJSampler}: (Step 1) use Algorithm \ref{alg:sampler} to sample a join result $s$ from $Q_O \coloneqq \Join_{F \in E_O} \pi_O(R_F)$, i.e., join the relations that contain any attribute in $O$ only. Here, if $Q_O$ is disconnected, we take a sample from each connected component. (Step 2) check if $Q_{\hypernodes \setminus O}(s) \coloneqq \Join_{F \in E_{\hypernodes \setminus O}}(\pi_{\hypernodes \setminus O}(R_F \ltimes s))$ is empty. If not, we return $s$ as a sampled result, otherwise, we repeat. Then, the final $s$ returned is a result in $\pi_O(Q)$. Using the same analysis of \cite{GJSampler}, sampling a tuple in $\pi_O(Q)$ takes $\O\big(\frac{AGM(Q)}{\OUT(\pi_O(Q))}\big)$ time in expectation for our method.}

\rerevision{Step 1 of our algorithm takes time $\O\big(\frac{AGM(Q_O)}{\OUT({Q_O})}\big)$ in expectation. Step 2 takes time $\O(AGM(Q_{\hypernodes \setminus O}(s)))$, using any worst-case optimal join algorithm, e.g., \GenericJoin. Because each $s \in Q_O$ is sampled with probability $\frac{1}{\OUT(Q_O)}$, the expected runtime of Step 2 is (the big-Oh of) $\sum_{s \in Q_O} \frac{AGM(Q_{\hypernodes \setminus O}(s))}{\OUT(Q_O)} \leq \frac{AGM(Q)}{\OUT(Q_O)}$ where the inequality follows from the query decomposition lemma (Lemma \ref{lemma:decomposition:genericjoin}). Please refer to Section 3 of \cite{GJSampler} for the remaining part of the analysis proving the $\O\big(\frac{AGM(Q)}{\OUT(\pi_O(Q))}\big)$ time.}

\rerevision{Here, we would like to address the complexity of computing $\pi_O(R_F)$ before calling Algorithm \ref{alg:sampler}. If there exists an index for $R_F$ with the attribute order such that the attributes in $O \cap F$ precede the attributes in $F \setminus O$, then $\pi_O(R_F)$ is readily available (i.e., in $\O(1)$ time) as explained in \cref{subsec:ours:data_model}. This results in total $\O\big(\frac{AGM(Q)}{\OUT(\pi_O(Q))}\big)$ time. Otherwise, we need an additional linear-time preprocessing to obtain $\pi_O(R_F)$ from $R_F$, resulting in total $\O\big(\frac{AGM(Q)}{\OUT(\pi_O(Q))} + \IN\big)$ time. Nevertheless, we emphasize that this is still lower than the complexity of \GJSampler for non-sequenceable join-project queries, which is $\O\big(\IN \cdot \frac{AGM(Q)}{\OUT(\pi_O(Q))} + \IN\big)$ \cite{GJSampler}.}

\ifFullVersion

\rerevision{We also note that the proposed algorithm in \cite{conjunctive1} requires computing $N(T'_i)$ in their ``High-Level Sampling Template'' which is the (approximate) size of the $i$-th subset $T'_i$ of a set $T$. This has to be computed \emph{for every possible} $i$ since $\frac{N(T'_i)}{\sum_{i} N(T'_i)}$ is used as the probability of sampling $T'_i$. This is analogous to computing $AGM(\hypergraph_{\sappend})$ \emph{for each candidate} $s_I$ in \GJSampler. A difference between \GJSampler and \cite{conjunctive1} is that computing $AGM(\hypergraph_{\sappend})$ takes a constant time in \GJSampler while computing $N(T'_i)$ takes a polynomial time according to Lemma 4.5 in the full version of \cite{conjunctive1}.}
\rerevision{Therefore, \cite{conjunctive1} inherently has the overhead of computing $N(T'_i)$'s analogous to \GJSampler. Removing such overhead by extending our idea to conjunctive queries, would be an interesting future work.}

\fi
\vspace*{-0.1cm}
\section{\revtext{Component-at-a-time Sampling}} \label{appendix:decomposition}


For each component, \SamplingFramework (Algorithm \ref{alg:sampler}) is called in two recursions, e.g., if $\hypergraph$ consists of four odd cycles and three stars, the maximum recursion depth is \revtext{$2 \cdot (4 + 3) + 1 = 15$, including the calls that reach Line \ref{alg:sampler:return1}.}
\revision{Since the components are vertex-disjoint, i.e., \second{variable}-disjoint, $R_F \ltimes s = R_F$ whenever $F$ and (the attributes of) $s$ are contained in different components. Therefore, we can safely drop $\ltimes s$ from the following explanations for the component-at-a-time methods.}

We first explain for a star $S$ of $n$ edges \revision{$F_1, F_2, ..., F_n$. Let $A$ be the center attribute, i.e., $\{A\} = \cap_{i \in [1,n]} F_i$.}
For \SSTE, in the first call to \SamplingFramework, \revision{$I$ becomes the (singleton set of) center attribute, i.e., $I = \{A\}$, $\SampleSpace_I = \pi_A R_{F_1}$} and $k = 1$.
\revision{To sample a vertex $a$ from $\SampleSpace_I$, \SSTE first samples an edge $e$ from \revision{$R_{F_1}$}, and let $a = \pi_A (e)$. Then, $P(a) = \frac{|\revision{R_{F_1}} \ltimes a|}{|\revision{R_{F_1}}|}$ since $|\revision{R_{F_1}} \ltimes a|$ edges in $\revision{R_{F_1}}$ have $a$ as their projection onto $A$.}
When \SamplingFramework is called for the second time, $I$ becomes the remaining attributes of $S$, i.e., \revision{$I = \cup_{i} F_i \setminus \{A\}$}, $\SampleSpace_I = \revision{\prod_{i} \pi_I (R_{F_i} \ltimes a)}$, and $k = 1$; a sequence of $n$ vertices $(b_1, b_2, ..., b_n)$ is sampled from $\SampleSpace_I$ once (i.e., $b_i$ is sampled from $\pi_I (R_{F_i} \ltimes a)$). If $(a, b_1, b_2, ..., b_n)$ forms a star in the database, $\I[\cdot]$ for these two calls become 1, and the recursion proceeds to the remaining components, \rerevision{with $\outputnodes \leftarrow \outputnodes \setminus (\cup_{i} F_i)$ and $s \leftarrow s \uplus (a, b_1, b_2, ..., b_n)$.}

For \SUST, in the first call to \SamplingFramework, $I$ is simply the set of all attributes of $S$ and \revision{$\SampleSpace_I = \prod_{i} R_{F_i}$}, $P(\cdot) = \frac{1}{|\SampleSpace_I|}$, and $k = 1$; a sequence of $n$ edges is sampled once. $\I[\cdot]$ is 1 iff they share a common vertex for the center. The second call to \SamplingFramework is a dummy that directly proceeds to the remaining components.

We next explain for an odd cycle $C$ of $2n+1$ edges \revtext{with the vertex (attribute) sequence $(u_1, v_1, u_2, v_2, ..., u_n, v_n, w, u_1)$ \revision{where $F_i = \{u_i, v_i\}$ for $i \in [1,n]$ and $F_0 = \{u_1, w\}$.} A constraint is that the sampled vertex (attribute value) for $u_1$ must have the smallest degree among the sampled vertices, where the degree of a vertex $a$ is $|R \ltimes a|$ \revision{(note that $R = R_{F_i}$ for every $i$)}.}
In the first call to \SamplingFramework, $I$ contains all \revtext{$u_i$'s and $v_i$'s} except \revtext{$w$}, \revision{$\SampleSpace_I = \prod_{i \in [1,n]} R_{F_i}$}, and $k = 1$; \revtext{a sequence of $n$ edges $(\{a_1, b_1\}, ..., \{a_n, b_n\})$ is sampled where $u_i$ and $v_i$ map to $a_i$ and $b_i$.} 
{When \SamplingFramework is called for the second time,} \revtext{$I = \{w\}$. \SSTE and \SUST differ here.}

For \SSTE, \revision{$\SampleSpace_I = \pi_I (R_{F_0} \ltimes a_1)$}, $P(\cdot) = \frac{1}{|\SampleSpace_I|}$, and $k$ is $\ceil{\frac{|\SampleSpace_I|}{\sqrt{|R|}}}$. Therefore, $k$ vertices are uniformly sampled from $\SampleSpace_I$ with replacement.
For \SUST, $k$ is fixed to 1, and $\SampleSpace_I$ is defined based on the degree of $a_1$. If the degree is smaller than $\sqrt{|R|}$, then \revision{$\SampleSpace_I = \pi_I (R_{F_0} \ltimes a_1)$} and $P(\cdot) = \frac{1}{\sqrt{|R|}}$. If the degree is larger than $\sqrt{|R|}$, then \revision{$\SampleSpace_I = \pi_I R_{F_0}$} and $P(s_I) = \frac{|R \ltimes s_I|}{|R|}$. \revtext{Then, at Line \ref{alg:sampler:reject}, 1) any $s_I$ with a smaller degree than $a_1$ is rejected due to the constraint, and 2) an accepted} $s_I$ is further kept with probability $\frac{\sqrt{|R|}}{|R \ltimes s_I|}$, which is less than 1 from $|R \ltimes s_I| \geq |R \ltimes a_1| \geq \sqrt{|R|}$. \revtext{Therefore, $P(s_I)$ is again $\frac{|R \ltimes s_I|}{|R|} \frac{\sqrt{|R|}}{|R \ltimes s_I|} = \frac{1}{\sqrt{|R|}}$.}
{For both \SSTE and \SUST,} if \revtext{$(a_1, b_1, ..., a_n, b_n, s_{\{w\}}, a_1)$} forms a cycle in the database, the recursion proceeds to the remaining components.

SUST guarantees the uniformity of sampling odd cycles and stars. \revtext{To sample an odd cycle, \SUST relies on the constraint that $a_1$ has the smallest degree.}
This is allowed in unlabeled graphs since other non-qualifying cycles \revtext{(e.g., $a_2$ having the smallest degree)} can be found by permuting the vertices and edges \revtext{(e.g., constrain w.r.t. $a_1$ and let $a_2 = a_1$).} The new proof of \SSTE also uses this constraint to bound the variance 
\ifFullVersion
\second{(see \cref{appendix:proof}).}
\else
\second{(see \S{E} in our full paper \cite{AAA}).}
\fi
\revtext{However,} in non-identical relations, \revtext{we cannot permute the edges with different labels.}


The essence of \revtext{component-at-a-time sampling} is to use the \emph{divide-and-conquer} approach by breaking a query $\hypergraph$ into smaller components \bluetext{(i.e., odd cycles and stars)}, solve simpler problems for these components, and estimate back the cardinality of the original query. Lemmas \ref{lemma:decomposition} and \ref{lemma:subproblem} allow us to set an induction hypothesis that bounds the variance of estimates in simpler problems. We explain these lemmas in \cref{appendix:decomposition:unlabeled} and extend them for labeled graphs in \cref{appendix:decomposition:labeled}. 



\ifFullVersion
\else
\fi
\subsection{For Unlabeled Graphs} \label{appendix:decomposition:unlabeled}

\begin{lemma2}
\label{lemma:decomposition}
(Decomposition Lemma) \emph{\cite{SSTE}:} Given a query hypergraph $\hypergraph(\hypernodes, \hyperedges)$ of identical binary relations (i.e., $R_F = R : \forall F \in \hyperedges$), the LP in Definition \ref{def:agm} admits a half-integral optimal solution $x$ where 1) $x_F \in \{0, \frac{1}{2}, 1\} : \forall F \in \hyperedges$ and 2) the support of $x$, $supp(x) \coloneqq \{F \in \hyperedges | x_F > 0\}$, forms a set of vertex-disjoint odd cycles (i.e., each cycle has an odd number of edges) and stars.
\end{lemma2}

\begin{lemma2} \label{lemma:subproblem}
\emph{\cite{SSTE}:} Suppose that $\hypergraph$ (with identical binary relations as $\hyperedges$) is decomposed into odd cycles and stars by a half-integral optimal solution $x$ in Lemma \ref{lemma:decomposition}.
For arbitrary subsets of odd cycles $\mathbb{C}$ and stars $\mathbb{S}$ of $\hypergraph$, let $\hypergraph_s = (\hypernodes_s, \hyperedges_s)$ be the induced subquery of $\hypergraph$ on vertices of $\mathbb{C}$ and $\mathbb{S}$. 
Then, for output size $\OUT_s$ of $\hypergraph_s$, $\OUT_s \leq |R|^{\sum_{F \in \hyperedges_s} x_F}$, and $\sum_{F \in \hyperedges_s} x_F = \sum_{C(\hypernodes_C, \hyperedges_C) \in \mathbb{C}} \sum_{F \in \hyperedges_C} x_F + \sum_{S(\hypernodes_S, \hyperedges_S) \in \mathbb{S}} \sum_{F \in \hyperedges_S} x_F$.
\end{lemma2}



\subsection{For Labeled Graphs} \label{appendix:decomposition:labeled}

We extend Lemma \ref{lemma:decomposition} to Lemma \ref{lemma:decomposition:extended} for labeled graphs.
\ifFullVersion
\second{First, we prove Proposition \ref{proposition:bipartite}.}
\else
\second{Please refer to our full paper \cite{AAA} for all proofs.}
\fi




\begin{proposition2}
\label{proposition:bipartite}
The LP in Definition \ref{def:agm} admits an integral optimal solution if $\hypergraph$ is a bipartite graph, where relations are \revtext{binary but} not necessarily identical.
\end{proposition2}


\ifFullVersion
\begin{proof}
Let $\hypernodes = \{v_1, v_2, ..., v_n\}$, $\hyperedges = \{F_1, F_2, ..., F_m\}$, and $A \in \mathbb{R}^{n \times m}$ be the \emph{incidence matrix} of $\hypergraph$, where $A_{ij} = \I[v_i \in F_j]$. It is well known that if $\hypergraph$ is bipartite, $A$ is so-called a \emph{totally unimodular matrix}\footnote{\url{https://en.wikipedia.org/wiki/Unimodular_matrix}}. 
We can rewrite the objective of LP, \bluetext{i.e., $\min_{x} \sum_{F \in \hyperedges} x_F \log_{\IN} \abs{R_F}$}, as $\min_{x}{cx}$ ($c$ is a vector of $(\log_{\IN}{|R_F|} : F \in \hyperedges) \in \mathbb{R}^{1 \times m}$, $x \in \mathbb{R}^{m \times 1}$) and constraint $Ax \geq b$ ($b$ is a vector of 1's $\in \mathbb{R}^{n \times 1}$).
Then, it is also known that an LP of objective $\min_{x}{cx}$ and constraint $Ax \geq b$ admits an integral optimal solution if $A$ is totally unimodular and $b$ is a vector of integers, which is exactly our case.
\end{proof}
\fi


Therefore, we can easily remove the identity assumption on relations if $\hypergraph$ is bipartite. 
The following proposition fills the remaining non-bipartite case. 
\ifFullVersion
\second{We omit the proof since it is the same as the proof by \revtext{Assadi et al.} \cite{SSTE}.}
\fi



\begin{proposition2}
\label{proposition:nonbipartite}
The LP in Definition \ref{def:agm} admits a half-integral optimal solution if $\hypergraph$ is a non-bipartite graph, where relations are not necessarily identical.
\end{proposition2}

Finally, we state Lemma \ref{lemma:decomposition:extended}. The extension of Lemma \ref{lemma:subproblem} for labeled graphs, can be directly proven by Lemma \ref{lemma:decomposition:extended} as the proof by \revtext{Assadi et al.} \cite{SSTE}.


\begin{lemma2} \label{lemma:decomposition:extended}
(Extended Decomposition Lemma) Given a query hypergraph $\hypergraph$ of binary relations, the LP in Definition \ref{def:agm} admits a \mbox{half-integral optimal solution that satisfies the conditions in Lemma \ref{lemma:decomposition}.}
\end{lemma2}

\ifFullVersion
\begin{proof}
The proof is similar to the proof by \revtext{Assadi et al.} \cite{SSTE}, but we consider $\log{|R_{F}|}$ which may be different over $F \in \hyperedges$.
Assume that any half-integral optimal solution $x$ is given.

We first show that any edge $F$ in a cycle (odd or even) in $supp(x)$ has $x_F = \frac{1}{2}$. Otherwise if $x_F = 1$ (cannot be 0 due to the definition of $supp(x)$), we can always reduce it into $\frac{1}{2}$ without breaking the feasibility of $x$, since both endpoints of $F$ have at least two edges in the cycle. This results in a smaller objective, contradicting that $x$ is an optimal solution. 

We then remove any even cycles.
If an even cycle $C$ with edges $\{F_1$, $F_2$, …, $F_{2k}\}$ exists in $supp(x)$, 
we argue that $N_1 + N_3 + ... + N_{2k-1}$ = $N_2 + N_4 + ... + N_{2k}$ where $N_i = \log{|R_{F_i}|}$. If LHS < RHS, we subtract $\frac{1}{2}$ from each $x_{F_{2i}}$ and add $\frac{1}{2}$ to each $x_{F_{2i-1}}$. Again, this does not break the feasibility of $x$ but decreases the objective by $\frac{1}{2} \cdot (\text{RHS} - \text{LHS})$, contradicting the optimality of $x$. Similarly, LHS > RHS does not hold.
Thus, LHS = RHS, \bluetext{indicating:}

\ifFullVersion
\vspace*{-0.2cm}
\fi
\begin{equation}\label{eq:LHSRHS}
    |R_{F_1}| |R_{F_3}| ... |R_{F_{2k-1}}| = |R_{F_2}| |R_{F_4}| ... |R_{F_{2k}}|
\end{equation}
\ifFullVersion
\vspace*{-0.2cm}
\else
\vspace*{0.2cm}
\fi

Then, we can break $C$ by safely subtracting $\frac{1}{2}$ from each $x_{F_{2i}}$ and adding $\frac{1}{2}$ to each $x_{F_{2i-1}}$ (or vice versa), preserving the objective.

We next remove any odd cycles connected to any other components.
If an odd cycle $C$ with edges $\{F_1$, $F_2$, …, $F_{2k-1}\}$ exists in $supp(x)$ such that a vertex $v$ in $C$ has an incident edge $F_{2k}$ outside $C$, we again argue that $N_1 + N_3 + ... + N_{2k-1}$ = $N_2 + N_4 + ... + N_{2k}$ using the same logic as we remove the even cycles. Then, we can remove such odd cycles connected to any other components.

\vspace{5pt}

We now have a set of vertex-disjoint odd cycles and trees in $supp(x)$. \bluetext{We will break trees into stars.} For each tree, let $\langle F_1, F_2, ..., F_j \rangle$ be any path connecting two \bluetext{leaf} vertices. Note that $x_{F_1} = x_{F_j} = 1$ to satisfy the constraint in the LP for these two vertices. $x_{F_2} = x_{F_3} = ... = x_{F_{j-1}} = \frac{1}{2}$ since any midpoint has two incident edges in the path, and from the same logic used for cycles.

If $j = 2k$ (even), we argue $N_2 + N_4 + ... + N_{2k-2}$ = $N_3 + N_5 + ... + N_{2k-1}$ using the same logic as above. Then, we can subtract $\frac{1}{2}$ from each $x_{F_{2i-1}}$ and add $\frac{1}{2}$ to each $x_{F_{2i-2}}$. If $j = 2k+1$ (odd), $N_2 + N_4 + ... + N_{2k}$ = $N_3 + N_5 + ... + N_{2k-1}$, and we subtract $\frac{1}{2}$ from each $x_{F_{2i-1}}$ and add $\frac{1}{2}$ to each $x_{F_{2i}}$ except $F_{2}$. In both cases, the path is decomposed into segments of at most two edges each. We repeat this for every path connecting \bluetext{two leaf} vertices, and the final forest consists of trees of a maximum diameter of two, i.e., stars.
\end{proof}
\fi

\revtext{\revtext{Assadi et al.} \cite{SSTE} do not present a concrete join size estimator for labeled graphs but prove a lower bound $\Omega\big(\frac{AGM(\hypergraph_U)}{\OUT}\big)$. Here, $\hypergraph_U$ represents $\hypergraph$ without labels, and $AGM(\hypergraph_U) \geq AGM(\hypergraph)$ since $|R_F| \leq |R|$ for each $F \in \hyperedges$; $R_F$'s are subsets of the set of all data edges $R$.}






\ifFullVersion
\section{Proofs of Lemmas and Propositions} \label{appendix:proof}

{In order to prove Lemmas \ref{lemma:unbiased}-\ref{lemma:covz} and Propositions \ref{prop:variance:sste}-\ref{prop:time:ours},
we first expand \revtext{$Z_{\hypergraph_s}$ and $\Var[Z_{\hypergraph_s}]$}, then derive inequalities on \revtext{$\Var[Z_{\hypergraph_s}]$}.
We define another random variable \revision{$Z_{\hypergraph_s}(s_I) \coloneqq \frac{1}{P(s_I)} \cdot {\I[s_I \in \, \Join_{F \in \hyperedges_I} \pi_{I} (R_F \ltimes s)]} \cdot \revtext{Z_{\hypergraph_{s \uplus s_I}}}$} \revision{where $s_I$ is a random variable in $Z_{\hypergraph_s}(s_I)$.}
Then, \revision{$Z_{\hypergraph_s} = \frac{1}{k} \sum_{s_I \in S_I} Z_{\hypergraph_s}(s_I)$}, i.e., $Z_{\hypergraph_s}$ and $Z_{\hypergraph_s}(s_I)$ are mutually recursively defined. Then, from \cite{cov}:}


\begin{equation}\label{eq:totalvariance}
\begin{split}
\Var[&Z_{\revision{\hypergraph_s}}] = \revision{\frac{\Var[Z_{\hypergraph_s}(s_I)]}{k}} \Big(1 - \frac{k - 1}{|\SampleSpace_I| - 1} \I[S_I \sim \SampleSpace_I \text{ w/o replacement}]\Big).
\end{split}
\end{equation}

{This holds since $S_I$ consists of \bluetext{identically distributed} samples $\{s_I\}$. Note that $s_I \in S_I$ may not be independent due to sampling without replacement, e.g., in \Alley. Then,
$\I[\cdot]$ above is 1 and the variance is decreased by $\frac{k-1}{|\SampleSpace_I| - 1}$ portion \cite{cov}.}

From the law of total variance:

\begin{equation}\label{eq:lawoftotal}
\begin{split}
\Var[Z_{\revision{\hypergraph_s}}(s_I)]
=& \E_{s_I}[\Var[Z_{\revision{\hypergraph_s}}|s_I]] + \Var_{s_I}[\E[Z_{\revision{\hypergraph_s}}|s_I]]. \\
\end{split}
\end{equation}


\bluetext{Here, $Z_{\revision{\hypergraph_s}}|s_I = \frac{1}{P(s_I)} \I[s_I] Z_{\revision{\hypergraph_{s \uplus s_I}}}$ as $Z_{\revision{\hypergraph_s}}(s_I)$,
but $s_I$ is a fixed value in $Z_{\revision{\hypergraph_s}}|s_I$ 
instead of a random variable.
In the second term above, we show that $\E[Z_{\revision{\hypergraph_s}}|s_I] =  \revtext{\frac{1}{P(s_I)} \cdot \I[s_I]} \cdot \revtext{\OUT({\revision{\hypergraph_{s \uplus s_I}}})}$ in the proof of Lemma \ref{lemma:unbiased}. Then,}

\begin{equation}\label{eq:varexp}
\begin{split}
\Var_{s_I}&[\E[Z_{\revision{\hypergraph_s}}|s_I]] \leq \E_{s_I}[\E[Z_{\revision{\hypergraph_s}}|s_I]^2] \\
=& \sum_{s_I} P(s_I) \cdot \frac{1}{P(s_I)^2} \cdot \revision{\I[s_I \in \, \Join_{F \in \hyperedges_I} \pi_{I} (R_F \ltimes s)]} \cdot \OUT(\revision{\hypergraph_{s \uplus s_I}})^2 \\
=& \sum_{s_I} \frac{1}{P(s_I)}\I[s_I] \OUT(\revision{\hypergraph_{s \uplus s_I}})^2.
\end{split}
\end{equation}

\revtext{The $\frac{1}{P(s_I)}$ (> 1) factor has been missing in the original paper \revtext{of Assadi et al. \cite{SSTE}}, reducing an upper bound of $\Var_{s_I}[\E[Z_{\revision{\hypergraph_s}}|s_I]]$ incorrectly.}





\begin{proof}[\textbf{Proof of Lemma \ref{lemma:unbiased}}]

We use an induction hypothesis that \SamplingFramework (Algorithm \ref{alg:sampler}) returns an unbiased estimate of \revision{$\OUT(\hypergraph_s)$}. For the base case where \revision{$I = \rerevision{\outputnodes}$}, Line \ref{alg:sampler:return2} returns {$Z_{\revision{\hypergraph_s}} = \frac{\sum_{s_I \in S_I} Z_{\revision{\hypergraph_s}}(s_I)}{k}$}, where $Z_{\revision{\hypergraph_s}}(s_I) = \frac{1}{P(s_I)} \cdot {\I[s_I]} \cdot \revision{Z_{\hypergraph_{s \uplus s_I}}}$. Here, $\revision{Z_{\hypergraph_{s \uplus s_I}}} = 1$ since the recursive call to \SamplingFramework \revision{reaches Line \ref{alg:sampler:return1}.}
{Since $s_I$'s follow an identical distribution, $\E[Z_{\revision{\hypergraph_s}}] = \E[Z_{\revision{\hypergraph_s}}(s_I)]$ and}
\vspace*{-0.2cm}

\begin{equation}
\begin{split}
\E[Z_{\revision{\hypergraph_s}}(s_I)] &= \sum_{s_I \in \SampleSpace_I} P(s_I) \cdot \frac{1}{P(s_I)} \cdot {\I[s_I]} \cdot 1 \\
&= \sum_{s_I \in \SampleSpace_I} {\I[s_I]} \\
&= \sum_{s_I \in  \, \revision{\Join_{F \in \hyperedges_I} \pi_{I} (R_F \ltimes s)}} 1 \,\,\, (\because \SampleSpace_I \supseteq \, \revision{\Join_{F \in \hyperedges_I} \pi_{I} (R_F \ltimes s)}) \\
&= \abs{\revision{\Join_{F \in \hyperedges_I} \pi_{I} (R_F \ltimes s)}} \\
&= \revtext{\OUT(\revision{\hypergraph_s})}. \,\,\,\,\,\,\,\,\,\,\,\,\,\,\,\,\,\,\, (\because \revision{I = \rerevision{\outputnodes}}) \\
\end{split}
\end{equation}

Note that the assumption $\SampleSpace_I \supseteq \, \revision{\Join_{F \in \hyperedges_I} \pi_{I} (R_F \ltimes s)}$ in Lemma \ref{lemma:unbiased} is used.
If $\SampleSpace_I \not\supseteq \, \revision{\Join_{F \in \hyperedges_I} \pi_{I} (R_F \ltimes s)}$, then $\E[Z_{\revision{\hypergraph_s}}(s_I)] =  \sum_{s_I \in \SampleSpace_I} \I[s_I \in \, \allowbreak \revision{\Join_{F \in \hyperedges_I} \pi_{I} (R_F \ltimes s)}]$ < $\abs{\revision{\Join_{F \in \hyperedges_I} \pi_{I} (R_F \ltimes s)}}$, leading to a negative bias.

For the inductive case where \revision{$I \subsetneq \rerevision{\outputnodes}$}, we use the law of total expectation and our induction hypothesis at Line \ref{alg:sampler:return2}, achieving 
\vspace*{-0.2cm}

\begin{equation}
\begin{split}
    \E[&Z_{\revision{\hypergraph_s}}(s_I)] = \E[\E[Z_{\revision{\hypergraph_s}}|s_I]] \\
    &= \E\Big[\E\Big[ \frac{1}{P(s_I)} \cdot {\I[s_I]} \cdot \revtext{Z_{\revision{\hypergraph_{s \uplus s_I}}}} \Big| s_I\Big]\Big] \\
    &= \E\Big[ \frac{1}{P(s_I)} \cdot {\I[s_I]} \cdot \revtext{\E[Z_{\revision{\hypergraph_{s \uplus s_I}}}} | s_I] \Big] \\
    &= \E\Big[ \frac{1}{P(s_I)} \cdot {\I[s_I]} \cdot \revtext{\OUT({\revision{\hypergraph_{s \uplus s_I}}})} \Big] \,\,\, (\because \text{induction hypothesis}) \\
    &= \sum_{s_I \in \SampleSpace_I} P(s_I) \cdot \frac{1}{P(s_I)} \cdot {\I[s_I]} \cdot \revtext{\OUT({\revision{\hypergraph_{s \uplus s_I}}})} \\
    &= \sum_{s_I \in \SampleSpace_I} {\I[s_I]} \cdot \revtext{\OUT({\revision{\hypergraph_{s \uplus s_I}}})} \\
    &= \sum_{s_I \in  \, \revision{\Join_{F \in \hyperedges_I} \pi_{I} (R_F \ltimes s)}} \revtext{\OUT({\revision{\hypergraph_{s \uplus s_I}}})} \,\,\, (\because \SampleSpace_I \supseteq \, \revision{\Join_{F \in \hyperedges_I} \pi_{I} (R_F \ltimes s)}) \\
    &= \revtext{\OUT(\revision{\hypergraph_s})}.
\end{split}
\end{equation}

The assumption $\SampleSpace_I \supseteq \, \revision{\Join_{F \in \hyperedges_I} \pi_{I} (R_F \ltimes s)}$ is used again.
The last equality holds from Lines \ref{alg:genericjoin:foreach}-\ref{alg:genericjoin:R} of \GenericJoin (Algorithm \ref{alg:genericjoin}); \revision{$s_I$ iterates over $\revision{\Join_{F \in \hyperedges_I} \pi_{I} (R_F \ltimes s)}$, not sampled from $\Omega_I$.}
Therefore, $\E[Z_{\revision{\hypergraph_s}}] = \E[Z_{\revision{\hypergraph_s}}(s_I)] = \revtext{\OUT({\revision{\hypergraph_s}})}$.


\end{proof}




\begin{proof}[\textbf{Proof of Lemma \ref{lemma:psmall}}]

We first rewrite the AGM bounds in LHS using their definitions. Given an optimal fractional edge cover $x$ of $\hypergraph$, the LHS is

\begin{equation}
\begin{split}
\revision{\frac{AGM(\hypergraph_{\sappend})}{AGM(\hypergraph_s)}} =
\revision{\frac{  \prod_{F \in \hyperedges_{\rerevision{\outputnodes} \setminus I}} \abs{R_F \ltimes \sappend}^{x_F} }{  \prod_{F \in \hyperedges_{\rerevision{\outputnodes}}} \abs{R_F \ltimes s}^{x_F}  }}.
\end{split}
\end{equation}


\revision{We partition 
1) $\hyperedges_{\rerevision{\outputnodes} \setminus I}$ into $(\hyperedges_I \cap \hyperedges_{\rerevision{\outputnodes} \setminus I}) \cup (\hyperedges_{\rerevision{\outputnodes} \setminus I} \setminus \hyperedges_I)$
and 2) $\hyperedges_{\rerevision{\outputnodes}}$ into $\hyperedges_I \cup (\hyperedges_{\rerevision{\outputnodes} \setminus I} \setminus \hyperedges_I)$ since any $F \in \hyperedges_{\rerevision{\outputnodes}}$ must contain an attribute in $I$ or $\rerevision{\outputnodes} \setminus I$.}

\second{Since $s_I \in \, \Join_{F \in \hyperedges_I} \pi_{I} (R_F \ltimes s)$, $\abs{R_F \ltimes \sappend} \geq 1 : \forall F \in \hyperedges_I$. Chen \& Yi \cite{GJSampler} also compute the probability of $s_I$ only if $s_I$ contributes to any join answer.}



\begin{equation}
\begin{split}\label{eq:reviewer2}
&\revision{\frac{  \prod_{F \in \hyperedges_{\rerevision{\outputnodes} \setminus I}} \abs{R_F \ltimes \sappend}^{x_F} }{  \prod_{F \in \hyperedges_{\rerevision{\outputnodes}}} \abs{R_F \ltimes s}^{x_F}  }} \\
&= \revision{\frac{  \prod_{F \in \hyperedges_I \cap \hyperedges_{\rerevision{\outputnodes} \setminus I}} \abs{R_F \ltimes \sappend}^{x_F} }{  \prod_{F \in \hyperedges_I} \abs{R_F \ltimes s}^{x_F}  }}
\cdot
\revision{\frac{  \prod_{F \in \hyperedges_{\rerevision{\outputnodes} \setminus I} \setminus \hyperedges_I} \abs{R_F \ltimes \sappend}^{x_F} }{  \prod_{F \in \hyperedges_{\rerevision{\outputnodes} \setminus I} \setminus \hyperedges_I} \abs{R_F \ltimes s}^{x_F}  }}
\\
&= \revision{\frac{  \prod_{F \in \hyperedges_I \cap \hyperedges_{\rerevision{\outputnodes} \setminus I}} \abs{R_F \ltimes \sappend}^{x_F} }{  \prod_{F \in \hyperedges_I} \abs{R_F \ltimes s}^{x_F}  }} \\
& \;\;\;\;\;\; (\revision{\because R_F \ltimes \sappend = R_F \ltimes s : \forall F \in \hyperedges_{\rerevision{\outputnodes} \setminus I} \setminus \hyperedges_I}) \\
&\leq \revision{\frac{  \prod_{F \in \hyperedges_I} \abs{R_F \ltimes \sappend}^{x_F} }{  \prod_{F \in \hyperedges_I} \abs{R_F \ltimes s}^{x_F}  }} \;\;\; (\revision{\because \abs{R_F \ltimes \sappend} \geq 1 : \forall F \in \hyperedges_I \setminus \hyperedges_{\rerevision{\outputnodes} \setminus I}}) \\
&= \revision{\prod_{F \in \hyperedges_I} rdeg_{F,s}(s_I)^{x_F}} \\
&\leq \revision{\prod_{F \in \hyperedges_I} rdeg_{F^*,s}(s_I)^{x_F}} \;\; (\revision{\because rdeg_{F,s}(s_I) \leq rdeg_{F^*,s}(s_I)}) \\
&= \revision{rdeg_{F^*,s}(s_I)^{\sum_{F \in \hyperedges_I} x_F}} \\
&\leq \revision{rdeg_{F^*,s}(s_I)} \;\; \Big( \because \sum_{F \in \hyperedges_I} x_F \geq 1, \revision{rdeg_{F^*,s}(s_I)} \leq 1 \Big).
\end{split}
\end{equation}

\end{proof}



\begin{proof}[\textbf{Proof of Proposition \ref{prop:variance:sste}}]


{We use an induction hypothesis $\Var[Z_\hypergraph] \leq 2^d \cdot AGM(\hypergraph) \cdot \OUT(\hypergraph)$ on $d$, the number of odd cycles and stars in $\hypergraph$.}


For the base case when $d = 1$, $\hypergraph$ consists of either a single odd cycle or a star.
\revision{As explained in \cref{appendix:decomposition}, the maximum recursion depth is 2 + 1. We use $I_1$ (and $k_1$) and $I_2$ (and $k_2$) to denote the $I$ (and $k$) values of the first two recursions, where \rerevision{$\outputnodes = \emptyset$} in the last recursion so Lines \ref{alg:sampler:if}-\ref{alg:sampler:return1} of Algorithm \ref{alg:sampler} are executed.}

If $\hypergraph$ is an odd cycle $C$ of $2n+1$ vertices $\{u_1, v_1, u_2, v_2, ..., u_n, v_n, w\}$, \bluetext{$AGM(\hypergraph) = |R|^{n+\frac{1}{2}}$ from Lemma \ref{lemma:decomposition},} \revision{$I_1 = \{u_1, v_1, ..., u_n, v_n\}, I_2 = \{w\}$, $\SampleSpace_{I_1} = \prod_{i \in [1,n]} R_{F_i}$ and $P(s_{I_1}) = \frac{1}{|R|^n} : \forall s_I \in \SampleSpace_{I_1}$, where $F_i = \{u_i, v_i\}$. Recall that $R_{F_i} = R$ for $i \in [1,n]$.}


\revtext{We explain \SSTE first.}
{Since \revision{$k_1 = 1$ and $s = \emptyset$}, $\Var[Z_\hypergraph] = \E[\Var[Z_\hypergraph | s_{I_1}]] + \Var[\E[Z_\hypergraph | s_{I_1}]]$ from (\ref{eq:lawoftotal}). 
For the first term,}

\begin{equation} \label{eq:sste:expvar:cycle}
\begin{split}
\E[\Var[&Z_\hypergraph|s_{I_1}] ] \\
=& \, \frac{1}{|R|^n} \sum_{s_{I_1}} \Var[Z_\hypergraph | s_{I_1}]   \,\,\, \Big(\because P(s_{I_1}) = \frac{1}{|R|^n} \Big) \\
=& \, \frac{1}{|R|^n} \sum_{s_{I_1}} \Var \Big[\frac{1}{P(s_I)} \I[s_{I_1}] Z_{\revtext{\hypergraph_{s_{I_1}}}} \Big] \\
=& \, |R|^n \sum_{s_{I_1}} \I[s_{I_1}] \Var[Z_{\revtext{\hypergraph_{s_{I_1}}}}] \\
=& \, |R|^n \sum_{s_{I_1}} \frac{\I[s_{I_1}] }{k_{I_2}} \Var[Z_{\revtext{\hypergraph_{s_{I_1}}}}(s_{I_2})] \,\,\, (\because (\ref{eq:totalvariance})) \\
\leq & \, |R|^n \sum_{s_{I_1}} \frac{\I[s_{I_1}] }{k_{I_2}} \E[Z^2_{\revtext{\hypergraph_{s_{I_1}}}}(s_{I_2})] \\
= & \, |R|^n \sum_{s_{I_1}} \frac{\I[s_{I_1}] }{k_{I_2}} \sum_{s_{I_2}} \frac{1}{|\SampleSpace_{I_2}|} Z^2_{\revtext{\hypergraph_{s_{I_1}}}}(s_{I_2})  \,\,\, \Big(\because P(s_{I_2}) = \frac{1}{|\SampleSpace_{I_2}|} \Big) \\
= & \, |R|^n \sum_{s_{I_1}} \frac{\I[s_{I_1}] }{k_{I_2}} \sum_{s_{I_2}} |\SampleSpace_{I_2}| \I[s_{I_2}] \,\,\, \Big(\because Z^2_{\hypergraph_{s_{I_1}}}(s_{I_2}) = |\SampleSpace_{I_2}| \I[s_{I_2}] 1 \Big) \\
\leq & \, |R|^{n+\frac{1}{2}}  \sum_{s_{I_1}} \sum_{s_{I_2}} \I[s_{I_1}] \I[s_{I_2}] \,\,\, \Bigg(\because k_{I_2} = \ceil{\frac{|\SampleSpace_{I_2}|}{|R|^\frac{1}{2}}} \geq \frac{|\SampleSpace_{I_2}|}{|R|^\frac{1}{2}} \Bigg) \\
= & \, |R|^{n+\frac{1}{2}} \OUT(\hypergraph) = AGM(\hypergraph) \cdot \OUT(\hypergraph) \,\,\, (\because \text{Lemma \ref{lemma:decomposition}}).
\end{split}
\end{equation}


\revtext{For the second term of the total variance, the new proof of \SSTE reduces the $\OUT({\hypergraph_{s_{I_1}}})$ factor in (\ref{eq:varexp}) into $\sqrt{|R|}$. We have $\OUT({\hypergraph_{s_{I_1}}}) \leq |\SampleSpace_{I_2}| = |\pi_w(R_{F_0} \ltimes a_1)|$ for $a_1$ \revision{denoting the sampled data vertex for $u_1$.} If $|\pi_w(R_{F_0} \ltimes a_1)| \leq \sqrt{|R|}$, then $\OUT({\hypergraph_{s_{I_1}}}) \leq \sqrt{|R|}$. Otherwise, recall that $a_1$ is chosen to have the smallest degree among the cycle vertices. Then, there can be at most $\sqrt{|R|}$ vertices in the data graph with degrees larger than $a_1$; using the proof by contradiction, the total number of edges will be larger than $|R|$ if more than $\sqrt{|R|}$ vertices have larger degrees than $a_1$. Hence, we have at most $\sqrt{|R|}$ results for the cycle given $s_{I_1}$, i.e., $\OUT({\hypergraph_{s_{I_1}}}) \leq \sqrt{|R|}$:}

\begin{equation} \label{eq:sste:varexp:cycle}
\begin{split}
\Var[\E[& Z_\hypergraph|s_{I_1}]] \leq \sum_{s_{I_1}} \frac{1}{P(s_{I_1})} \I[s_{I_1}] \OUT({\hypergraph_{s_{I_1}}})^2 \,\,\,\,\, (\because (\ref{eq:varexp})) \\
= & \, |R|^n \sum_{s_{I_1}} \I[s_{I_1}] \OUT({\hypergraph_{s_{I_1}}})^2 \leq |R|^n \sum_{s_{I_1}} \I[s_{I_1}] \OUT_{\hypergraph_{s_{I_1}}} \sqrt{|R|} \\
= & \, |R|^{n+\frac{1}{2}} \OUT(\hypergraph) = AGM(\hypergraph) \cdot \OUT({\hypergraph}).
\end{split}
\end{equation}

\revtext{Adding (\ref{eq:sste:expvar:cycle}) and (\ref{eq:sste:varexp:cycle}) completes the proof for \SSTE given an odd cycle.}

\revtext{For \SUST, $\Var[\E[Z_\hypergraph|s_{I_1}]$ is bounded by $AGM(\hypergraph) \cdot \OUT(\hypergraph)$ from (\ref{eq:sste:varexp:cycle}) and the following bound for $\E[\Var[Z_\hypergraph|s_{I_1}]]$ completes the proof for \SUST given an odd cycle.}

\vfill\null

\begin{equation}
\begin{split}
\E[\Var[&Z_\hypergraph|s_{I_1}]] \\
=& \, |R|^n \sum_{s_{I_1}} \I[s_{I_1}] \Var[Z_{\revtext{\hypergraph_{s_{I_1}}}}] \,\,\, (\because (\ref{eq:sste:expvar:cycle})) \\
\leq & \, |R|^n \sum_{s_{I_1}} \I[s_{I_1}] \E[Z^2_{\revtext{\hypergraph_{s_{I_1}}}}] = |R|^n \sum_{s_{I_1}} \I[s_{I_1}] \sum_{s_{I_2}} \frac{1}{P(s_{I_2})} \I[s_{I_2}] \\
= & \, |R|^{n+\frac{1}{2}} \sum_{s_{I_1}} \sum_{s_{I_2}} \I[s_{I_1}] \I[s_{I_2}] \Big(\because P(s_{I_2}) = \frac{1}{\sqrt{|R|}}\Big) \\
= & \, AGM(\hypergraph) \cdot \OUT(\hypergraph).
\end{split}
\end{equation}


{If $\hypergraph$ is a star $S$ of $n$ edges $\{F_1, F_2, ..., F_n\}$, $AGM(\hypergraph) = |R|^n$ from Lemma \ref{lemma:decomposition}. \revtext{For \SSTE,} $I_1$ contains the center attribute $A$ of $S$ only, $\SampleSpace_{I_1} = \pi_A R$, $I_2$ is the set of attributes of $S$ except $I_1$, and $\SampleSpace_{I_2} = \prod_i \pi_{I_2} (R_{F_i} \ltimes s_{I_1})$ for a sampled center vertex $s_{I_1}$.}

{Again, $\Var[Z_\hypergraph] = \E[\Var[Z_\hypergraph | s_{I_1}]] + \Var[\E[Z_\hypergraph | s_{I_1}]]$ since $k_{I_1} = 1$. For the first term,}

\begin{equation}\label{eq:sste:expvar:star}
\begin{split}
\E[\Var[&Z_\hypergraph|s_{I_1}]] \\
= & \sum_{s_{I_1}} \frac{|R \ltimes s_{I_1}|}{|R|} \Var[Z_\hypergraph|s_{I_1}] \,\,\, \Big(\because P(s_{I_1}) = \frac{|R \ltimes s_{I_1}|}{|R|}\Big) \\
\leq & \sum_{s_{I_1}} \frac{|R \ltimes s_{I_1}|}{|R|} \E[Z^2_\hypergraph|s_{I_1}] \\
= & \sum_{s_{I_1}} \frac{|R \ltimes s_{I_1}|}{|R|} \sum_{s_{I_2}} \frac{1}{|R \ltimes s_{I_1}|^n} Z^2_\hypergraph|s_{I_1} \,\,\, \Big(\because P(s_{I_2}) = \frac{1}{|R \ltimes s_{I_1}|^n} \Big) \\
= & \sum_{s_{I_1}} \frac{|R \ltimes s_{I_1}|}{|R|} \sum_{s_{I_2}} \frac{1}{|R \ltimes s_{I_1}|^n}  \Big( \frac{1}{P(s_{I_1})} \I[s_{I_1}] Z_{\revtext{\hypergraph_{s_{I_1}}}} \Big)^2 \\
= & \sum_{s_{I_1}} \frac{|R \ltimes s_{I_1}|}{|R|} \sum_{s_{I_2}} \frac{1}{|R \ltimes s_{I_1}|^n}  \Big( \frac{1}{P(s_{I_1})} \I[s_{I_1}] \frac{1}{P(s_{I_2})} \I[s_{I_2}] \Big)^2 \\
= & \sum_{s_{I_1}} \frac{|R \ltimes s_{I_1}|}{|R|} \sum_{s_{I_2}} \frac{1}{|R \ltimes s_{I_1}|^n}  \Big( \frac{|R|}{|R \ltimes s_{I_1}|} \I[s_{I_1}] |R \ltimes s_{I_1}|^{n} \I[s_{I_2}] \Big)^2 \\
= & \, |R| \sum_{s_{I_1}} |R \ltimes s_{I_1}|^{n-1} \sum_{s_{I_2}} \I[s_{I_1}] \I[s_{I_2}] \\
\leq & \, |R| |R|^{n-1} \sum_{s_{I_1}} \sum_{s_{I_2}} \I[s_{I_1}] \I[s_{I_2}] \\
= & \, |R|^{n} \OUT(\hypergraph) = AGM(\hypergraph) \cdot \OUT(\hypergraph) \,\,\, (\because \text{Lemma \ref{lemma:decomposition}}).
\end{split}
\end{equation}

{For the second term of the total variance, $\Var[\E[Z_\hypergraph|s_{I_1}]] \leq \OUT(\hypergraph)^2 \leq AGM(\hypergraph) \cdot \OUT(\hypergraph)$ from (\ref{eq:varexp}).
Therefore, we have $\Var[Z_\hypergraph] \leq 2 \cdot AGM(\hypergraph) \cdot \OUT(\hypergraph)$ \revtext{for \SSTE given a star.}}

\revtext{For \SUST, $I_1$ is the set of attributes of the star, $\SampleSpace_{I_1} = R^n$, and $P(s_{I_1}) = \frac{1}{|R|^n}$. Thus, $\Var[Z_\hypergraph] \leq \E[Z^2_\hypergraph] = \sum_{s_{I_1}} \frac{1}{P(s_{I_1})} \cdot \I[s_{I_1}] \cdot 1 = |R|^n \cdot \OUT(\hypergraph) = AGM(\hypergraph) \cdot \OUT(\hypergraph)$.}

\revtext{For the inductive case when $d > 1$, $\hypergraph$ consists of multiple odd cycles and stars. The proofs for both \SSTE and \SUST are naive expansions of (\ref{eq:sste:expvar:cycle})-(\ref{eq:sste:expvar:star}) conditioned on multiple components instead of a part of an odd cycle or a star. Please refer to \cite{SSTE} for more details.}

\end{proof}

\begin{proof}[\textbf{Proof of Proposition \ref{prop:variance:alley}}]


\revision{We use an induction hypothesis $\Var[Z_{\revision{\hypergraph_s}}] \leq \frac{t^{|\rerevision{\outputnodes}|} - t}{t - 1} \cdot {\OUT(\hypergraph_s)^2}$.}

For the base case \revision{$I = \rerevision{\outputnodes}$} and $|I| = 1$, $s_I \in \SampleSpace_I = \revision{\Join_{F \in \hyperedges_I} \pi_{I} (R_F \ltimes s)}$, and $Z_{\revision{\hypergraph_s}}(s_I) = \frac{1}{P(s_I)} = |\Omega_I|$ regardless of $s_I$. Therefore, $\Var[Z_{\revision{\hypergraph_s}}] = 0 \leq \revision{\frac{t^{|\rerevision{\outputnodes}|} - t}{t - 1} \cdot \revtext{\OUT(\hypergraph_s)^2}}$.

For the inductive case \revision{$I \subsetneq \rerevision{\outputnodes}$}, 

\vspace*{-0.2cm}
\begin{equation}\label{eq:temp3}
\begin{split}
\E[&\Var[ Z_{\revision{\hypergraph_s}}|s_I]] \\
&= \sum_{s_I} \frac{1}{|\SampleSpace_I|} \Var[Z_{\revision{\hypergraph_s}}|s_I] \\
&= \sum_{s_I} \frac{1}{|\SampleSpace_I|} \Var[|\SampleSpace_I| \cdot \I[s_I] \cdot \revtext{Z_{\revision{\hypergraph_{\sappend}}}}] \\
&= \sum_{s_I} |\SampleSpace_I| \Var[\revtext{Z_{\revision{\hypergraph_{\sappend}}}}] \,\,\, (\because \I[s_I] = 1) \\
&\leq \sum_{s_I} |\SampleSpace_I| \revision{\frac{t^{|\rerevision{\outputnodes} \setminus I|} - t}{t - 1} \revtext{\OUT({\hypergraph_{\sappend}})^2}} \,\,\, (\because \text{induction hypothesis}).
\end{split}
\end{equation}

\bluetext{From (\ref{eq:varexp}), $\Var[\E[Z_{\revision{\hypergraph_s}}|s_I]] \leq \sum_{s_I} \revtext{|\SampleSpace_I|} \OUT(\revision{\hypergraph_{\sappend}})^2$. By plugging this and (\ref{eq:temp3}) into (\ref{eq:lawoftotal}), we have

\vspace*{-0.2cm}
\begin{equation}\label{eq:temp100}
\begin{split}
\Var&[Z_{\revision{\hypergraph_s}}(s_I)] \\
&\leq \sum_{s_I} \Big(|\SampleSpace_I| \revision{\frac{t^{|\rerevision{\outputnodes} \setminus I|} - t}{t - 1}} + \revtext{|\SampleSpace_I|}\Big) \OUT(\revision{\hypergraph_{\sappend}})^2 \\
&= |\SampleSpace_I| \revision{\frac{t^{|\rerevision{\outputnodes} \setminus I|} - 1}{t - 1}} \sum_{s_I} \OUT(\revision{\hypergraph_{\sappend}})^2.
\end{split}
\end{equation}

Therefore, from (\ref{eq:totalvariance}),}

\vspace*{-0.2cm}
\begin{equation}\label{eq:temp4}
\begin{split}
  \Var&[Z_{\revision{\hypergraph_s}}] = \frac{1}{k} \Big(1 - \frac{k - 1}{|\SampleSpace_I| - 1} \Big) \Var[Z_{\revision{\hypergraph_s}}(s_I)] \\
  &\leq \frac{1}{k} \Big(1 - \frac{k - 1}{|\SampleSpace_I| - 1}\Big) |\SampleSpace_I| \revision{\frac{t^{|\rerevision{\outputnodes} \setminus I|} - 1}{t - 1}} \sum_{s_I} \OUT(\revision{\hypergraph_{\sappend}})^2 \\
  &= \frac{|\SampleSpace_I| - k}{k (|\SampleSpace_I| - 1)} |\SampleSpace_I| \revision{\frac{t^{|\rerevision{\outputnodes} \setminus I|} - 1}{t - 1}} \sum_{s_I} \OUT(\revision{\hypergraph_{\sappend}})^2 \\
  &\leq \frac{1-b}{b} \frac{|\SampleSpace_I|}{|\SampleSpace_I| - 1} \revision{\frac{t^{|\rerevision{\outputnodes} \setminus I|} - 1}{t - 1}} \sum_{s_I} \OUT(\revision{\hypergraph_{\sappend}})^2 \\ 
  & \;\;\;\;\;\; (\because k = \ceil{b \cdot |\SampleSpace_I|} \geq b |\SampleSpace_I|) \\
  &\leq \frac{2(1-b)}{b} \revision{\frac{t^{|\rerevision{\outputnodes} \setminus I|} - 1}{t - 1}} \sum_{s_I} \OUT(\revision{\hypergraph_{\sappend}})^2 \,\,\, \Big(\because \frac{|\SampleSpace_I|}{|\SampleSpace_I| - 1} \leq 2 \Big) \\
  &= t \revision{\frac{t^{|\rerevision{\outputnodes} \setminus I|} - 1}{t - 1}} \sum_{s_I} \OUT(\revision{\hypergraph_{\sappend}})^2 \,\,\, \Big(\because t = \frac{2(1-b)}{b}\Big) \\
  &\leq t \revision{\frac{t^{|\rerevision{\outputnodes} \setminus I|} - 1}{t - 1}} \Big( \sum_{s_I} \OUT(\revision{\hypergraph_{\sappend}}) \Big)^2 \\
  &= \revision{\frac{t^{|\rerevision{\outputnodes}|} - t}{t - 1}} \OUT(\revision{\hypergraph_s})^2.
\end{split}
\end{equation}

\end{proof}


\vfill\null

\begin{proof}[\textbf{Proof of Proposition \ref{prop:variance:gjsampler}}]

\revision{We use an induction hypothesis $\Var[Z_{\revision{\hypergraph_s}}] \leq |\rerevision{\outputnodes}| \cdot  AGM(\hypergraph_s) \cdot \OUT(\hypergraph_s)$.}

For the base case \revision{$I = \rerevision{\outputnodes}$} and $|I| = 1$, $AGM(\revision{\hypergraph_{\sappend}}) = 1$, for \revision{$\rerevision{\outputnodes} \setminus I = \emptyset$}. Hence,

\begin{equation}\label{eq:gjsampler:base}
\begin{split}
\Var[Z_{\revision{\hypergraph_s}}] &\leq \E[Z^2_{\revision{\hypergraph_s}}] \\
&= \sum_{s_I} \frac{AGM({\revision{\hypergraph_{\sappend}}})}{AGM({\revision{\hypergraph_s}})} \Big(\frac{AGM({\revision{\hypergraph_s}})}{AGM({\revision{\hypergraph_{\sappend}}})}\Big)^2 \I[s_I] \\
&= \sum_{s_I} \frac{AGM({\revision{\hypergraph_s}})}{AGM({\revision{\hypergraph_{\sappend}}})} \I[s_I] \\
&= AGM({\revision{\hypergraph_s}}) \sum_{s_I} \I[s_I] \;\;\; (\because AGM(\revision{\hypergraph_{\sappend}}) = 1) \\
&= AGM({\revision{\hypergraph_s}}) \cdot \OUT({\revision{\hypergraph_s}}).
\end{split}
\end{equation}

For the inductive case \revision{$I \subsetneq \rerevision{\outputnodes}$}, we again use the induction hypothesis:

\begin{equation}\label{eq:gjsampler:evar}
\begin{split}
\E&[\Var[Z_{\revision{\hypergraph_s}}|s_I]] \\
&= \sum_{s_I} \frac{AGM({\revision{\hypergraph_{\sappend}}})}{AGM({\revision{\hypergraph_s}})} \Var[Z_{\revision{\hypergraph_s}}|s_I] \\
&= \revtext{\sum_{s_I} \frac{AGM({\revision{\hypergraph_{\sappend}}})}{AGM({\revision{\hypergraph_s}})} \Var\Big[\frac{1}{P(s_I)} \cdot \I[s_I] \cdot \revtext{Z_{\revision{\hypergraph_{\sappend}}}}\Big]} \\
&= \sum_{s_I} \frac{AGM({\revision{\hypergraph_s}})}{AGM({\revision{\hypergraph_{\sappend}}})} \I[s_I] \Var[Z_{\revision{\hypergraph_{\sappend}}}] \;\; \Big( \because P(s_I) = \frac{AGM({\revision{\hypergraph_{\sappend}}})}{AGM({\revision{\hypergraph_s}})} \Big) \\
&\leq \sum_{s_I} \frac{AGM({\revision{\hypergraph_s}})}{AGM({\revision{\hypergraph_{\sappend}}})} \I[s_I] \revision{|\rerevision{\outputnodes} \setminus I|} AGM({\revision{\hypergraph_{\sappend}}}) \OUT({\revision{\hypergraph_{\sappend}}}) \\
& \;\;\;\;\; (\because \text{induction hypothesis}) \\
&= \revision{|\rerevision{\outputnodes} \setminus I|} AGM({\revision{\hypergraph_s}}) \sum_{s_I} \I[s_I] \OUT({\revision{\hypergraph_{\sappend}}}) \\
&= \revision{|\rerevision{\outputnodes} \setminus I|} \cdot AGM({\revision{\hypergraph_s}}) \cdot \OUT({\revision{\hypergraph_s}}).
\end{split}
\end{equation}

From (\ref{eq:varexp}), \revtext{we have}

\begin{equation}\label{eq:gjsampler:vare}
\begin{split}
\revtext{\Var[\E[& Z_{\revision{\hypergraph_s}}|s_I]]} \\
&\leq \revtext{\sum_{s_I} \frac{AGM({\revision{\hypergraph_s}})}{AGM({\revision{\hypergraph_{\sappend}}})}\I[s_I] \OUT({\revision{\hypergraph_{\sappend}}})^2} \\
&\leq \revtext{\sum_{s_I} \frac{AGM({\revision{\hypergraph_s}})}{AGM({\revision{\hypergraph_{\sappend}}})}\I[s_I] AGM({\revision{\hypergraph_{\sappend}}}) \OUT({\revision{\hypergraph_{\sappend}}})} \\
&= \revtext{AGM({\revision{\hypergraph_s}}) \sum_{s_I} \I[s_I] \OUT({\revision{\hypergraph_{\sappend}}})} \\
&= \revtext{AGM({\revision{\hypergraph_s}}) \cdot \OUT({\revision{\hypergraph_s}})}.
\end{split}
\end{equation}

\revtext{Therefore, plugging these into (\ref{eq:lawoftotal}) and setting $k = 1$ in (\ref{eq:totalvariance}) give $\Var[Z_{\revision{\hypergraph_s}}] \leq \revision{|\rerevision{\outputnodes}|} \cdot AGM({\revision{\hypergraph_s}}) \cdot \OUT({\revision{\hypergraph_s}})$.}
In addition, $|\hypernodes|$ in the proposition can be removed with a simple proof by \revtext{Chen \& Yi} \cite{GJSampler} \revtext{using the variance of Binomial distribution}.

\end{proof}


\begin{proof}[\textbf{Proof of Proposition \ref{prop:variance:ours}}]

\revision{We use an induction hypothesis $\Var[Z_{\revision{\hypergraph_s}}] \leq |\rerevision{\outputnodes}| \cdot \prod_{I \subset \revision{\rerevision{\outputnodes}}, |I| = 1} |\hyperedges_I| \cdot AGM(\hypergraph_s) \cdot \OUT(\hypergraph_s)$,} similar to the proof of Proposition \ref{prop:variance:gjsampler}. 

For the base case \revision{$I = \rerevision{\outputnodes}$} and $|I| = 1$,

\vspace*{-0.2cm}
\begin{equation}\label{eq:temp:ours}
\begin{split}
\Var[&Z_{\revision{\hypergraph_s}}] \leq \E[Z^2_{\revision{\hypergraph_s}}] \\
&= \revtext{\sum_{s_I} \frac{AGM({\revision{\hypergraph_{\sappend}}})}{|\hyperedges_I| AGM({\revision{\hypergraph_s}})} \Big(\frac{|\hyperedges_I| AGM({\revision{\hypergraph_s}})}{AGM({\revision{\hypergraph_{\sappend}}})}\Big)^2 \I[s_I]} \\
&= \revtext{\sum_{s_I} \frac{|\hyperedges_I| AGM({\revision{\hypergraph_s}})}{AGM({\revision{\hypergraph_{\sappend}}})} \I[s_I] = |\hyperedges_I| \cdot AGM({\revision{\hypergraph_s}}) \cdot \OUT({\revision{\hypergraph_s}}).}
\end{split}
\end{equation}

\revtext{Note that, compared to (\ref{eq:gjsampler:base}) of \GJSampler, the upper bound of $\Var[Z_{\revision{\hypergraph_s}}]$ has increased by a constant factor $\prod_{I} |\hyperedges_I|$.
Similarly, we can easily show that the upper bounds of both (\ref{eq:gjsampler:evar}) and (\ref{eq:gjsampler:vare}) increase by $\prod_{I \subset \revision{\rerevision{\outputnodes}}, |I| = 1} |\hyperedges_I|$ times, proving the inductive case of Proposition \ref{prop:variance:ours}.
By using the proof by \revtext{Chen \& Yi} \cite{GJSampler}, we can remove the $|\hypernodes|$ factor from the proposition as well.}

\end{proof}


Now, we prove the bounds for $T_\hypergraph$ (or $\E[T_\hypergraph]$). 
\second{In Algorithm \ref{alg:sampler}, Lines \ref{alg:sampler:if}-\ref{alg:sampler:I} take $\O(1)$. Line \ref{alg:sampler:samplespace} takes $\O(\min_{F \in \hyperedges_I} |\pi_I \revision{(R_F \ltimes s)}|)$ time for \AlleyS due to the intersection and $\O(1)$ for the others, \revision{i.e., $\SampleSpace_I$ is readily available from the query model in \cref{subsec:ours:query_model}.}
Line \ref{alg:sampler:P} takes $\O(\IN)$ for \GJSampler for non-sequenceable queries and $\O(1)$ for other cases. 
Line \ref{alg:sampler:sample} takes $\O(k)$. Line \ref{alg:sampler:reject} \revision{takes $\O(1)$; computing relative degrees takes $\O(|\hyperedges_I|) = \O(1)$ in \second{DRS}.} Line \ref{alg:sampler:return2} takes $\O(k) + \sum_{s_I \in S_I} \I[s_I \in \, \Join_{F \in \hyperedges_I} \pi_{I} \revision{(R_F \ltimes s)}] \revtext{T_{\hypergraph_{\sappend}}}$. 
\revision{Note that $\I[s_I \in \, \Join_{F \in \hyperedges_I} \pi_{I} \revision{(R_F \ltimes s)}] = \prod_{F \in \hyperedges_I} \I[s_I \in \pi_{I} \revision{(R_F \ltimes s)}]$ takes $\O(|\hyperedges_I|) = \O(1)$ time to evaluate.}}

\second{From this,} we can recursively define $T_{\revision{\hypergraph_s}}$, the runtime of computing $Z_{\hypergraph_s}$:

\vspace*{-0.3cm}
\begin{equation}\label{eq:time:recursion}
T_{\revision{\hypergraph_s}} = \left\{
\begin{array}{ll}
    \O(1) & \text{if } I = \revision{\rerevision{\outputnodes}} \\
    \O(\IN) +\I[s_I] T_{\revision{\hypergraph_{\sappend}}} & \text{if } I \subsetneq \revision{\rerevision{\outputnodes}}, \\
    & \text{non-sequenceable } \\
    & \text{queries for \GJSampler} \\
    \O(k) + \sum_{s_I \in S_I} \I[s_I] T_{\revision{\hypergraph_{\sappend}}} & \text{otherwise} \\
\end{array} 
\right.    
\end{equation}



\vspace*{-0.2cm}
\begin{proof}[\textbf{Proof of Proposition \ref{prop:time:sste}}]


In \SSTE, $k = \ceil{\frac{|\SampleSpace_I|}{\sqrt{|R|}}}$ if $I$ consists of a single vertex $w$ of an odd cycle and $k = 1$ otherwise.
We prove $\E[k] = \O(1)$ for the first case. Then, if there are $n$ cycles in $\hypergraph$ with $k = k_1, k_2, ..., k_n$ each, $\E[T_\hypergraph] = \O(\prod_i \E[k_i]) = \O(1)$ from (\ref{eq:time:recursion}).

\revision{$k$ for $I = \{w\}$ depends solely on the sampled data edge $\{a_1, b_1\}$ for $F_1 = \{u_1, u_2\}$. Recall that $\SampleSpace_{I} = \pi_I (R_{F_0} \ltimes a_1)$ and $|\pi_I (R_{F_0} \ltimes a_1)| \leq |\pi_I (R_{F_0} \ltimes b_1)|$ since $a_1$ has the smallest degree among the cycle vertices.} The last equality below holds from a well-known fact in graph theory \cite{SSTEFull}.

\vspace*{-0.4cm}
\begin{equation}
\vspace*{-0.2cm}
\begin{split}
\E[k] &= \sum_{\{a_1, b_1\} \in R} \frac{1}{|R|} \ceil{\frac{\min(| \pi_I (R \ltimes a_1) |, | \pi_I (R \ltimes b_1)  |)}{\sqrt{|R|}}} \\ 
&\leq 1 + \sum_{\{a_1, b_1\}} {\frac{1}{|R|} \frac{\min(| \pi_I (R \ltimes a_1) |, | \pi_I (R \ltimes b_1)  |)}{\sqrt{|R|}}} \\
&= \O(1) \,\,\, (\because \text{Proposition 2.2 in Assadi et al. \cite{SSTEFull}})
\end{split}
\end{equation}
\vspace*{-0.3cm}

\end{proof}



\begin{proof}[\textbf{Proof of Proposition \ref{prop:time:alley}}]

Simply, \AlleyS takes $b^{|\hypernodes|}$ portion of all paths searched by \GenericJoin in expectation. Therefore, $\E[T_\hypergraph] = \O(b^{|\hypernodes|} AGM(\hypergraph))$. $T_\hypergraph$ is $\O(AGM(\hypergraph))$ in the worst case, \revtext{since the computation cost of \SamplingFramework for \AlleyS is subsumed by \GenericJoin.}

\end{proof}


\begin{proof}[\textbf{Proof of Proposition \ref{prop:time:gjsampler}}]

\bluetext{In \GJSampler, $k$ is fixed to 1 regardless of $s_I$.
Therefore, $T_{\revision{\hypergraph_s}} = \O(1) + T_{\revision{\hypergraph_{\sappend}}}$ for sequenceable queries and $T_{\revision{\hypergraph_s}} = \O(\IN) + T_{\revision{\hypergraph_{\sappend}}}$ for non-sequenceable queries from (\ref{eq:time:recursion}). Since the maximum recursion depth is $|\hypernodes|$, $T_\hypergraph = \O(|\hypernodes|) = \O(1)$ for sequenceable queries \revtext{(note the excluded preprocessing cost)} and $T_\hypergraph = \O(|\hypernodes| \cdot \IN) = \O(\IN)$ for non-sequenceable queries.}


\end{proof}



\begin{proof}[\textbf{Proof of Proposition \ref{prop:time:ours}}]

\revtext{In \second{DRS}, $k$ is also fixed to 1, and $T_{\revision{\hypergraph_s}} = \O(|\hyperedges_I|) + T_{\revision{\hypergraph_{\sappend}}}$. Hence, $T_\hypergraph = \O(\sum_{I \subset \hypernodes, |I| = 1} |\hyperedges_I|) = \O(1)$. For \SUST, $k$ is also fixed to 1, and $T_{\revision{\hypergraph_s}} = \O(1) + T_{\revision{\hypergraph_{\sappend}}}$. Hence, $T = \O(1)$.}

\end{proof}

\begin{proof}[\textbf{Proof of Lemma \ref{lemma:ghdsampler:unbiased}}]

The unbiasedness of \GHDSampler can be explained by the following (\ref{eq:succinct}), since $R_t$ is an unbiased estimate of
\second{$\hypergraph^{G(t)}_t = \sum_{\bag(t) \setminus G(t)} \hypergraph_t = \sum_{\bag(t) \setminus G(t)} {\Join_{F \in \hyperedges_{\bag(t)}} \pi_{\bag(t)} R_F}$.}
Note that if we replace \SamplingFramework with \GenericJoin \revision{in Algorithm \ref{alg:genericsampler}}, then \GHDSampler will return $\OUT$.

\begin{equation}\label{eq:succinct}
\begin{split}
\OUT &= \sum_{\hypernodes} \Join_{F \in \hyperedges} R_F \\
&= \sum_{\hypernodes} \Join_{t \in \hypernodes_\GHD} \big( {\Join_{F \in \hyperedges_{\bag(t)}} \pi_{\bag(t)} R_F} \big) \,\,\, (\because \text{using GHD } (\GHD, \bag)) \\
&= \sum_{G(\GHD)} \Join_{t \in \hypernodes_\GHD} \Big(\sum_{\bag(t) \setminus G(t)} \Join_{F \in \hyperedges_{\bag(t)}} \pi_{\bag(t)} R_F \Big) \\
&= \sum_{G(\GHD)} \Join_{t} \E[R_t].
\end{split}
\end{equation}

\end{proof}

\begin{proof}[\textbf{Proof of Lemma \ref{lemma:varz}}]

We use an induction hypothesis on $|f|$.
For the base case \bluetext{$|f| = 1$ so $f = \{g_t\}$ for a node $t$}, $\Var[Z[f]]$ = \\
\vspace*{-0.25cm} \\ $\Var[Z[g_t]]$. Therefore, $\Var[Z[f]]$ = $\O(\OUT(\second{\hypergraph_{g_t}})^2)$. \\ \vspace*{-0.2cm}


For the inductive case $|f| > 1$, we use an induction hypothesis that $\Var[Z[f]] = \O(\prod_{g_t \in f} \OUT(\second{\hypergraph_{g_t}})^2)$ for every $|f| \leq n$ for \\
\vspace*{-0.25cm} \\ some $n \geq 1$. For any $f$ with $|f| = n + 1$, we take any $g_t \in f$. Since $g_t$ and $f \setminus g_t$ are independent,


\begin{equation}
\begin{split}
\Var&[Z[f]] \\
&= (\Var[Z[g_t]] + \E[Z[g_t]]^2)(\Var[Z[f \setminus g_t]] + \E[Z[f \setminus g_t]]^2) \\
            & \;\;\;\;\;\;\; - \E[Z[g_t]]^2 \E[Z[f \setminus g_t]]^2 \\
            &= \O(\E[Z[g_t]]^2) \O(\E[Z[f \setminus g_t]]^2) - \E[Z[g_t]]^2 \E[Z[f \setminus g_t]]^2 \\
            & \;\;\;\;\;\; (\because \text{induction hypothesis}) \\
            &= \O(\E[Z[g_t]]^2 \cdot \E[Z[f \setminus g_t]]^2) \\
            &= \O(\E[Z[f]]^2) \,\,\, (\because \text{(\ref{eq:multiexp:prod})}).
\end{split}
\end{equation}


\end{proof}


\begin{proof}[\textbf{Proof of Lemma \ref{lemma:covz}}]

If $f_1$ and $f_2$ are independent ($f_1 \cap f_2 = \emptyset$), then $\Cov(Z[f_1], Z[f_2]) = 0$.
If not, $f_1 \cap f_2$, $f_1 \setminus f_2$, and $f_2 \setminus f_1$ are independent, \bluetext{and $Z[f_1] = Z[f_1 \cap f_2] \cdot Z[f_1 \setminus f_2]$ and $Z[f_2] = Z[f_1 \cap f_2] \cdot Z[f_2 \setminus f_1]$. Therefore, }

\begin{equation}\label{eq:ghdsampler:cov}
\begin{split}
\Cov(Z[f_1], \,& Z[f_2]) = \E[Z[f_1] \cdot Z[f_2]] - \E[Z[f_1]] \cdot \E[Z[f_2]] \\
=& \E[Z[f_1 \cap f_2]^2 \cdot Z[f_1 \setminus f_2] \cdot Z[f_2 \setminus f_1]] \\
&- \E[Z[f_1 \cap f_2]]^2 \cdot \E[Z[f_1 \setminus f_2]] \cdot \E[Z[f_2 \setminus f_1]] \\
=& \E[Z[f_1 \cap f_2]^2] \cdot \E[Z[f_1 \setminus f_2]] \cdot \E[Z[f_2 \setminus f_1]] \\
&- \E[Z[f_1 \cap f_2]]^2 \cdot \E[Z[f_1 \setminus f_2]] \cdot \E[Z[f_2 \setminus f_1]] \\
=& \Var[Z[f_1 \cap f_2]] \cdot \E[Z[f_1 \setminus f_2]] \cdot \E[Z[f_2 \setminus f_1]].
\end{split}
\end{equation}

From Lemma \ref{lemma:varz} and (\ref{eq:multiexp:prod}), the last term is

\begin{equation}\label{eq:ghdsampler:cov:internal}
\begin{split}
\O(& \E[Z[f_1 \cap f_2]]^2) \cdot \E[Z[f_1 \setminus f_2]] \cdot \E[Z[f_2 \setminus f_1]] \\
&= \O\Big(\prod_{g_t \in f_1 \cap f_2} \E[Z[g_t]]^2\Big) \cdot \prod_{g_t \in f_1 \setminus f_2} \E[Z[g_t]] \cdot \prod_{g_t \in f_2 \setminus f_1} \E[Z[g_t]] \\
&= \O\Big(\prod_{g_t \in f_1 \cap f_2} \E[Z[g_t]]^2 \cdot \prod_{g_t \in f_1 \setminus f_2} \E[Z[g_t]] \cdot \prod_{g_t \in f_2 \setminus f_1} \E[Z[g_t]] \Big) \\
&= \O\Big(\prod_{g_t \in f_1} \E[Z[g_t]] \cdot \prod_{g_t \in f_2} \E[Z[g_t]]\Big).
\end{split}
\end{equation}

\end{proof}

\begin{proof}[\textbf{Proof of Proposition \ref{prop:converge}}]

\bluetext{If $\Var[Z[g_t]]$ approaches to 0 for every $g_t \in f$, $\Var[Z[f]]$ approaches to 0 from (\ref{eq:multivar:prod}). 
For any $f_1$ and $f_2$, if $\Var[Z[g_t]]$ approaches to 0 for every $g_t \in f_1 \cap f_2$, $\Var[Z[f_1 \cap f_2]]$ approaches to 0.
If $\Var[Z[f_1 \cap f_2]]$ approaches to 0, $\Cov(Z[f_1], Z[f_2])$ approaches to 0 from (\ref{eq:ghdsampler:cov}). Therefore, $\Var[Z]$ also approaches to 0 from (\ref{eq:ghdsampler:var}).}


\end{proof}

\fi

\end{document}